\documentclass[11pt,a4paper]{article}
\pdfoutput=1
\usepackage{cancel}
\usepackage{enumitem}
\usepackage{etex}
\usepackage{amsthm}
\usepackage{geometry}
\usepackage{dcolumn}
\usepackage{epsf}
\usepackage{mathrsfs}
\usepackage{multirow}
\usepackage{booktabs}
\usepackage{tabularx}
\usepackage{array}
\usepackage{slashed}
\usepackage{float}
\usepackage{leftidx}
\usepackage{setspace}
\usepackage{verbatim}
\usepackage{adjustbox}
\usepackage{tikz}
\usepackage[caption=false]{subfig}
\usepackage{arydshln}
\usepackage{jheppub}
\usepackage{shuffle}
\usepackage{amsmath}
\usepackage{cases}
\usepackage{lipsum}
\usepackage{tabularx}
\usepackage{graphicx}
\usepackage{lscape}
\numberwithin{equation}{section}
\usetikzlibrary{arrows,decorations.markings,shapes.arrows,patterns,positioning}

\newcommand{\bea}{\begin{eqnarray}}
\newcommand{\eea}{\end{eqnarray}}
\newcommand{\bean}{\begin{eqnarray*}}
\newcommand{\eean}{\end{eqnarray*}}
\newcommand{\nn}{\nonumber\\}
\newcommand{\Sl}{\sum\limits}

\def\W #1{\widetilde{#1}}

\def\Label#1{\label{#1}%
  \smash{\hbox to0pt{\raise1ex\hbox{\tiny[#1]}\hss}}}
\def\Label#1{\label{#1}}
\renewcommand{\eqref}[1]{eq.~(\ref{#1})}
\newcommand{\figref}[1]{Fig.~\ref{#1}}
\newcommand{\tabref}[1]{table~\ref{#1}}
\newcommand{\secref}[1]{section~\ref{#1}}
\newcommand{\appref}[1]{appendix~\ref{#1}}

\def\Sl{\sum\limits}

\newcommand{\ctobedelete}[1]{}

\allowdisplaybreaks

\title{Extracting quadratic propagators by refined graphic rule}

\author[a]{Chongsi Xie} \author[a,b]{Yi-Jian Du\footnote{Corresponding author}}

\affiliation[a]{Department of Physics, School of Physics and Technology,
Wuhan University, \\
No.299 Bayi Road, Wuhan 430072, China}
\affiliation[b]{Hubei Key Laboratory of Nuclear Solid Physics, School of Physics and Technology, Wuhan University,\\
No.299 Bayi Road, Wuhan 430072, China}
\emailAdd{chongsi.xie@whu.edu.cn} \emailAdd{yijian.du@whu.edu.cn}

\date{\today}
\abstract{One-loop integrands in Cachazo-He-Yuan (CHY) formula, which is based on the forward limit of tree-level amplitudes, involves linear propagators  that are different from quadratic ones in traditional Feynman diagrams. In this paper, we provide a general approach to converting linear propagators in one-loop CHY formula into quadratic propagators, by refined graphic rule stemming from the recursive expansion of tree-level Einstein-Yang-Mills amplitudes. Particularly, we establish the correspondence between refined graphs and 
bi-adjoint scalar (BS) Feynman diagrams with linear propagators. Using this correspondence and graph-based relations of Berends-Giele currents in BS theory, the nonlocal terms accompanied by refined graphs can either be canceled out or be collected into local ones. Once the locality has been achieved, the integrand with linear propagators can be directly arranged into that with quadratic propagators. Following this approach, we first convert the linear propagators in single-trace Yang-Mills-scalar (YMS) integrands (with a pure-scalar loop) into quadratic ones. This result is then demonstrated to match the traditional one-loop Feynman diagrams. The discussions on single-trace YMS integrands are generalized to multi-trace YMS and Yang-Mills integrands.  }


\begin{document}
\maketitle \flushbottom

\section{Introduction}


In recent decades, there have been many significant progresses on perturbative scattering amplitudes, among which, Cachazo-He-Yuan (CHY) formula \cite{Cachazo:2013gna,Cachazo:2013hca,Cachazo:2013iea,Cachazo:2014nsa,Cachazo:2014xea} provides a compact form of tree-level scattering amplitudes. In particular, according to CHY formula, a tree-level amplitude is expressed as an integral over scattering variables that are constrained by scattering equations. The CHY integrand, which contains the kinematic information of external particles, relies on theories and can be written as a product of two half integrands. CHY formula for various theories, such as bi-adjoint  scalar (BS) theory, Yang-Mills-scalar (YMS) theory, Yang-Mills (YM) theory, Einstein-Yang-Mills (EYM) theory and gravity (GR) have already been established \cite{Cachazo:2013gna,Cachazo:2013hca,Cachazo:2013iea,Cachazo:2014nsa,Cachazo:2014xea}. Moreover, it has also been shown to exist in effective theories \cite{Cachazo:2014xea}.

 A crucial feature of CHY formula is that the color-ordered BS amplitudes, whose CHY integrands are presented as a product of two Parke-Taylor (PT) \cite{Parke:1986gb} factors, play as the backbone of the CHY formula.  CHY half integrands for other theories can be decomposed in terms of a proper  combination of the PT factors. The expansion coefficients are constructed systematically by refined graphic rule \cite{Hou:2018bwm,Du:2019vzf} which benefits from the recursive expansion formula of EYM amplitudes \cite{Stieberger:2016lng,Nandan:2016pya,Schlotterer:2016cxa,Fu:2017uzt,Chiodaroli:2017ngp,Teng:2017tbo,Du:2017gnh}.  As a consequence of the expansion of half integrands, the full amplitudes in other theories are finally written in terms of BS amplitudes. This has been effectively applied to the construction of local Bern-Carrasco-Johansson (BCJ) numerators that satisfy color-kinematics duality \cite{Du:2017kpo}.

The CHY formula was generalized to one-loop amplitudes \cite{Mason:2013sva,He:2015yua,Cachazo:2015aol}, based on an analysis of worldsheet structure. As pointed out in \cite{He:2015yua,Cachazo:2015aol}, the CHY half integrands for an $n$-point one-loop amplitude could be obtained via taking the forward limit of $(n+2)$-point half integrands at tree-level. This observation allows one to convert properties at tree-level to one-loop level. Following the forward limit approach, the one-loop BCJ numerators  have been expressed by tree-level ones \cite{He:2016mzd}. Although the forward limit provides a rather straightforward one-loop generalization of the tree-level CHY formula, the two $(n+2)$-point PT factors with forward limits in fact can only produce propagators whose denominators are linear functions of the loop momentum \cite{He:2015yua,Cachazo:2015aol,Geyer:2015jch,Geyer:2017ela,Edison:2020uzf}. This conflicts with traditional Feynman diagrams, in which the loop propagators involve quadratic function of loop momentum in the denominator. Many efforts have been made on converting the linear propagators  into quadratic ones \cite{He:2015yua,Cachazo:2015aol,Geyer:2015jch,Geyer:2017ela,Edison:2020uzf,Feng:2022wee,Baadsgaard:2015hia,Cardona:2016bpi,Cardona:2016wcr,Gomez:2016cqb,Gomez:2017lhy,Ahmadiniaz:2018nvr,Agerskov:2019ryp,Farrow:2020voh,Porkert:2022efy}. In \cite{Feng:2022wee} among these efforts, a tensorial PT factor was supposed to produce quadratic propagators. As further pointed out in a recent work \cite{Dong:2023stt}, once the CHY half integrands are decomposed into these tensorial PT factors, one can construct BCJ \cite{Bern:2008qj,Bern:2010ue} numerators at one-loop level. Along this line, the critical issue for constructing quadratic propagators becomes decomposing CHY half integrands in terms of tensorial PT factors. Nevertheless, it is still lack of a general way to decompose an arbitrary CHY half integrand into tensorial PT factors.

 In this work, we provide a generic approach to converting one-loop CHY formula for YMS (with a pure scalar loop) and YM with linear propagators into expressions with quadratic propagators, by the help of (refined) graphic rule \cite{Hou:2018bwm,Du:2019vzf,Wu:2021exa}. In particular, since the CHY half integrands at one-loop are expressed by forward limits of $(n+2)$-point half integrands at tree-level, while one of the tree-level YMS half integrands can further be expanded in terms of PT factors according to graphic rule, we express one of the one-loop half integrands as a sum over graphs. Together with the other half integrand  of YMS, the previous step in fact expresses the loop integrand as a combination of linear-propagator Feynman diagrams (LPFD), where Berends-Giele (BG) subcurrents of BS are attached to the line with linear propagators. The combination coefficients are determined by graphs, while a LPFD is associated  with a partition of external particles which is related to decomposing graphs into subgraphs. For a given LPFD accompanied by a given graph, there may exist subgraphs which are contracted with each other but are separated by linear propagators in the LPFD. Such contraction of subgraphs contribute nonlocal terms. The critical point for extracting quadratic propagators is to achieve the locality. We find that the nonlocal terms can either be canceled out by the help of graph-based identities \cite{Hou:2018bwm,Du:2019vzf,Du:2022vsw}, or be collected into local terms. Once the nonlocality has been treated, the integrand turns into sums of structures where BG subcurrents of YM, YMS, BS and/or contractions of two YM subcurrents are planted to the linear-propagator line. This expression of integrand is further arranged into a formula with quadratic propagators, straightforwardly. We show that the formula  of YMS integrand matches the Feynman diagrams with quadratic propagators. As a special case of this approach, contributions of a certain class of graphs, i.e., the graphs with no $\epsilon\cdot\epsilon$ factor, are written in terms of the tensorial PT factors, therefore naturally induce quadratic propagators. For the contributions of more general graphs that are treated in this work, our approach does not provide the tensorial PT factor decomposition, but can  produce quadratic propagators. The extension to multi-trace YMS amplitudes with pure scalar loop are straightforwardly obtained by involving more types of components in the graphs \cite{Du:2019vzf}.  Considering the relationship between YMS and YM, we further provide an expansion formula for YM integrand. In the current work, we only study the YMS amplitudes with a scalar loop. {\it It is worth pointing out that the case considered in this paper does not cover the full single-trace (multi-trace) sector.} There also exist single-trace (multi-trace) integrands which involve gluon propagators on the loop but correspond to different orders of coupling constants. This situation was studied in \cite{Porkert:2022efy}. The integrands proposed in \cite{Porkert:2022efy} are reviewed in \appref{App:KKOneLoop}. We leave the general discussion on the integrands, which contain gluon propagators on the loop, in a future work. 

 The  structure of this paper is following. In \secref{sec:review}, we review helpful results including tree-level and one-loop CHY formulas, the refined graphic rule as well as properties of off-shell BG current in BS. In \secref{sec:SimpleExample}, we study the integrand with two scalars and two gluons, and further propose two approaches for treating the nonlocalities. Helpful patterns, X- and BCJ-patterns,  are observed in \secref{sec:SimpleExample} and are extended to more general cases in \secref{sec:XBCJ}. More complicated examples are studied in \secref{sec:5ptExample}, where pairs of spurious graphs with opposite signs are introduced to cancel the nonlocalities. A general approach to a quadratic-propagator formula of YMS integrand is presented in  \secref{Sec:GenYMS}. In \secref{sec:FeynDiagrams}, we show the formula obtained via graphic rule matches with the result derived from the Feynman diagrams in YMS. We further extend this discussion to multi-trace YMS (with pure scalar loop) and pure YM theory. A summary and further discussions are presented in \secref{sec:Conclusions}. Suppliments of graphic rule, explicit examples, construction of subgraphs, Feynman rule in YMS and a byproduct identity can be found in the appendix.

%

%

\section{CHY formula, refined graphic rule, tensorial PT factors and properties of BS currents} \label{sec:review}
In this section, we review the main results of the tree-level and one-loop CHY formula, the refined graphic rule, the tensorial PT factors as well as properties of BG current in BS theory. These results provide a preparation for the coming discussions.

\subsection{Tree-level CHY}

CHY formula expresses tree-level amplitude with $n$ massless external particles by an integral over scattering variables $\{z_i\}$ $(i=1,...,n)$, as follows
\bea
M_n^{\text{tree}}=\int \text{d}\mu_{\,n}^{\,\text{tree}}\,I_{\,\text{L}}^{\,\text{tree}}\,I_{\,\text{R}}^{\,\text{tree}}.
\Label{Eq:TreeCHY}
\eea
The integration measure in the above is given by 
\bea
\text{d}\mu_{\,n}^{\,\text{tree}}=\frac{\text{d}z_1\dots \text{d}z_n}{\text{vol SL(2,C)}}\,{\prod\limits_{i=1}^{n}}'\delta\left(\Sl_{j=1,j\neq i}^n\frac{s_{ij}}{z_{ij}}\right),~~~~~z_{ij}\equiv z_i-z_j,
\eea
where $s_{ij}\equiv (k_i+k_j)^2=2k_i\cdot k_j$ ($k_i$, $k_j$ are the momenta of the massless particles $i$ and $j$). The ${\prod\limits_{i=1}^{n}}'$ is defined as 
\bea
{\prod\limits_{i=1}^{n}}'\delta\left(\Sl_{j=1,j\neq i}^n\frac{s_{ij}}{z_{ij}}\right)=\prod\limits_{i\neq a,b,c}z_{ab}z_{bc}z_{ca}\delta\left(\Sl_{j=1,j\neq i}^n\frac{s_{ij}}{z_{ij}}\right),
\eea
where $a$, $b$ and $c$ are three arbitrarily chosen particles. The delta functions imposes the scattering equation constraints on the scattering variables
\bea
\Sl_{j=1,j\neq i}^n\frac{s_{ij}}{z_{ij}}=0.
\eea
\begin{table}[]
    \centering
    {\begin{tabular}{|c|c|c|c|c|c|}
        \hline
       Theories   & BS  & YM & YMS & EYM & GR        \\ \hline
      Amplitudes & \small $A^{\text{tree}}_{\text{BS}}(\pmb{\sigma}|\pmb{\rho})$ &\small $A_{\text{YM}}^{\text{tree}}(\pmb{\sigma})$ &\small  $A_{\text{YMS}}^{\text{tree}}\big(\pmb{\sigma}_1;...;\pmb{\sigma}_m||\mathsf{G}\big|\pmb{\rho}_{\mathsf{S}\cup \mathsf{G}}\big)$  & \small $A_{\text{EYM}}^{\text{tree}}\big(\pmb{\sigma}_1;...;\pmb{\sigma}_m||\mathsf{H}\big)$ & \small $M_{\text{GR}}^{\text{tree}}(1,...,n)$  \\ \hline
    $I^{\text{tree}}_{\text{L}}$ &\small $\text{PT}(\pmb{\sigma})$ & \small$\text{PT}(\pmb{\sigma})$ & \small$\text{PT}(\pmb{\sigma}_1)...\text{PT}(\pmb{\sigma}_m)\,\mathcal{P}$ & \small$\text{PT}(\pmb{\sigma}_1)...\text{PT}(\pmb{\sigma}_m)\,\mathcal{P}$&\small $\text{Pf}\,'[\Psi]$\\ \hline
      $I^{\text{tree}}_{\text{R}}$ &\small $\text{PT}(\pmb{\rho})$ & \small $\text{Pf}\,'[\Psi]$ & \small $\text{PT}\left(\pmb{\rho}_{\mathsf{S}\cup \mathsf{G}}\right)$ & \small
    $\text{Pf}\,'[\Psi]$ &\small $\text{Pf}\,'[\Psi]$\\ \hline
    \end{tabular} }
    \caption{Tree-level CHY integrands for BS, YM, YMS, EYM and GR ~~In the single-trace case, $m=1$ and the YMS (EYM) half integrand $I^{\text{tree}}_{\text{L}}$ turns into a single PT factor $\text{PT}(\pmb{\sigma}_1)$ associated with  a Pfaffian $\text{Pf}\,[\Psi]_{\mathsf{G};\mathsf{G}}=\text{Pf}\,[\Psi_{\mathsf{G}}]$  ($\text{Pf}\,[\Psi]_{\mathsf{H};\mathsf{H}}=\text{Pf}\,[\Psi_{\mathsf{H}}]$). } \label{table:TreeIntegrands}
\end{table}
External information including the polarization vectors and momenta, is involved in the half integrands $I_{\text{L}}^{\text{tree}}$ and $I_{\text{R}}^{\text{tree}}$, which rely on theory. For BS, YM, YMS, EYM and GR, the half integrands are presented in \tabref{table:TreeIntegrands}, where the Parke-Taylor factor $\text{PT}(\pmb{\sigma})$ ($\pmb{\sigma}\equiv (\sigma_1...\sigma_n)$ is a permutation of external particles) is defined as follows 
\bea
\text{PT}(\pmb{\sigma})\equiv{1\over {z_{\sigma_1\sigma_2}z_{\sigma_2\sigma_3}...z_{\sigma_{n-1}\sigma_n}z_{\sigma_n\sigma_1}}}.~~~\Label{Eq:PTtree}
\eea
The reduced Pfaffian $\text{Pf}\,'[\Psi]$ is given by
\bea
\text{Pf}\,'[\Psi]\equiv{(-1)^{i+j}\over z_{ij}}\text{Pf}\,\left[\,\Psi_{ij}^{ij}\,\right],~~~~~\Label{Eq:ReducedPf}
\eea
where $\Psi$ is a $2n\times 2n$ antisymmetric matrix 
\[\Psi=\begin{pmatrix}
A&-C^T\\
C&B\\
\end{pmatrix},~~~\Label{Eq:Psi}
\]
where $\Psi_{ij}^{ij}$ means the $i$-, $j$-th rows and columns are deleted, $A$, $B$ and $C$ are $n\times n$ matrices which contain the external kinematic deta
\bea
A_{ab}=\Bigg\{
            \begin{array}{cc}
              \frac{k_a\cdot k_b}{z_{ab}} &, a\ne b \\
               0 &,a=b \\
            \end{array}\,,~~~~
B_{ab}=\Bigg\{
            \begin{array}{cc}
               \frac{\epsilon_a\cdot \epsilon_b}{z_{ab}} &, a\ne b \\
               0 &,a=b \\
            \end{array}\,,~~~~
C_{ab}=\Bigg\{
            \begin{array}{cc}
              \frac{\epsilon_a\cdot k_b}{z_{ab}} &, a\ne b \\
              -\Sl_{c\ne a}\frac{\epsilon_a\cdot k_c}{z_{ac}} &,a=b \\
            \end{array}\,.
\Label{Eq:ABC}
\eea
The $\mathcal{P}$, which exists in the $I_L$ for EYM amplitudes, in \tabref{table:TreeIntegrands} is given by
\bea
\mathcal{P}\equiv\Sl_{\substack{a_1\prec b_1 \in \pmb{1}\\\dots\\a_{m-1}\prec b_{m-1}\in \pmb{m-1}}}sgn(\{a,b\})\,z_{a_1b_1}\dots z_{a_{m-1}b_{m-1}}\text{Pf}\,[\Psi]_{\mathsf{H},a_1,b_1,\dots,a_{m-1},b_{m-1};\mathsf{H}}\,,\Label{Eq:P}
\eea
where $\text{Pf}\,[\Psi]_{\mathsf{H},a_1,b_1,\dots,a_{m-1},b_{m-1};\mathsf{H}}$ stands for the Pfaffian of a matrix which is obtained from $\Psi$ in the following way:  Keep the rows and columns with respect to (i). elements in the graviton set $\mathsf{H}$ and $a_i, b_i$ ($i=1,...,m-1$) pairs for gluon traces, among the first $n$ rows and columns, (ii). elements in $\mathsf{H}$, among the second $n$ rows and columns. The $\mathcal{P}$ for YMS has the same expression (\ref{Eq:P}) but replacing gluons and the graviton set $\mathsf{H}$ by scalars and the gluon set $\mathsf{G}$, respectively.
{\small\begin{table}[]
    \centering
  
   \scalebox{0.95}{\begin{tabular}{|c|c|c|c|c|c|}
        \hline
        \small Theories   & \small BS  & \small YM &\small YMS & \small EYM &\small GR        \\ \hline
      \small Amplitudes & \small $A^{\text{1-loop}}_{\text{BS}}(\pmb{\sigma}|\pmb{\rho})$ &\small $A_{\text{YM}}^{\text{1-loop}}(\pmb{\sigma})$ &\small $A_{\text{YMS}}^{\text{1-loop}}\big(\pmb{\sigma}_1;...;\pmb{\sigma}_m||\mathsf{G}\big|\pmb{\rho}_{\mathsf{S}\cup \mathsf{G}}\big)$  & \small $A_{\text{EYM}}^{\text{1-loop}}\big(\pmb{\sigma}_1;...;\pmb{\sigma}_m||\mathsf{H}\big)$ & \small $M_{\text{GR}}^{\text{1-loop}}(1,...,n)$  \\ \hline
    \small$I^{\text{1-loop}}_{\text{L}}$ &\begin{tabular}{c}\small $\text{PT}(+,\pmb{\sigma},-)$\nn $\small+\text{cyc}(\pmb{\sigma})$\end{tabular} & \begin{tabular}{c}\small $\text{PT}(+,\pmb{\sigma},-)$\\\small$+\text{cyc}(\pmb{\sigma})$ \end{tabular}& \begin{tabular}{c}\small $\text{PT}(+,\pmb{\sigma}_1,-)...$\\\small$\text{PT}(\pmb{\sigma}_m)\,\mathcal{P}+\text{cyc}(\pmb{\sigma}_1)$\end{tabular} & \begin{tabular}{c}\small$\text{PT}(+,\pmb{\sigma}_1,-)...$\\\small$\text{PT}(\pmb{\sigma}_m)\,\mathcal{P}+\text{cyc}(\pmb{\sigma}_1)$\end{tabular}&\small $\text{Pf}\,'[\Psi_{n+2}]$\\ \hline
      \small$I^{\text{1-loop}}_{\text{R}}$ &\begin{tabular}{c}\small $\text{PT}(+,\pmb{\rho},-)$\\\small$+\text{cyc}(\pmb{\rho})$\end{tabular} & \small $\text{Pf}\,'[\Psi_{n+2}]$ & \begin{tabular}{c}\small$\text{PT}\left(+,\pmb{\rho}_{\mathsf{S}\cup \mathsf{G}},-\right)$\\\small$+\text{cyc}(\pmb{\rho}_{\mathsf{S}\cup \mathsf{G}})$\end{tabular} & \small
    $\text{Pf}\,'[\Psi_{n+2}]$ &\small $\text{Pf}\,'[\Psi_{n+2}]$\\ \hline
    \end{tabular} }
    \caption{One-loop level CHY integrands for BS, YM, YMS, EYM and GR~~~Here, we only provide the YMS and EYM half-integrands that result in pure scalar loop and pure gluon loop, respectively.} \label{table:LoopIntegrands}
\end{table}}

\subsection{One-loop CHY formula from forward limit}


The CHY formula for one-loop $n$-point amplitudes is obtained from the forward limit of the CHY formula for tree-level $(n+2)$-point amplitudes
\bea
M_n^{\text{1-loop}}=\int {\text{d}^Dl\over l^2}\lim\limits_{k_{\pm}\to \pm l}\int \text{d}\mu_{\,n+2}^{\,\text{tree}}\,I_{\,\text{L}}^{\,\text{1-loop}}\,I_{\,\text{R}}^{\,\text{1-loop}},
\Label{Eq:LoopCHY}
\eea
where $l^{\mu}$ denotes the loop momentum in $D$ dimensions. The above integral is performed under the constraint of the one-loop scattering equations 
\bea
{l\cdot k_i\over z_i}+\Sl_{\substack{j=1\\ j\neq i}} {k_i\cdot k_j\over z_{ij}}=0,
\eea
in which, $k^{\mu}_i$ ($i=1,...,n$) refer to the external momenta. The half integrands $I_{\,\text{L}}^{\,\text{1-loop}}$ and $I_{\,\text{R}}^{\,\text{1-loop}}$  displayed explicitly by \tabref{table:LoopIntegrands}, are functions of loop momentum and external data, and are obtained from the forward limits of the $(n+2)$-point tree-level half integrands in \tabref{table:TreeIntegrands}. A PT factor $\text{PT}(+,\pmb{\sigma},-)$ in \tabref{table:LoopIntegrands} is defined by the $(n+2)$-point PT factor when $+$ and $-$ (which are introduced by forward limit) are considered as the first and the last elements, respectively. The $\text{cyc}(\pmb\sigma)$ in \tabref{table:LoopIntegrands} for a given permutation $\pmb\sigma$ stands for all cyclic permutations of  $\pmb\sigma$.
As shown in \cite{Porkert:2022efy,Feng:2022wee}, once the half integrands at one-loop level are expanded in terms of PT factors, the full integrand is then expanded as a combination of $(n+2)$-point tree-level BS amplitudes with the forward limit $k^{\mu}_{\pm}\to \pm l^{\mu}$. In addition, the polarizations $\epsilon^{\mu}_{-}$ and $\epsilon^{\nu}_+$ in the reduced Pfaffian should be replaced according to the following rule
\bea
\Sl_{+,-}\epsilon^{\mu}_{-}\epsilon^{\nu}_{+}=\Delta^{\mu\nu},~~\eta_{\mu\nu}\Delta^{\mu\nu}=D-2,~~V_{\mu}W_{\nu}\Delta^{\mu\nu}=V\cdot W, \Label{Eq:Polarizations}
\eea
where $V^{\mu}$ and $W^{\nu}$ are two vectors.

 In the next subsection, we introduce the refined graphic rule for expanding the half integrands, which comes from the expansion formula of tree-level amplitudes.

\subsection{Expanding the half integrands by refined graphic rule} \label{sec:RefinedGraph}


The expansion formula of the tree-level CHY half integrands totally inherits the expansion of tree amplitudes {which was proposed in \cite{Hou:2018bwm,Du:2019vzf}}, while the half integrands for $n$-point amplitude at one-loop level are obtained from the forward limit of the corresponding $(n+2)$-point tree-level half integrands. 
In the following, we present the refined graphic rules for expanding the $\text{PT}(+,\pmb{\sigma},-)\,\text{Pf}[{\,\Psi_{\mathsf{G}}}]$ for single-trace YMS (and EYM), the $\text{PT}(+,\pmb{\sigma}_1,-)\,\text{PT}(\pmb{\sigma}_2)\dots\text{PT}(\pmb{\sigma}_{m}) \,\mathcal{P}$ for multi-trace YMS (and EYM), and the $\text{Pf}\,'[\,\Psi\,]$ for YM, EYM and GR. We do not consider the integrands where the forward limit is taken for particles  $+$, $-$ belonging to distinct PT factors, in the current paper.

\subsubsection{Expansion of $\text{PT}(+,\pmb{\sigma},-)\,\text{Pf}\,[\,\Psi_{\mathsf{G}}\,]$} \label{sec:RefinedGraph1}
\begin{figure}
\centering
\includegraphics[width=0.4\textwidth]{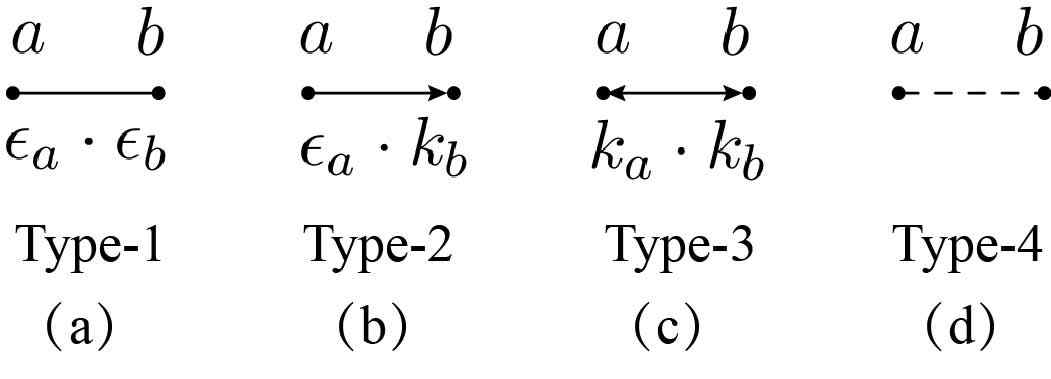}
\caption{Line styles in the refined graphic rule}
\label{Fig:LineStyles}
\end{figure}

 The $\text{PT}(+,\pmb{\sigma},-)\,\text{Pf}\,[\,\Psi_{\mathsf{G}}\,]$ can be expanded in terms of PT factors in Kleiss-Kuijf (KK) basis \cite{Kleiss:1988ne}, where $+$ and $-$ play as the first and the last elements, respectively
\bea
\text{PT}(+,\pmb{\sigma},-)\,\text{Pf}\,[\,\Psi_{\mathsf{G}}\,]=\Sl_{\mathcal{F}}\,\mathcal{C}^{\mathcal{F}}\,\left[\,\Sl_{\pmb{\rho}^{\mathcal{F}}}\text{PT}\left(+,\pmb{\rho}^{\mathcal{F}},-\right)\,\right]. \Label{Eq:Expansion1}
\eea
In the above, we have summed over all possible graphs $\mathcal{F}$ which are generated by considering all external particles as nodes, the Lorentz contractions $\epsilon_a\cdot \epsilon_b$, $\epsilon_a\cdot k_b$ and $k_a\cdot k_b$ as lines between nodes (as shown by \figref{Fig:LineStyles}, the type-4 line with no kinematic factor is introduced for recording the relative order between two nodes), and the set $\{+, \sigma_1,...,\sigma_r\}$ as the root set.  Each graph $\mathcal{F}$ is associated with a kinematic coefficient $\mathcal{C}^{\mathcal{F}}$ and a set of permutations $\pmb{\rho}^{\mathcal{F}}$ (where the relative ordering of scalars is $\pmb{\sigma}$). The second summation means that we should sum over all  permutations $\pmb{\rho}^{\mathcal{F}}$. Now we demonstrate  {\bf\emph {refined graphic rule}} for constructing $\mathcal{F}$:
\newline
\newline
\noindent{\bf Step-1} Define a reference order $\mathsf{R}=\{\gamma_1,...,\gamma_s\}$. Here, $\pmb{\gamma}$ is a permutation of all $s$ elements in the set $\mathsf{G}$. We define the position of an element of $\mathsf{G}$ in $\mathsf{R}$ as its {\it weight}. 

%
\begin{figure}
\centering
\includegraphics[width=0.45\textwidth]{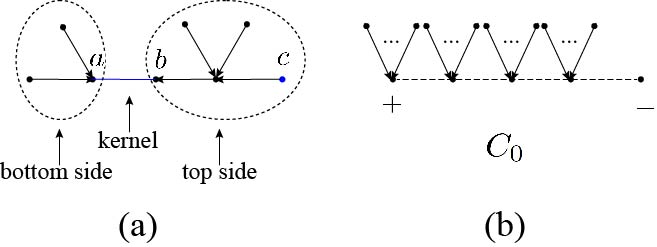}
\caption{Graphs (a) and (b) are typical components in single-trace case. In (a), the kernel is the type-1 line between $a$ and $b$, while the side containing the highest-weight node in this component, say $c$, is defined as the top side. The opposite side is the bottom side.  Graph (b) is the component $C_0$ containing scalar trace.}
\label{Fig:SpriousGraph}
\end{figure}

\noindent{\bf Step-2} Graphs $\mathcal{F}$ can be classified according to the number of  type-1 lines (which are mentioned as {\it the kernels}). A graph with $k$ ($0\leq k\leq \lfloor{s\over 2}\rfloor$) type-1 lines, is constructed by the following steps:

\begin{itemize}
\item (i). Pick out $k$ pairs of elements arbitrarily from the set $\mathsf{G}$, and construct $k$ type-1 lines (kernels) between each pair of nodes. Connect the nodes in the ordered set $\{+,\pmb{\sigma},-\}$ (i.e. the trace) by type-4 lines in accordance to the relative order.

\item (ii). Connect other elements in $\mathsf{G}$ towards either the elements in the set $\{+,\pmb{\sigma}\}$  or the $2k$ end nodes of the $k$ kernels which already have been used in the previous step, via type-2 lines. Now we have $k+1$ mutually disconnected {\it components} $\{C_0,C_1,C_2,...,C_k\}$ (structure of components are shown by \figref{Fig:SpriousGraph}), where the $C_0$ refers to the one involving the trace.

\end{itemize}
\noindent{\bf Step-3} Now we connect the components $C_0,C_1,C_2,...,C_k$ together via type-3 lines into a fully connected graph $\mathcal{F}$ as follows:

\begin{itemize}
 \item (i). Define the {\it weight of a component} $C_i$ (for $i\neq 0$) by the weight of its highest-weight node, and then we arrange the components in the reference order $\mathsf{R}_C=\{C_{\beta_1},...,C_{\beta_k}\}$ as the increasing order of their weights. According to the definition, a component $C_{\,i}$ ($i\neq 0$) is always divided into two parts by the type-1 line. We further define the part which involves the highest-weight node of $C_{\,i}$ as the top side $C^{\,\text{t}}_{\,i}$, while the opposite side as the bottom side,  $C^{\,\text{b}}_{\,i}$. A component $C_{\,i}$ is sometimes presented by the symbol $C_{\,i}=\left[C^{\,\text{b}}_{\,i}\right]-\left[C^{\,\text{t}}_{\,i}\right]$, in which a ``$-$'' is used to denote the type-1 line.

\item (ii). Define the root set $\mathcal{R}_C\equiv C_{\,0}$ (Here, the node $-$ is always excluded from the root set). Construct a \emph{chain of components} (COC) via type-3 lines
\bea
\left[C^{\,\text{t}}_{\,a}\right]-\left[C^{\,\text{b}}_{\,a}\right]\leftrightarrow \left[C^{\,\text{t}(\text{b})}_{\,l_1}\right]-\left[C^{\,\text{b}(\text{t})}_{\,l_1}\right]\leftrightarrow \left[C^{\,\text{t}(\text{b})}_{\,l_2}\right]-\left[C^{\,\text{b}(\text{t})}_{\,l_2}\right]\leftrightarrow\cdots\leftrightarrow \left[C^{\,\text{t}(\text{b})}_{\,l_j}\right]-\left[C^{\,\text{b}(\text{t})}_{\,l_j}\right]\leftrightarrow C_{\,0}.\Label{Eq:ChangeOfComponents}
\eea
 which starts from the top side of the highest-weight component $C_{a}=C_{\beta_k}$, passes through some other components $C_{\,l_1},C_{\,l_2},...,C_{\,l_j}$ and ends up at $\mathcal{R}_C$. In the above expression, the notation ``$\leftrightarrow$'' denotes a type-3 line. The superscripts in an internal component $\left[C^{\,\text{t}(\text{b})}_{\,l_i}\right]-\left[C^{\,\text{b}(\text{t})}_{\,l_i}\right]$ mean that once the left side is chosen as the top (bottom) one, the right side is always chosen as the bottom (top) side.

\item (iii). Redefine the ordered set $\mathsf{R}_C$  and the root set $\mathcal{R}_C$ by 
\bea
\mathsf{R}_C\to \mathsf{R}_C\setminus\left\{C_a,C_{l_1},...,C_{l_j}\right\},~~~~\mathcal{R}_C=C_{\,0}\cup C_a\cup C_{l_1}\cup ...\cup C_{l_j}.
\eea

Repeat step (ii) until the ordered set $\mathsf{R}_C$ becomes empty, then we get a fully connected refined graph $\mathcal{F}$ with $k$ type-1 lines.
\end{itemize}

\begin{figure}
\centering
\includegraphics[width=0.65\textwidth]{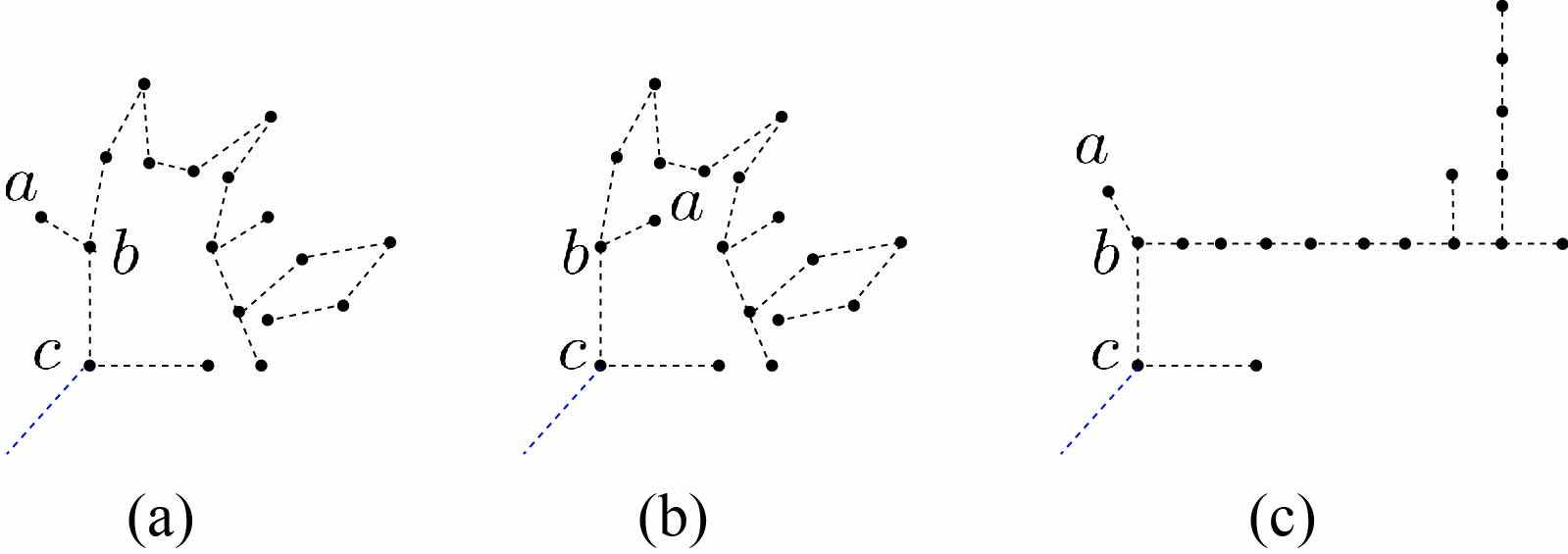}
\caption{The permutations established by a connected tree graph only depend on the topology of a graph and the choice of the nearest-to-root node. Graphs (a), (b) and (c) have the same structure. Once the node $c$ is chosen as the nearest-to-root one, these graphs correspond to the same set of permutations. Here we should note that two graphs are the same when they are related by exchanging two tree structures attached to a same node. Although, the lines used in these graphs are type-4 ones, those graphs with other line styles but the same topology with (a), (b) and (c) establish the same set of permutations. }
\label{Fig:ExampleRefinedGF}
\end{figure}

\noindent {\bf Kinematic coefficients $\mathcal{C}^{\mathcal{F}}$ and permutations $\pmb{\rho}^{\mathcal{F}}$} To each graph $\mathcal{F}$ constructed by the above steps, we associate a kinematic factor $\mathcal{C}^{\mathcal{F}}$ which is given by the product of the Lorentz contractions implied by the lines. An extra minus is introduced if there is an arrow pointing away from the direction of roots. Concretely, each type-3 line produces a minus sign, as does each type-2 line pointing away from the root. The possible permutations established by the graph $\mathcal{F}$ are defined as follows: (i). The first and the last elements in the permutations are always $+$ and $-$. (ii). If two nodes $a$ and $b$ live on a same path towards  $+$, and the node $a$  is nearer to $+$ than $b$, we have $a\prec b$ in $\pmb{\rho}^{\mathcal{F}}$ (the notation $a\prec b$  means $a$ appears to the left of $b$ for a given permutation). (iii). If there are two branches attached to a node, we should shuffle the relative orders corresponding to the two branches together. For example, if we have two branches attached to node $a$, one of them contains nodes $b$, $c$ in turn, the other has a single node $d$, the relative orders between these nodes in $\pmb{\rho}^{\mathcal{F}}$ are
\bea
\{a,\{\{b,c\}\shuffle\{d\}\}\} \to \{a,b,c,d\},~\{a,b,d,c\},~\{a,d,b,c\}.
\eea

Finally, we sum over all possible graphs $\mathcal{F}$, which include all possible graphs with $k$ type-1 lines for $k=1,...,\lfloor{s\over 2}\rfloor$, then arrive at the expansion of $\text{PT}(+,\pmb{\sigma},-)\,\text{Pf}\,[\,\Psi_{\mathsf{G}}\,]$. It is clear that the permutations corresponding to the graphs established by the refined graphic rule, which have the same chosen nearest-to-root node, are related only to the topology of the graphs, as shown by \figref{Fig:ExampleRefinedGF}. This graphic rule has an equivalent version which is expressed by strength tensors, see \appref{sec:ExpansionNew}.

\subsubsection{Expansion of $\text{Pf}\,'[\,\Psi_{n+2}\,]$}

As pointed in \cite{Fu:2017uzt,Du:2017kpo}, the reduced Pfaffian $\text{Pf}\,'[\,\Psi_{n+2}\,]$ (which includes the $n$ external particles and $+,-$) can be decomposed in terms of $\text{PT}(+,...,-)\,\text{Pf}\,[\,\Psi\,]$ with less elements in the Pfaffian and more elements in the PT factor. Specifically,
\bea
\text{Pf}\,'[\,\Psi_{n+2}\,]&=&~~(\epsilon_{+}\cdot\epsilon_{-})\,\text{PT}(+,-)\,\text{Pf}\,\left[\,\Psi_{\{1,...,n\}}\,\right]\nn
&&-\Sl_{i}(\epsilon_{+}\cdot F_i\cdot\epsilon_{-})\,\text{PT}(+,i,-)\,\text{Pf}\,\left[\,\Psi_{\{1,...,n\}\setminus i}\,\right]\nn
&&+\Sl_{\{i_1,i_2\}\in \text{S}_2}(\epsilon_{+}\cdot F_{i_1}\cdot F_{i_2}\cdot\epsilon_{-})\,\text{PT}(+,i_1,i_2,-)\,\text{Pf}\,\left[\,\Psi_{\{1,...,n\}\setminus \{i_1,i_2\}}\,\right]\nn
&&-\cdots\nn
&&+(-1)^n\Sl_{\{i_1,i_2,...,i_n\}\in \text{S}_{n}}(\epsilon_{+}\cdot F_{i_1}\cdot F_{i_2}\cdot...\cdot F_{i_n}\cdot\epsilon_{-})\,\text{PT}(+,i_1,i_2,...,i_n,-).\Label{Eq:ExpandReducedPfaffian1}
\eea
where in each summation $\Sl_{\{i_1,i_2,...,i_k\}\in \text{S}_{k}}$, we have summed over all possible permutations of $k$ elements $i_1,i_2,...,i_k$ for all possible choices of  $i_1,i_2,...,i_k\in \{1,...,n\}$. The $F^{\mu\nu}_i$ is the strength tensor that is defined by $k_i^{\mu} \epsilon_i^{\nu}-\epsilon_i^{\mu} k_i^{\nu}$. When the rule (\ref{Eq:Polarizations}) for forward limit is taken into account, the above expansion turns into
\bea
\text{Pf}\,'[\,\Psi_{n+2}\,]&=&(D-2)\,\text{PT}(+,-)\,\text{Pf}\,\left[\,\Psi_{\{1,...,n\}}\,\right]\Label{Eq:ExpandReducedPfaffian2}\\
%
%
&&+\Sl_{\{i_1,i_2\}\in\text{S}_{2}\setminus\text{Z}_{2}}\text{Tr}[F_{i_1}\cdot F_{i_2}]\Big[\,\text{PT}(+,i_1,i_2,-)+\text{cyc}(i_1,i_2)\Big]\,\text{Pf}\,\left[\,\Psi_{\{1,...,n\}\setminus \{i_1,i_2\}}\,\right]\nn
&&-\cdots\nn
&&+(-1)^n\Sl_{\{i_1,i_2,...,i_n\}\in \text{S}_{n}\setminus\text{Z}_{n}}\text{Tr}[ F_{i_1}\cdot F_{i_2}\cdot...\cdot F_{i_n}]\,\Big[\,\text{PT}(+,i_1,i_2,...,i_n,-)+\text{cyc}(i_1,i_2,...,i_n)\,\Big],\nonumber
\eea
where each $\text{Tr}[F_{i_1}\cdot F_{i_2}\cdot...\cdot F_{i_l}]$ is defined by 
\bea
\text{Tr}[F_{i_1}\cdot F_{i_2}\cdot...\cdot F_{i_l}]\equiv {{(F_{i_1})}^{\mu_1}_{~\,\mu_2} {(F_{i_2})}^{\mu_2}_{~\,\mu_3}... {(F_{i_l})}^{\mu_{l}}_{~\,\mu_1}},
\eea
which has cyclic symmetry with respect to $i_1,i_2,...,i_l$, and $\text{Tr}[F_{i}]=0$ due to the antisymmetry of the strength tensor $F_{i}^{\mu\nu}$.  The property (\ref{Eq:ExpandReducedPfaffian2}) will be applied to obtain the quadratic propagator form of YM integrand.

\subsection{Tensorial PT factors and the problem of quadratic propagators}

In the previous section, we have already expanded CHY half integrands in terms of the tree-level PT factors of the form $\text{PT}(+,...,-)$, thus any one-loop integrand is expressed by a combination of 
\bea
{1 \over l^2}\int \text{d}\mu_{\,n+2}^{\,\text{tree}}\text{PT}(+,\pmb{\sigma},-)\text{PT}(+,\pmb{\rho},-),\Label{Eq:One-loopBasis}
\eea
where $\pmb{\sigma}$ and $\pmb{\rho}$ are two permutations of $n$ external particles. The above expression, except for the factor ${1\over l^2}$, is nothing but a tree-level $(n+2)$-point BS amplitude, in which $+$ and $-$ are fixed as the two ends. Thus such a term can be rewritten as \cite{Feng:2022wee}
\bea
{1 \over l^2}\int \text{d}\mu_{\,n+2}^{\,\text{tree}}\text{PT}(+,\pmb{\sigma},-)\text{PT}(+,\pmb{\rho},-)&=&\Sl_{\substack{(A_1A_2...A_i)={\pmb{\sigma}}\\(\W A_1\W A_2...\W A_i)={\pmb{\rho}}\\ A_j=\W A_j}}{1\over l^2}\,{1\over s_{A_1,l}}\,{1\over s_{A_1A_2,l}}\,\cdots\,{1\over s_{A_1A_2\cdots A_{i-1},l}}\phi_{A_1|\W A_1}\phi_{A_2|\W A_2}\cdots \phi_{A_i|\W A_i}.\Label{Eq:One-loopBasis-1}\nn
\eea
Here, the summation over $(A_1A_2...A_i)={\pmb{\sigma}}$ and $(\W A_1\W A_2...\W A_i)={\pmb{\rho}}$ means we sum over all divisions of $\pmb{\sigma}$ and $\pmb{\rho}$ including the case $i=1$. The $\phi_{A_i|\widetilde{A_i}}$ is the BG current of BS theory (which will be introduced later). The ${1\over s_{A_1\cdots A_j,l}}$ is defined by
\bea
{1\over s_{A_1\cdots A_j,l}}={1\over {2 l\cdot (k_{A_1}+\cdots +k_{A_j})+(k_{A_1}+\cdots +k_{A_j})^2}}.\Label{Eq:LinearPropagator}
\eea
Being different from the propagators in traditional Feynman diagrams whose denominator is a quadratic function of $l^{\mu}$, the denominator of the above propagator is a linear function, thus is called {\it linear propagator}.

In \cite{Feng:2022wee}, tensorial PT factors  $\text{PT}_{\text{t}}^{\mu_1\mu_2...\mu_r}(1,2,...,n)$ were introduced
\bea
\text{PT}_{\text{t}}^{\mu_1\mu_2...\mu_r}(1,2,...,n)\equiv\Sl^{n}_{i=1}\text{PT}(+,i,i+1,...,1,2,...,i-1,-)\prod^{r}_{j=1}(l^{\mu_j}-k^{\mu_j}_{1,2,...,i-1}).\Label{Eq:TensorialPT}
\eea
The subscript “t”of $\text{PT}_{\text{t}}^{\mu_1\mu_2...\mu_r}(1,2,...,n)$ is used to distinguish tensorial PT factors from those PT factors we have used before. When there is no Lorentz index, the tensorial PT factor turns into  a sum over cyclic permutations of $1,...,n$ and is called a scalar PT factor
\bea
\text{PT}_{\text{t}}(1,2,...,n)\equiv\Sl^{n}_{i=1}\text{PT}(+,i,i+1,...,i-1,-).\Label{Eq:ScalarPT}
\eea
The product of a scalar PT factor (\ref{Eq:ScalarPT}) and a general tensorial PT factor (\ref{Eq:TensorialPT}) produces quadratic propagators when the integral over scattering variables are performed 
\bea
&&~~{1\over l^2}\int \text{d}\mu_{n+2}^{\text{tree}}\text{PT}_{\text{t}}^{\mu_1\mu_2...\mu_r}\left(1,\pmb{\rho}\,(2,...,n)\right)\text{PT}_{\text{t}}\left(1,\pmb{\sigma}\,(2,...,n)\right)\nn
&\cong&\Sl_{\small\substack{(A_1A_2...A_m)=(1\,\pmb\rho)\\(\widetilde{A_1}\widetilde{A_2}...\widetilde{A_m})=(1\,\pmb\sigma)\\A_j=\widetilde{A_j}\\2\leq m\leq n}}\left(l+k_{A}\right)^{\,\mu_1}...\,\left(l+k_{A}\right)^{\,\mu_r}\,gon\left(A_1,A_2,...,A_m\right)\,\prod_{i=1}^m\,\phi_{A_i|\widetilde{A_i}}.
\Label{Eq:gon1}
\eea
\begin{figure}
\centering
\includegraphics[width=0.35\textwidth]{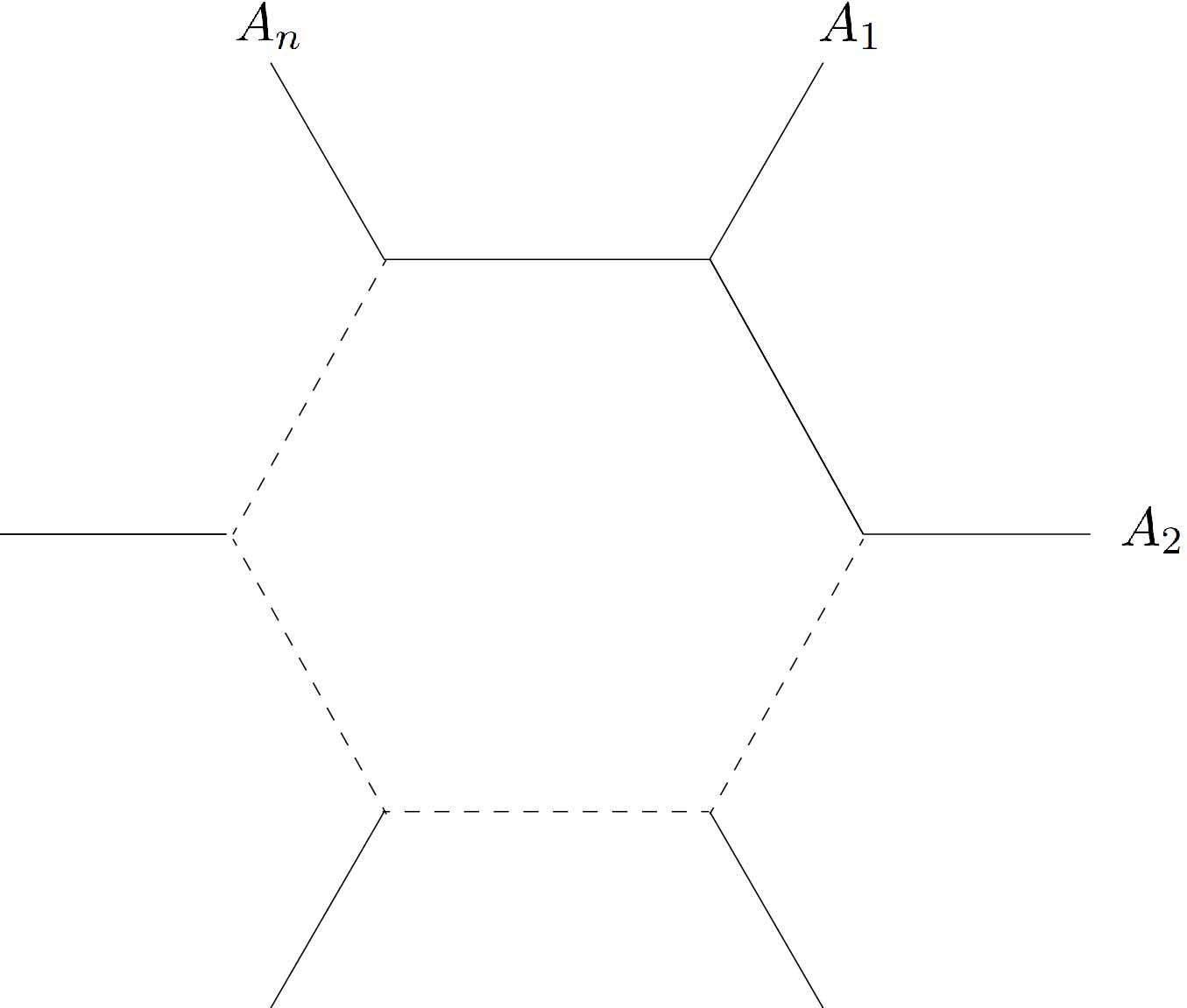}
\caption{An n-gon diagram with $n$ subtrees which represent the BG currents $\phi_{A_i|\widetilde{A_i}}$ ($i=1,...,n$).}
\label{Fig:n-gon}
\end{figure}
The $\cong$ means that the LHS gives the same result with the RHS when the integral over loop momentum $l^{\mu}$ is taken.
In the above expression, $(A_1A_2...A_m)$ and $(\widetilde{A_1}\widetilde{A_2}...\widetilde{A_m})$ are the cyclic partitions of color orderings $(1\,\pmb\rho)$ and $(1\,\pmb\sigma)$, where $A_1=A1\bar{A}$, $A$ and $\bar{A}$ denote the ordered sets of elements in $A_1$ on the LHS of $1$ and on the RHS of $1$, respectively. The $A_j=\widetilde{A_j}$ implies that $A_j$ and $\widetilde{A_j}$ have the same elements but may not be in the same order. The $m$-gon,  $gon(A_1,A_2,...,A_m)$, is defined by
\bea
gon(A_1,A_2,...,A_m)\,\equiv\,{1\over {l^2 l^2_{A_1} l^2_{A_{12}}...l^2_{A_{12...m-1}}}}\,,
\eea
where $l^{\mu}_{A_{12...m-1}}=l^{\mu}_{A_1A_2...A_{m-1}}=l^{\mu}+k^{\mu}_{A_1}+k^{\mu}_{A_2}+...+k^{\mu}_{A_{m-1}}$, $k^{\mu}_A$ and $k^{\mu}_{\W A}$ denote the sum of the momenta of the elements in the ordered sets $A$ and $\W A$, respectively. The $\phi_{A_i|\widetilde{A_i}}$ is the BS current which represents the subtree planted on the $i$-th corner of the $m$-gon,  as shown by \figref{Fig:n-gon}. The case $m=1$ is excluded from the summation since it corresponds to tadpole diagrams which contain a combination of BS currents
\bea
\phi_{A|\widetilde{A}}+\text{cyclic permutations of $\W A$}
\eea
vanishing due to a  U(1)-decoupling identity. There is a crucial relation we used to obtain \eqref{Eq:gon1}
\bea
&&{N(l)\over l^2s_{A_1,l}s_{A_{12},l}\dots s_{A_{12\dots m-1},l}}+{N(l+k_{A_m})\over l^2s_{A_m,l}s_{A_{m1},l}\dots s_{A_{m1\dots m-2},l}}+\dots+{N(l+k_{A_{23\dots m}})\over l^2s_{A_2,l}s_{A_{23},l}\dots s_{A_{23\dots m},l}}\nn
&\cong& N(l)\,gon\left(A_1,A_2,...,A_m\right)
\Label{Eq:partial}
\eea
where $N(l)$ is an $l$-dependent numerator and the partial fraction identity have been applied.

When the scalar PT factor in \eqref{Eq:gon1}, is replaced by another copy of tensorial PT factor, one arrives a more general result
\bea
&&~~~{1\over l^2}\int d \mu_{n+2}^{\text{tree}}\,\text{PT}^{\mu_1\mu_2...\mu_r}_{\text{t}}\left(1,\pmb{\rho}\,(2,...,n)\right)\,\text{PT}^{\nu_1\nu_2...\nu_t}_{\text{t}}\left(1,\pmb{\sigma}\,(2,...,n)\right)\nn
&\cong&l^{\mu_1}\cdots l^{\mu_r}\Sl_{\small\substack{(A_1A_2...A_m)=(1\,\pmb\rho)\\(\widetilde{A_1}\widetilde{A_2}...\widetilde{A_m})=(1\,\pmb\sigma)\\A_j=\widetilde{A_j}\\2\leq m\leq n}}\left(l-\Delta_{A\W A}\right)^{\,\nu_1}...\,\left(l-\Delta_{A\W A}\right)^{\,\nu_t}\,gon^{A}\left(A_1,A_2,...,A_m\right)\,\prod_{i=1}^m\,\phi_{A_i|\widetilde{A_i}}.~\Label{Eq:gon2}
\eea
The $\Delta^{\mu}$ is defined by  $\Delta^{\mu}_{A\W A}\equiv k^{\mu}_A-k^{\mu}_{\W A}$, tadpole terms have been removed by hand. The $gon^{A}\left(A_1,A_2,...,A_m\right)\equiv gon\left(A_1,A_2,...,A_m\right)|_{l^{\mu}\to l^{\mu}-k_{A}^{\mu}}$. Eq. (\ref{Eq:gon2}) also produces quadratic propagators.

We  will see later that in some special cases, a CHY integrand can be arranged into a proper format such that it can be decomposed into combinations of (\ref{Eq:gon1}) (as a special case of (\ref{Eq:gon2})) and thus naturally reproduce quadratic propagators. However, a more general integrand in \tabref{table:LoopIntegrands} may not be directly decomposed into (\ref{Eq:gon2}). In this work, we show that the quadratic propagators arise when the CHY half integrands are decomposed according to the refined graphic rule which has been reviewed in the previous subsection.

\subsection{Properties of the BG currents in BS theory}
In this subsection, we review the definition of the BG \cite{Berends:1987me} currents in BS theory and provide helpful properties of the BG currents.


The tree-level BG current $\phi_{A|\widetilde{A}}$ is recursively defined by \cite{Mafra:2016ltu}
\bea
\phi_{A|\widetilde{A}}={1\over s_A}\Sl_{\substack{A=A_LA_R\\ \widetilde{A}=\widetilde{A}_L\widetilde{A}_R}}\Bigl[\,\phi_{A_L|\widetilde{A}_L}\phi_{A_R|\widetilde{A}_R}-\phi_{A_R|\widetilde{A}_L}\phi_{A_L|\widetilde{A}_R}\,\Bigr],~\Label{Eq:BScurrent}
\eea
where $s_A\equiv k_A^2$ and we have summed over divisions $A=A_LA_R$, $\widetilde{A}=\widetilde{A}_L\widetilde{A}_R$. The starting point of this recursion is $\phi_{a|a}=1$, $\phi_{a|b}=0$ $(a\neq b)$. If the elements in $A$ and $\widetilde{A}$ in $\phi_{A|\widetilde{A}}$ are not identical to each other, the current must vanish. The BS current is apparently symmetric under $A\Leftrightarrow \widetilde{A}$ and satisfies the generalized $U(1)$-decoupling identity \cite{Du:2011js}
\bea
\phi_{A\shuffle B|\widetilde{C}}=0, \Label{Eq:GenU1}
\eea
where $\W C$ is an arbitrary permutation of elements in $A\cup B$, and the summation over all shuffle permutations of $A$ and $B$ is implied.
Furthermore, as pointed in \cite{Wu:2021exa}, the BG current (\ref{Eq:BScurrent}) satisfies relations characterized by graphs.
%
%
\begin{figure}
\centering
\includegraphics[width=0.65\textwidth]{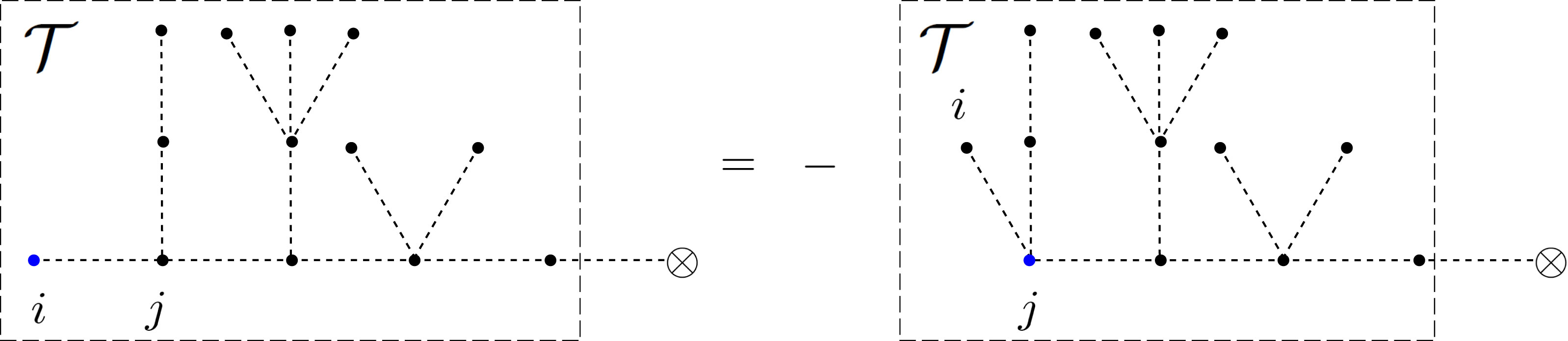}
\caption{A typical graph demonstrating the graph-based identity (\ref{Eq:GraphBased1}) for the BG currents of BS. The cross node $\otimes$ in each graph is introduced to remind that there exists an off-shell line of the BG current. Assuming $i$, $j$ are two adjacent nodes in a connected tree graph $\mathcal{T}$,  the combination of BS currents with respect to permutations when $i$ is considered as the leftmost element is the minus of the combination when $j$ is considered as the leftmost one.  Although there are only type-4 lines in the two graphs, we can replace the type-4 lines by any other type of lines and provide the corresponding factors.}
\label{Fig:GraphBasedProperty1}
\end{figure}
The first graph-based relation is 
\bea
\Sl_{\pmb{\sigma}\in\pmb{\rho}^{\mathcal{T}}|_{i}}\phi_{\pmb{\sigma}\,|\,\W A}=-\Sl_{\pmb{\gamma}\in\pmb{\rho}^{\mathcal{T}}|_{j}}\phi_{\pmb{\gamma}\,|\,\W A},~\Label{Eq:GraphBased1}
\eea
where $\pmb{\rho}^{\mathcal{T}}|_{i}$ denotes permutations established by the connected tree graph $\mathcal{T}$, with the leftmost element $i$.   The node $j$ on the RHS is a node adjacent to $i$. The above relation is shown by \figref{Fig:GraphBasedProperty1}.

\begin{figure}
\centering
\includegraphics[width=1\textwidth]{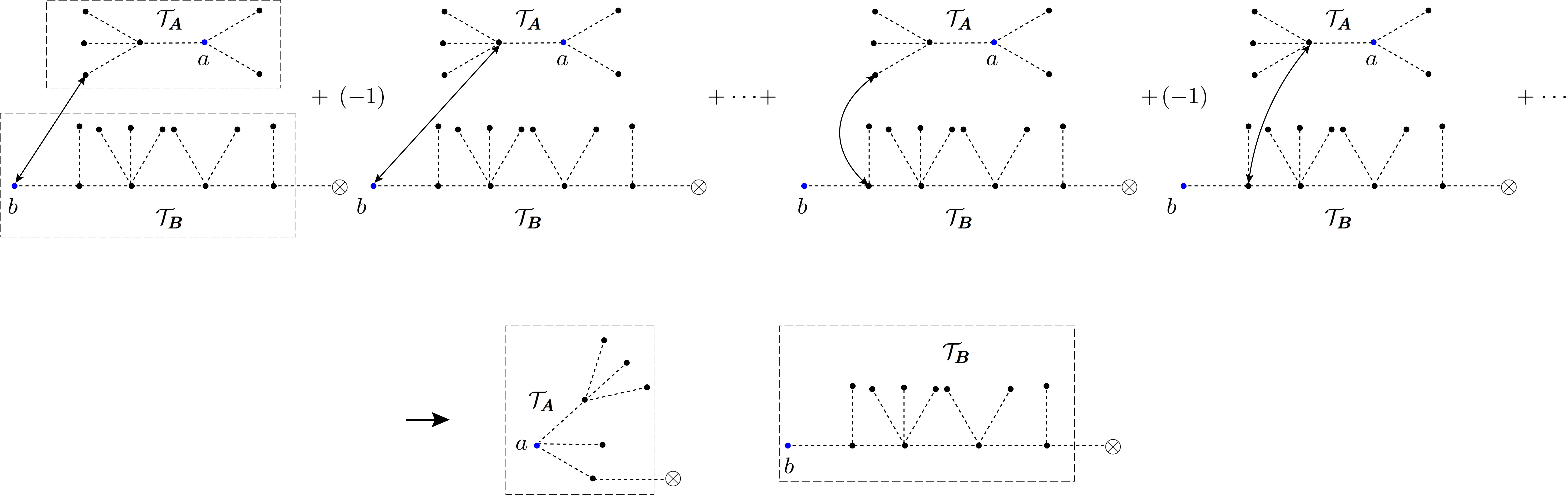}
\caption{Graphs on the first line establish the LHS of the off-shell BCJ relation (\ref{Eq:OffBCJ1}). Here, we have two connected tree graphs $\mathcal{T}_A$ and $\mathcal{T}_B$. A graph $\mathcal{T}_A\oplus\mathcal{T}_B$ is constructed by connecting a pair of nodes in $\mathcal{T}_A$ and $\mathcal{T}_B$, via a type-3 line. The two graphs on the second line define the RHS of off-shell BCJ relation, where $a$ and $b$ are chosen as the first elements in the corresponding currents. Similar to \figref{Fig:GraphBasedProperty1}, all type-4 lines can be substituted by other types of lines.}
\label{Fig:GraphBasedProperty2}
\end{figure}
\begin{figure}
\centering
\includegraphics[width=0.9\textwidth]{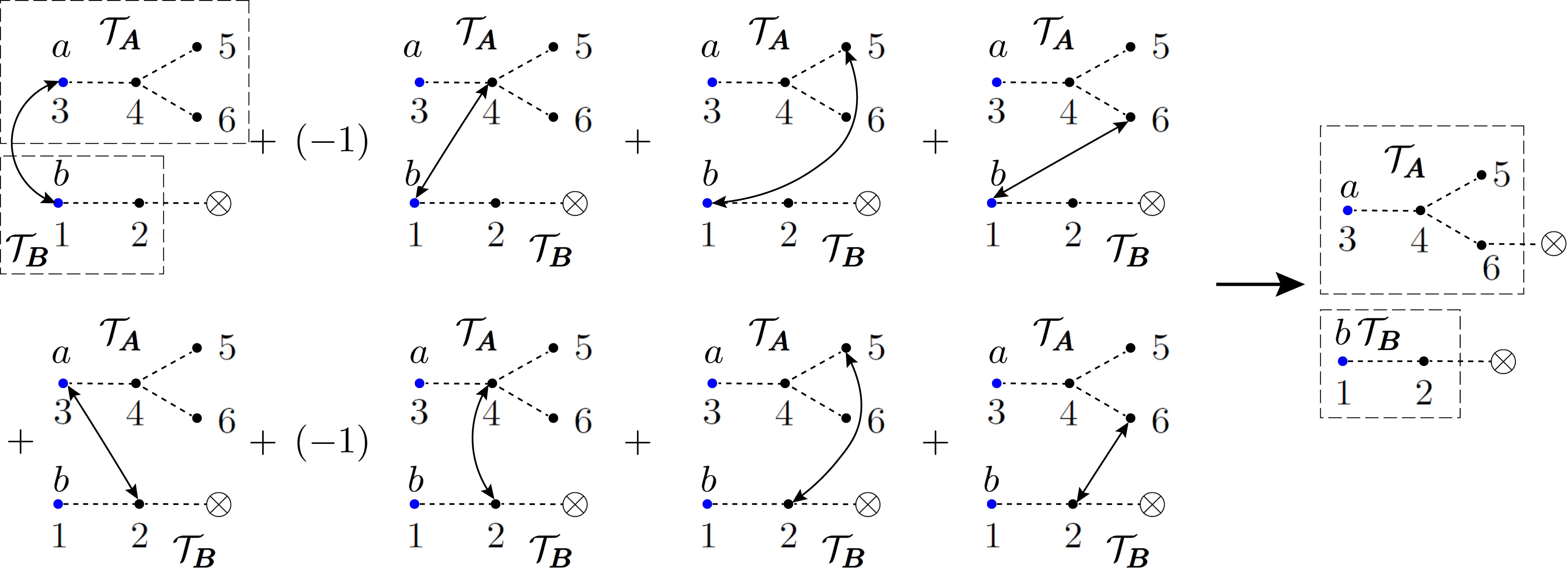}
\caption{An explicit example for the off-shell BCJ relation (\ref{Eq:OffBCJ1}), where the two tree graphs $\mathcal{T}_A$ and $\mathcal{T}_B$ contain elements $3,4,5,6$ and $1,2$, respectively. The nodes $a\in\mathcal{T}_A$ and $b\in\mathcal{T}_B$ are chosen as $3$ and $1$.}
\label{Fig:GraphBasedProperty2a}
\end{figure}

A kind of relation satisfied by the BG currents with more complicated expression is the {\it off-shell BCJ relation} \cite{Du:2011js,Frost:2020eoa,Wu:2021exa,Du:2022vsw}. These relations have nontrivial kinematic coefficients associating to the BG currents. Each coefficient is a function of $s_{ij}\equiv 2k_i\cdot k_j$, where $k^{\mu}_i$ and $k^{\mu}_j$ are momenta of on-shell external particles. The off-shell BCJ relations turn into amplitude relations under the on-shell limit. In \cite{Wu:2021exa,Du:2022vsw}, the following off-shell BCJ relation based on graphs was proposed
\bea
\Sl_{x\in \mathcal{T}_A, y\in \mathcal{T}_{{B}}}(-1)^{|ax|}\Sl_{\pmb{\gamma}\in\pmb{\rho}^{\left(\mathcal{T}_{{A}}\oplus\mathcal{T}_{{B}}\right)}\left.\right|_b}s_{xy}\phi_{\pmb{\gamma}|\pmb{\sigma}}&=&\Sl_{\pmb{\alpha}\in \pmb{\rho}^{\mathcal{T}_{A}}|_a}\Sl_{\pmb{\beta}\in \pmb{\rho}^{\mathcal{T}_{B}}|_b} \Big[\phi_{\pmb{\beta}|\pmb{\sigma}_{1,i}}\phi_{\pmb{\alpha}|\pmb{\sigma}_{i+1,l}} -\phi_{\pmb{\alpha}|\pmb{\sigma}_{1,{l-i}}}\phi_{\pmb{\beta}|\pmb{\sigma}_{l-i+1,l}}\Big], \Label{Eq:OffBCJ1}
\eea
in which, we have two connected tree graphs $\mathcal{T}_{{A}}$ and $\mathcal{T}_{{B}}$, as shown by \figref{Fig:GraphBasedProperty2}. When connecting two nodes $x\in \mathcal{T}_A$ and $y\in \mathcal{T}_{{B}}$ by a type-3 line, we get a graph $\mathcal{T}_{{A}}\oplus\mathcal{T}_{{B}}$. The first summation on the LHS means we sum over such graphs $\mathcal{T}_{{A}}\oplus\mathcal{T}_{{B}}$ corresponding to all possible choices of $(x,y)$ pair. The $a\in \mathcal{T}_{{A}}$ and $b\in \mathcal{T}_{{B}}$ are two arbitrarily chosen nodes. Here, the sign $(-1)^{|ax|}$ depends on the distance $|ax|$ between the nodes $a$ and $x$. The second summation is taken over permutations established by the graph $\mathcal{T}_{{A}}\oplus\mathcal{T}_{{B}}$, while $b$ is considered as the leftmost element. On the RHS, $\pmb{\alpha}\in \pmb{\rho}^{\mathcal{T}_{A}}|_a$ and $\pmb{\beta}\in \pmb{\rho}^{\mathcal{T}_{B}}|_b$ are the permutations established by $\mathcal{T}_{A}$ and $\mathcal{T}_{B}$ when $a$ and $b$ are considered as the leftmost elements, respectively. 
Permutation $\pmb{\sigma}=(\sigma_1\dots\sigma_l)$ in $\phi_{\,\pmb{\gamma}\,|\,\pmb{\sigma}\,}$ on the LHS is divided into two parts $\pmb{\sigma}_{1,i}=({\sigma}_{1}\dots{\sigma}_{i})$ and $\pmb{\sigma}_{i+1,l}=({\sigma}_{i+1}\dots{\sigma}_{l})$, where the number of nodes in $\mathcal{T}_{B}$ is assumed to be $i$ and the number of nodes in $\mathcal{T}_{{A}}\oplus\mathcal{T}_{{B}}$ is $l$. In the second term on the RHS, $\pmb{\sigma}$ is divided into $\pmb{\sigma}_{1,l-i}=({\sigma}_{1}\dots{\sigma}_{l-i})$ and $\pmb{\sigma}_{l-i+1,l}=({\sigma}_{1-i+1}\dots{\sigma}_{l})$. 
\figref{Fig:GraphBasedProperty2a} provides an example of (\ref{Eq:OffBCJ1}). Explicitly, this relation is displayed as
\bea
&&s_{13}\phi_{1\{2\}\shuffle\{34\{5\}\shuffle\{6\}\}|\pmb{\sigma}}-s_{14}\phi_{1\{2\}\shuffle\{4\{3\}\shuffle\{5\}\shuffle\{6\}\}|\pmb{\sigma}}+s_{15}\phi_{1\{2\}\shuffle\{54\{3\}\shuffle\{6\}\}|\pmb{\sigma}}+s_{16}\phi_{1\{2\}\shuffle\{64\{3\}\shuffle\{5\}\}|\pmb{\sigma}}\nn
&&+s_{23}\phi_{1234\{5\}\shuffle\{6\}|\pmb{\sigma}}-s_{24}\phi_{124\{3\}\shuffle\{5\}\shuffle\{6\}|\pmb{\sigma}}+s_{25}\phi_{1254\{3\}\shuffle\{6\}|\pmb{\sigma}}+s_{26}\phi_{1264\{3\}\shuffle\{5\}|\pmb{\sigma}}\nn
&=&\phi_{12|\sigma_1\sigma_2}\phi_{34\{5\}\shuffle\{6\}|\sigma_3\sigma_4\sigma_5\sigma_6}-\phi_{34\{5\}\shuffle\{6\}|\sigma_1\sigma_2\sigma_3\sigma_4}\phi_{12|\sigma_5\sigma_6},
\eea
where the sums over shuffle permutations are implied. We point out that the relation (\ref{Eq:OffBCJ1}) is satisfied by off-shell BG currents with nontrivial kinematic coefficients. When we multiply $k^2_{123456}$ to both sides of the above example and take the on-shell limit $k^2_{123456}=0$, each BS current on the LHS turns into an on-shell BS amplitude, while the RHS has to vanish. Thus, from this example, we can see the on-shell limit of the off-shell BCJ relation (\ref{Eq:OffBCJ1}) gives the corresponding BCJ relation at amplitude level (see \cite{Hou:2018bwm,Du:2019vzf}).

%
%
%
%

In the coming sections, (\ref{Eq:BScurrent})-(\ref{Eq:OffBCJ1}) will be used to extract quadratic propagators from one-loop CHY formula.

\section{Four-point integrand with two gluons}\label{sec:SimpleExample}
  In this section, we study the single-trace four-point YMS integrands with two external gluons.  Inspired by these examples,  we find that the crucial point for extracting quadratic propagators is to cancel the nonlocal terms, where two adjacent subgraphs contracted together are separated by linear propagators on the Feynman diagram side. Once this cancellation is achieved, one can always find terms related by cyclic permutations. Therefore, we can extract the quadratic propagators according to \eqref{Eq:partial}. The examples in this section show two typical approaches to canceling the nonlocalities.


\subsection{Four-point integrand with two gluons}\label{sec:TwoGluonEG}
\begin{figure}
\centering
\includegraphics[width=1\textwidth]{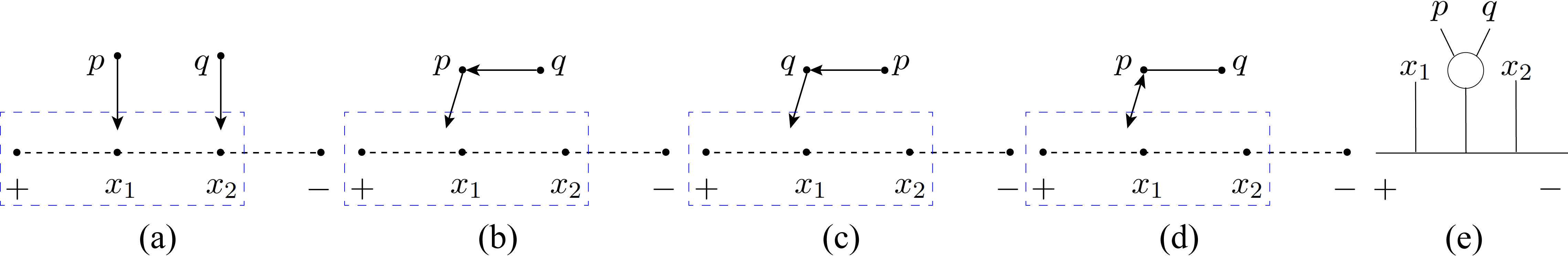}
\caption{ All possible graphs for the integrand with two gluons $p$, $q$ and two scalars $x_1$, $x_2$ are given by (a), (b), (c) and (d). The end node of an arrow inside the boxed region can be any one of $+$, $x_1$, $x_2$. The reference order is $p\prec q$ (i.e., the weight of $p$ is lower than that of $q$). A Feynman diagram with a linear propagator structure ${1\over s_{x_1,l}}{1\over s_{x_1pq,l}}$ for the partition $\{x_1,\{p,q\},x_2\}$ is given by (e).}
\label{Fig:4ptAll}
\end{figure}

 As shown by \tabref{table:LoopIntegrands}, the CHY half integrands of  one-loop single-trace YMS amplitudes with scalar trace $(x_1x_2)$ and two gluons $p$, $q$ are given by
\bea
I^{\text{1-loop}}_{\text{L}}\big(x_1,x_2||\{p,q\}\big)&=&\text{PT}(+,x_1,x_2,-)\text{Pf}\,[\,\Psi_{pq}\,]+\text{cyc}(x_1x_2),\, \nn
I^{\text{1-loop}}_{\text{R}}\big(a_1,a_2,a_3,a_4\big)&=&\text{PT}(+,a_1,a_2,a_3,a_4,-)+\text{cyc}(a_1a_2a_3a_4)\,,
\Label{Eq:4pt2glu}
\eea
where $(a_1a_2a_3a_4)$ is an arbitrary permutation of $x_1$, $x_2$, $p$, $q$. By applying (\ref{Eq:Expansion1}), the left half integrand in the above expression can  further be expanded as
\bea
I^{\text{1-loop}}_{\text{L}}\big(x_1,x_2||\{p,q\}\big)=\Sl_{\mathcal{F}}\,\mathcal{C}^{\mathcal{F}}\,\left[\,\Sl_{\pmb{\sigma}^{\mathcal{F}}}\text{PT}\left(+,\pmb{\sigma}^{\mathcal{F}},-\right)\,\right]+\text{cyc}(x_1x_2)\, .
\Label{Eq:4pt2glu1}
\eea
Here all the graphs $\mathcal{F}$, i.e., the graphs \figref{Fig:4ptAll} (a), (b), (c) and (d), generated by the refined graphic rule are summed up.   The coefficients  $\mathcal{C}^{\mathcal{F}}$ for \figref{Fig:4ptAll} (a), (b), (c) and (d) are respectively given by
\bea
(\epsilon_p\cdot k_i)(\epsilon_q\cdot k_j),~~~(\epsilon_p\cdot k_i)(\epsilon_q\cdot k_p),~~~(\epsilon_q\cdot k_i)(\epsilon_p\cdot k_q),~~~(-k_p\cdot k_i)(\epsilon_p\cdot \epsilon_q),
\eea
where each of $i$ and $j$ can be $x_1$, $x_2$ or $+$. If $i=x_1$, $j=x_2$, permutations $\pmb{\sigma}^{\mathcal{F}}$ for the graphs \figref{Fig:4ptAll} (a), (b), (c) and (d) are presented as 
\bea
\{x_1,\{p\}\shuffle\{x_2,q\}\},~\{x_1,\{p, q\}\shuffle\{x_2\}\},~\{x_1,\{q, p\}\shuffle\{x_2\}\},~\{x_1,\{p, q\}\shuffle\{x_2\}\}.
\eea
When the integral over scattering variables is taken, according to (\ref{Eq:One-loopBasis-1}), the one-loop integrand becomes
\bea
&&\frac{1}{l^2}\int \text{d}\mu_{\,6}^{\,\text{tree}}I_{\,\text{L}}^{\,\text{1-loop}}I_{\,\text{R}}^{\,\text{1-loop}}\Label{Eq:LoopCHYRefinedGraphs2}\\
&=&\Bigg[\Sl_{\mathcal{F}}\mathcal{C}^{\mathcal{F}}\Sl_{\substack{(A_1...A_i)={\pmb{\sigma}^{\mathcal{F}}}\\ (\W A_1...\W A_i)={(a_1a_2a_3a_4)}\\ |A_j|=|\W A_j|}}{1\over l^2}{1\over s_{A_1,l}}\cdots{1\over s_{A_1\cdots A_{i-1},l}}\phi_{A_1|\W A_1}\cdots \phi_{A_i|\W A_i}+\text{cyc}(x_1x_2)\Bigg]+\text{cyc}(a_1a_2a_3a_4).\nonumber
\eea
In the above expression, we divided each permutation $\pmb{\sigma}^{\mathcal{F}}$ into nonempty ordered subsets $A_1, A_2,...,A_i$ and then summed over all possible such divisions. The divisions $(\W A_1\W A_2...\W A_i)$ of $(a_1a_2a_3a_4)$ were also summed over. The $|A_j|=|\W A_j|$ for each $j$ means $A_j$ and $\W A_j$ have the same number of elements. Noting that $\phi_{A_j|\W A_j}=0$ when the elements of $A_j$ and $\W A_j$ are not identical to each other, the condition  $|A_j|=|\W A_j|$ is equivalent to the condition $A_j=\W A_j$ in (\ref{Eq:One-loopBasis-1}).


To transform the linear propagators in (\ref{Eq:LoopCHYRefinedGraphs2}) into quadratic propagators, we may rearrange (\ref{Eq:LoopCHYRefinedGraphs2}) into an expression with a cyclic sum so that (\ref{Eq:partial}) can be applied. However, the graphs $\mathcal{F}$ in general do not directly satisfy such cyclic property. To get a cyclic form, we first rewrite (\ref{Eq:LoopCHYRefinedGraphs2}) into the following form by collecting terms corresponding to a given propagator structure:
\bea
&&\frac{1}{l^2}\int \text{d}\mu_{\,6}^{\,\text{tree}}I_{\,\text{L}}^{\,\text{1-loop}}I_{\,\text{R}}^{\,\text{1-loop}}\Label{Eq:RearrangeLoopCHYRefinedGraphs2}\\
&=&\Bigg[\Sl_{\substack{\{A_1,A_2,...,A_i\}\\(\W A_1...\W A_i)={(a_1a_2a_3a_4)}\\ |A_j|=|\W A_j|}}{1\over l^2}{1\over s_{A_1,l}}\cdots{1\over s_{A_1\cdots A_{i-1},l}}\Sl_{\mathcal{F}}\mathcal{C}^{\mathcal{F}}\phi_{\pmb{\sigma}^{\mathcal{F}_1}|\W A_1}\cdots \phi_{\pmb{\sigma}^{\mathcal{F}_i}|\W A_i}+\text{cyc}(x_1x_2)\Bigg]+\text{cyc}(a_1a_2a_3a_4).\nonumber
\eea
In the above expression, for a fixed relative order of scalars $x_1$, $x_2$, we summed over all possible {\it partitions} $\{A_1,A_2,...,A_i\}$ ($i\leq 4$) of external particles $x_1$, $x_2$ and $p$, $q$. {\it Each partition  $\{A_1,A_2,...,A_i\}$ corresponds to a given product of linear propagators in the Feynman diagram (as shown by \figref{Fig:4ptAll} (e)), and is defined by splitting the external particles into $i$ disjoint nonempty subsets $A_1$, ..., $A_i$, which are arranged in a given order such that the relative order between scalars is preserved.} For example, in the case with two scalars $x_1$, $x_2$ (in the relative order $(x_1x_2)$) and two gluons $p$, $q$, all possible partitions $\{A_1,A_2\}$ with two subsets are given by
\bea
&&\{\{x_1\},\{x_2,p,q\}\},\{\{x_1,p,q\},\{x_2\}\},\{\{x_1,x_2,p\},\{q\}\},\{\{q\},\{x_1,x_2,p\}\},\{\{x_1,x_2,q\},\{p\}\},\nn
&&\{\{p\},\{x_1,x_2,q\}\},\{\{x_1,x_2\},\{p,q\}\},\{\{p,q\},\{x_1,x_2\}\},\{\{x_1,p\},\{x_2,q\}\},\{\{x_1,q\},\{x_2,p\}\}.
\eea
\begin{figure}
\centering
\includegraphics[width=0.65\textwidth]{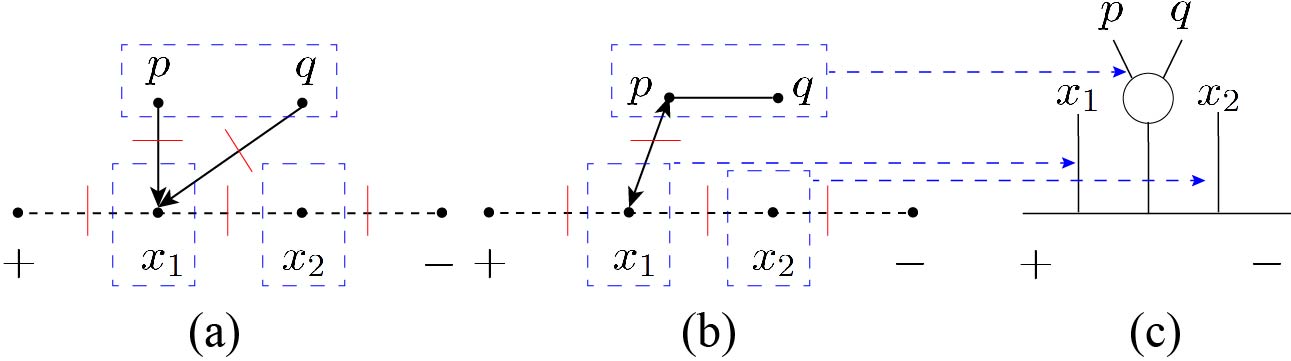}
\caption{In graphs (a) and (b), the red lines separate subgraphs (the boxed regions) from each other. Each subgraph (except for the $+$ and $-$) is associated with a subcurrent in the Feynman diagram (c). Graph (a) involves a disconnected subgraph that corresponds to a subcurrent $\phi_{p\shuffle q|...}$ on the Feynman diagram side. Thus (a) vanishes due to the $U(1)$-decoupling identity  (\ref{Eq:GenU1}).}
\label{Fig:4ptAll1}
\end{figure}
For a given partition $\{A_1,A_2,...,A_i\}$, we summed over all possible graphs $\mathcal{F}$ contributing to it. Each $\mathcal{F}$  can be decomposed into subgraphs $\mathcal{F}_1$, ..., $\mathcal{F}_i$ corresponding to subsets $A_1$, ..., $A_i$ in the partition, which means 
\begin{itemize}
\item (i). Each subgraph  $\mathcal{F}_j$ contains elements identical to those in a certain subset $A_j$.
 \item (ii). These subgraphs can be arranged in the same order as subsets, as shown by \figref{Fig:4ptAll1} (b) and (c).
\end{itemize}
The $\pmb{\sigma}^{\mathcal{F}_j}$ ($j=1,...,i$) in (\ref{Eq:RearrangeLoopCHYRefinedGraphs2}) denotes permutations established by subgraph $\mathcal{F}_j$. The summation over all possible $\pmb{\sigma}^{\mathcal{F}_j}$ is implied.
A subgraph $\mathcal{F}_j$ can either be {\it connected} or {\it disconnected}. A {\it disconnected subgraph}, e.g., \figref{Fig:4ptAll1} (a) contains disjoint parts $\mathcal{F}_A, \mathcal{F}_B, ...$ and produces $\phi_{\pmb{\sigma}^{\mathcal{F}_A}\shuffle\pmb{\sigma}^{\mathcal{F}_B}\shuffle...|\W C}$ which has to cancel out due to the generalized U(1)-decoupling identity (\ref{Eq:GenU1}). Thus, {\it all surviving subgraphs are connected ones}.

In the coming subsections, we show that (\ref{Eq:RearrangeLoopCHYRefinedGraphs2}) can be rearranged into a cyclic sum in (\ref{Eq:partial}), thus produces quadratic propagators. Our discussions will be carried out for three classes of graphs, according to the way the subgraphs are interconnected.

\subsection{The first class of graphs}
\begin{figure}
\centering
\includegraphics[width=0.94\textwidth]{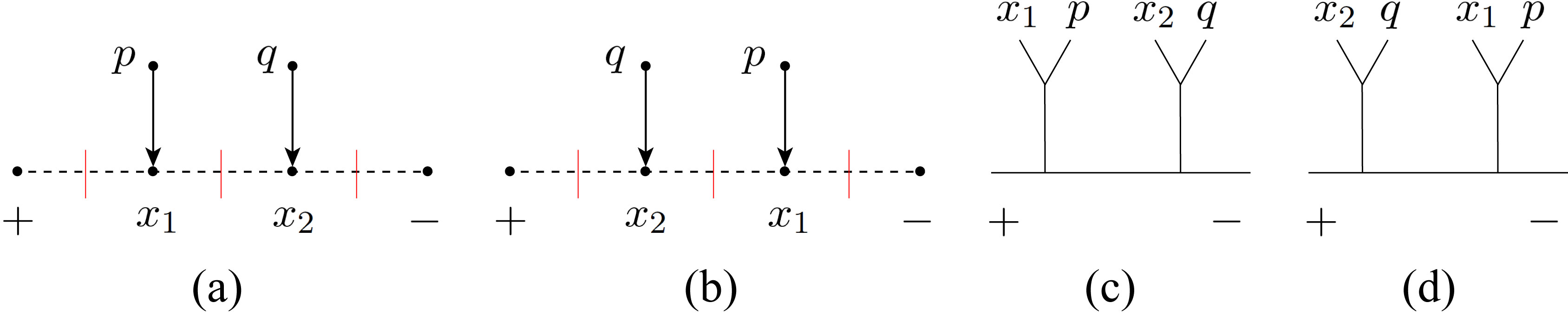}
\caption{Graph (a) is a typical graph whose two subgraphs are interconnected via type-4 lines, and graph (b) is the cyclic form of the two subgraphs in (a).  The Feynman diagrams under consideration which accompany to (a) and (b) are respectively (c) and (d), corresponding to  partitions of particles $\{\{x_1,p\},\{x_2,q\}\}$ and $\{\{x_2,q\},\{x_1,p\}\}$.  }
\label{Fig:4pt2glu1}
\end{figure}

 Graphs of the first class are those with all subgraphs interconnected by type-4 lines. The type-4 line is the line that only records the relative positions between particles and does not provide any kinematic factor. For the two-gluon case, a typical graph that produces subgraphs interconnected via type-4 lines is shown by \figref{Fig:4pt2glu1} (a), and the corresponding linear-propagator contribution (see the Feynman diagram \figref{Fig:4pt2glu1} (c)), can be written as
\bea
(\epsilon_p\cdot k_{x_1})(\epsilon_q\cdot k_{x_2}){ 1\over l^2  s_{x_1p,l} }\phi_{x_1p|a_1a_2}\phi_{x_2q|a_3a_4}\,.
\eea
\begin{figure}
\centering
\includegraphics[width=0.6\textwidth]{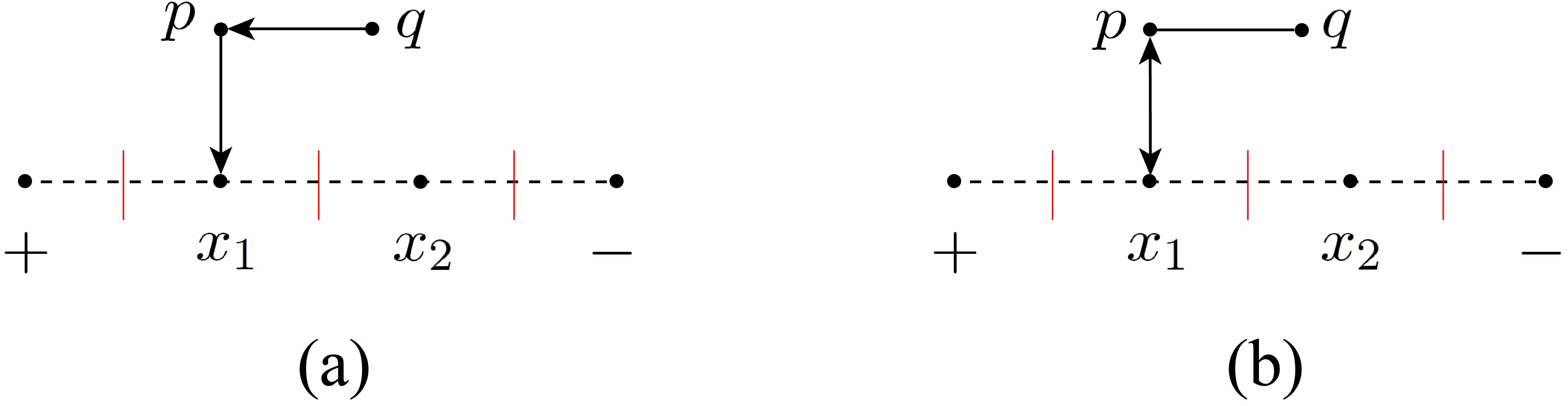}
\caption{In graphs (a) and (b), $p$, $q$ and $x_1$ belong to a same subgraph.}
\label{Fig:4pt2glu2}
\end{figure}
For the above term, there always exists another graph \figref{Fig:4pt2glu1} (b) which comes from  $\text{cyc}(x_1x_2)$ in (\ref{Eq:4pt2glu1}) and is accompanied by the Feynman diagram  \figref{Fig:4pt2glu1} (d).  The two subgraphs of \figref{Fig:4pt2glu1} (b) are exactly the cyclic form of the subgraphs in \figref{Fig:4pt2glu1} (a). Thus we can directly obtain the quadratic propagators by summing the contributions of the two graphs in \figref{Fig:4pt2glu1}
\bea
&&(\epsilon_p\cdot k_{x_1})(\epsilon_q\cdot k_{x_2}){ 1\over l^2  s_{x_1p,l} }\phi_{x_1p|a_1a_2}\phi_{x_2q|a_3a_4}+(\epsilon_p\cdot k_{x_1})(\epsilon_q\cdot k_{x_2}){ 1\over l^2  s_{x_2q,l} }\phi_{x_2q|a_3a_4}\phi_{x_1p|a_1a_2}\nn
&\cong&(\epsilon_p\cdot k_{x_1})(\epsilon_q\cdot k_{x_2}){ 1\over l^2  l^2_{x_1p} }\phi_{x_1p|a_1a_2}\phi_{x_2q|a_3a_4}\,,
\Label{Eq:4pt2glu3}
\eea
where the RHS permutations of subcurrents in the second term come from the $\text{cyc}(a_1a_2a_3a_4)$ in (\ref{Eq:4pt2glu}).

This class of graphs always provides quadratic propagators directly, since no kinematic factors are separated by linear propagators, i.e., there is no nonlocal term. Similarly, the graphs in \figref{Fig:4pt2glu2} give the quadratic propagators as follows
\bea
\text{(a)}.\,\,(\epsilon_p\cdot k_{x_1})(\epsilon_q\cdot k_{p}){ 1\over l^2  l^2_{x_1pq} }\phi_{x_1pq|a_1a_2a_3}\phi_{x_2|a_4}\,\,,\,\,\,\text{(b)}.\,\,(\epsilon_p\cdot\epsilon_q )(-k_p\cdot k_{x_1}){ 1\over l^2  l^2_{x_1pq} }\phi_{x_1pq|a_1a_2a_3}\phi_{x_2|a_4}\,.
\eea

\subsection{The second class of graphs and approach-1 to canceling nonlocal terms}\label{sec:TwoGluonEG1}
\begin{figure}
\centering
\includegraphics[width=0.85\textwidth]{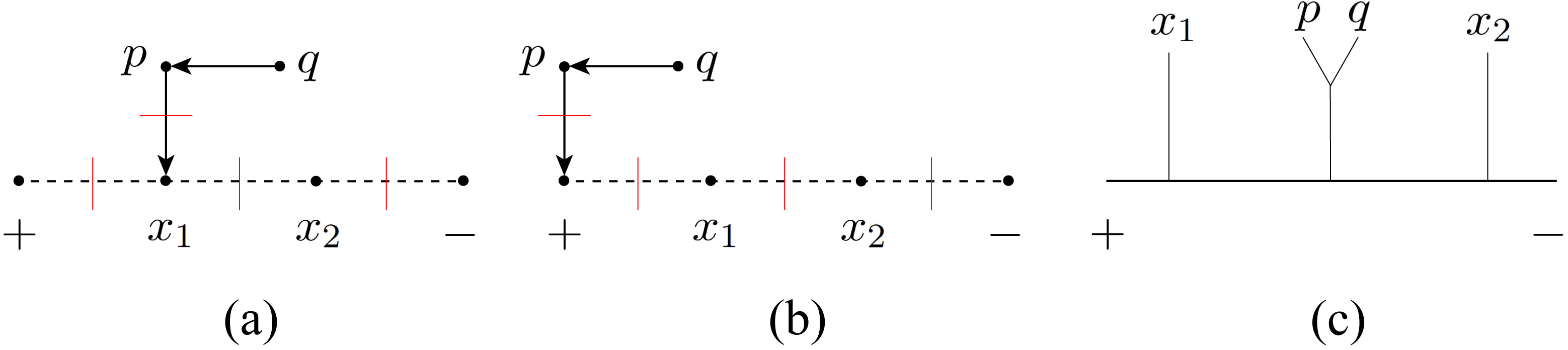}
\caption{Each of the  graphs (a) and (b), which involve $\epsilon_{p}\cdot k_{x_1}$ and $\epsilon_{p}\cdot l$ respectively, provides a nonlocal contribution to the Feynman diagram (c).}
\label{Fig:4pt2glu3}
\end{figure}
%


The second class of graphs for the four-point example involves subgraphs $A$ and $B$ interconnected via a $C_a\cdot k_b$ line, where $C_a^{\mu}$ can represent the polarization vector or momentum of a node in the subgraph $A$, and $k^{\mu}_b$ is the momentum of a node in subgraph $B$, e.g., \figref{Fig:4pt2glu3} (a). {\it Nonlocality} arises due to the nontrivial contraction between the kinematic factors of these two subgraphs, which are separated by linear propagators in the corresponding Feynman diagram. {\it For this class of graphs, we use approach-1 to treat these nonlocal terms: collecting graphs with the coefficients  $C_a\cdot k$ together into  $C_a\cdot X_A$, where $X^{\mu}_A$ is the momentum of the linear propagator to the left of the subcurrent associated with  $A$. Let us demonstrate this by examples.}

We first take \figref{Fig:4pt2glu3} as an example, where $p$, $q$ belong to the same subgraph in  \figref{Fig:4pt2glu3} (a), (b) and the same subcurrent in \figref{Fig:4pt2glu3} (c).  The graph \figref{Fig:4pt2glu3} (a) together with the Feynman diagram  \figref{Fig:4pt2glu3} (c) provide
\bea
(\epsilon_p\cdot k_{x_1})(\epsilon_{q}\cdot k_p){1\over l^2 s_{x_1,l} s_{x_1pq,l}}\phi_{x_1|a_1}\phi_{pq|a_2a_3}\phi_{x_2|a_4}.\Label{Eq:2gluonNonlocal1}
\eea
\begin{figure}
\centering
\includegraphics[width=1\textwidth]{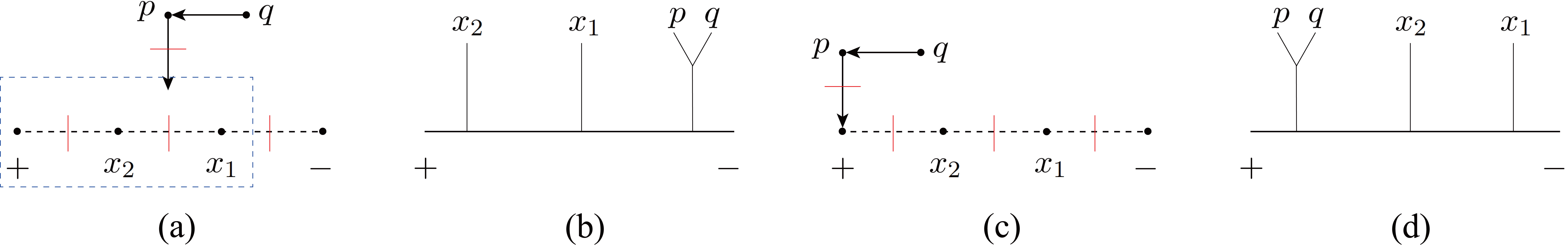}
\caption{Graphs (a), (c) accompanying to the Feynman diagrams (b), (d) provide the cyclic permutations of the subsets in \figref{Fig:4pt2glu3} (c) with graphs \figref{Fig:4pt2glu3} (a), (b). }
\label{Fig:4pt2glu4}
\end{figure}
Apparently, the subgraph with $p,q$ is contracted to the subgraph with $x_1$, via a type-2 line $\epsilon_p\cdot k_{x_1}$. In the Feynman diagram \figref{Fig:4pt2glu3} (c), these two subgraphs are separated by the linear propagator ${1\over s_{x_1,l}}$. Thus this is a nonlocal term. To deal with this nonlocality, let us consider \figref{Fig:4pt2glu3} (b) which provides another nonlocal term for the Feynman diagram \figref{Fig:4pt2glu3} (c)
\bea
(\epsilon_p\cdot l)(\epsilon_{q}\cdot k_p){1\over l^2 s_{x_1,l} s_{x_1pq,l}}\phi_{x_1|a_1}\phi_{pq|a_2a_3}\phi_{x_2|a_4}.\Label{Eq:2gluonNonlocal2}
\eea
The sum of (\ref{Eq:2gluonNonlocal1}) and (\ref{Eq:2gluonNonlocal2}) becomes
\bea
(\epsilon_p\cdot X_p)(\epsilon_{q}\cdot k_p){1\over l^2 s_{x_1,l} s_{x_1pq,l}}\phi_{x_1|a_1}\phi_{pq|a_2a_3}\phi_{x_2|a_4},\Label{Eq:2gluonNonlocal3}
\eea
where $X^{\mu}_p=l^{\mu}+k^{\mu}_{x_1}$ is the momentum of the linear propagator $1\over s_{x_1,l}$ attached to the subset $\{p,q\}$. Thus the nonlocal terms (\ref{Eq:2gluonNonlocal1}) and (\ref{Eq:2gluonNonlocal2})  have been summed into a local term (\ref{Eq:2gluonNonlocal3}). The cyclic permutations of $(x_1x_2)$ and $(a_1a_2a_3a_4)$ allow \figref{Fig:4pt2glu4} (a) and  (c) for the Feynman diagrams \figref{Fig:4pt2glu4} (b) and (d). According to (\ref{Eq:partial}), these contributions together with (\ref{Eq:2gluonNonlocal3}) produce a term with quadratic propagators
\bea
(\epsilon_p\cdot X_p)(\epsilon_{q}\cdot k_p){1\over l^2 l^2_{x_1} l^2_{x_1pq}}\phi_{x_1|a_1}\phi_{pq|a_2a_3}\phi_{x_2|a_4}.
\eea
\begin{figure}
\centering
\includegraphics[width=0.8\textwidth]{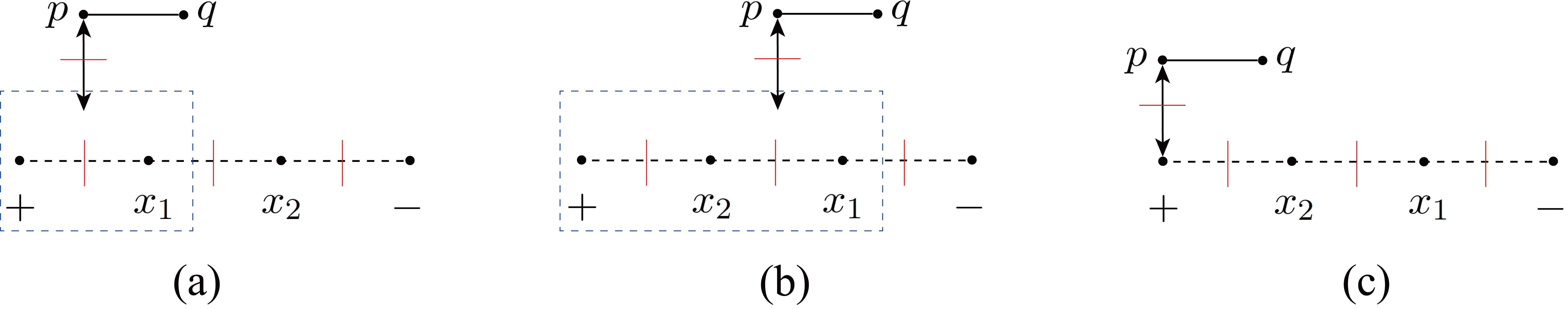}
\caption{Each of the graphs (a), (b) and (c) contains $p$, $q$ in a same subgraph and the kinematic factor is given by $(\epsilon_p\cdot\epsilon_q)(-k_p\cdot X_p)$. These three graphs contribute to the Feynman diagrams \figref{Fig:4pt2glu3} (c), \figref{Fig:4pt2glu4} (b) and (e), respectively.}
\label{Fig:4pt2glu5}
\end{figure}
\begin{figure}
\centering
\includegraphics[width=0.8\textwidth]{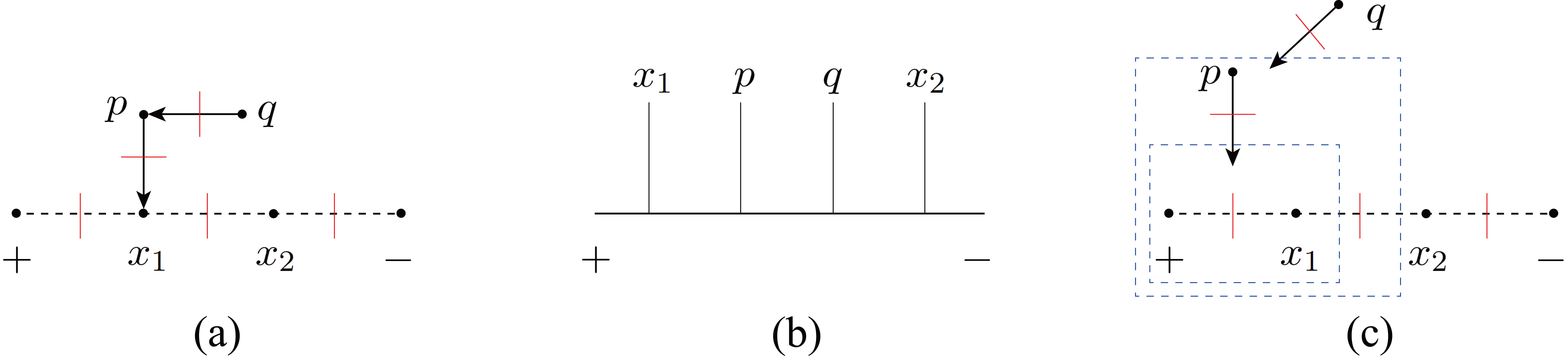}
\caption{Graph (a) which relates to the Feynman diagram (b), contains $p$ and $q$ in different subgraphs. The type-2 lines associating to $p$ and $q$ induce nonlocalities. These nonlocalities are overcomed by collecting all related graphs together, as shown by (c). }
\label{Fig:4pt2glu6}
\end{figure}
Similarly, the quadratic-propagator contribution of the graphs  shown by \figref{Fig:4pt2glu5} (a), (b) and (c) for the Feynman diagrams \figref{Fig:4pt2glu3} (c), \figref{Fig:4pt2glu4} (b) and (d) can be written as
\bea
(-k_p\cdot X_p)(\epsilon_{q}\cdot \epsilon_p){1\over l^2 l^2_{x_1} l^2_{x_1pq}}\phi_{x_1|a_1}\phi_{pq|a_2a_3}\phi_{x_2|a_4}.
\Label{Eq:4pt2glu4}
\eea

 The second example is \figref{Fig:4pt2glu6} (a), which is related to the Feynman diagram \figref{Fig:4pt2glu6} (b) and thus contains $p$ and $q$ in different subgraphs. For this case, two nonlocal contractions between $p$ and $x_1$, $q$ and $p$ need to be dealt with separately. So for the Feynman diagram  \figref{Fig:4pt2glu6} (b), we consider the contribution provided by the graphs in \figref{Fig:4pt2glu6} (c)
\bea
(\epsilon_p\cdot X_p)(\epsilon_{q}\cdot X_q){1\over l^2 s_{x_1,l} s_{x_1p,l}s_{x_1pq,l}}\phi_{x_1|a_1}\phi_{p|a_2}\phi_{q|a_3}\phi_{x_2|a_4},
\eea
where $X^{\mu}_p=l^{\mu}+k^{\mu}_{x_1}$ and $X^{\mu}_q=l^{\mu}+k^{\mu}_{x_1}+k^{\mu}_p$ are the momenta of the propagators attached to the subsets $\{p\}$ and $\{q\}$, respectively. The nonlocal terms are collected to give the local linear-propagator contribution. Thus, after a cyclic summation of the scalar particles, we obtain the quadratic-propagator term 
\bea
(\epsilon_p\cdot X_p)(\epsilon_{q}\cdot X_q){1\over l^2 l^2_{x_1} l^2_{x_1p}l^2_{x_1pq}}\phi_{x_1|a_1}\phi_{p|a_2}\phi_{q|a_3}\phi_{x_2|a_4}.
\Label{Eq:4pt2glu5}
\eea

For graphs with neither $\epsilon\cdot\epsilon$ nor $k\cdot k$ line, the $p$ and $q$ can belong to either the same subgraph or distinct subgraphs. These two situations can always be treated following a similar discussion with the first and the second example.
 The quadratic-propagator contribution of all possible graphs with neither $\epsilon\cdot\epsilon$ nor $k\cdot k$ line is given by 
\bea
&&{(\epsilon_p\cdot l_{x_1})(\epsilon_{q}\cdot l_{x_1px_2})\over l^2 l^2_{x_1} l^2_{x_1p}l^2_{x_1px_2}}\phi_{x_1|a_1}\phi_{p|a_2}\phi_{x_2|a_3}\phi_{q|a_4}
+{(\epsilon_p\cdot k_{x_1})(\epsilon_{q}\cdot k_{x_2})\over l^2 l^2_{x_1p} }\phi_{x_1p|a_1a_2}\phi_{x_2q|a_3a_4}\nn
&+&\bigg[{(\epsilon_p\cdot l_{x_1x_2})(\epsilon_{q}\cdot l_{x_1x_2p})\over l^2 l^2_{x_1} l^2_{x_1x_2}l^2_{x_1x_2p}}\phi_{x_1|a_1}\phi_{x_2|a_2}\phi_{p|a_3}\phi_{q|a_4}
+{(\epsilon_p\cdot l_{x_1x_2})(\epsilon_{q}\cdot l_{x_1x_2p})\over l^2 l^2_{x_1x_2} l^2_{x_1x_2p}}\phi_{x_1x_2|a_1a_2}\phi_{p|a_3}\phi_{q|a_4}\nn
&+&{(\epsilon_p\cdot k_{x_1})(\epsilon_{q}\cdot l_{x_1p})\over l^2 l^2_{x_1p} l^2_{x_1pq}}\phi_{x_1p|a_1a_2}\phi_{q|a_3}\phi_{x_2|a_4}
+{(\epsilon_p\cdot k_{x_1})(\epsilon_{q}\cdot l_{x_1px_2})\over l^2 l^2_{x_1p} l^2_{x_1px_2}}\phi_{x_1p|a_1a_2}\phi_{x_2|a_3}\phi_{q|a_4}\nn
&+&{(\epsilon_p\cdot l_{x_1})(\epsilon_{q}\cdot k_{p})\over l^2 l^2_{x_1} l^2_{x_1pq}}\phi_{x_1|a_1}\phi_{pq|a_2a_3}\phi_{x_2|a_4}
+{(\epsilon_p\cdot k_{x_1})(\epsilon_{q}\cdot l_{x_1px_2})\over l^2 l^2_{x_1px_2} }\phi_{x_1px_2|a_1a_2a_3}\phi_{q|a_4}\nn
&+&{(\epsilon_p\cdot k_{x_1x_2})(\epsilon_{q}\cdot l_{x_1x_2p})\over l^2 l^2_{x_1x_2p} }\phi_{x_1x_2p|a_1a_2a_3}\phi_{q|a_4}
+{(\epsilon_p\cdot k_{x_1})(\epsilon_{q}\cdot k_{x_1p})\over l^2 l^2_{x_1pq} }\phi_{x_1pq|a_1a_2a_3}\phi_{x_2|a_4}\nn
&+&{(\epsilon_p\cdot l_{x_1x_2})(\epsilon_{q}\cdot k_{p})\over l^2 l^2_{x_1x_2} }\phi_{x_1x_2|a_1a_2}\phi_{pq|a_3a_4}+\text{cyc}(x_1x_2)\bigg]
+(p\leftrightarrow q)+\text{cyc}(a_1a_2a_3a_4),\Label{Eq:FourPointResult}
\eea
where the $\epsilon\cdot k$ factors within each subgraph are preserved, and the $\epsilon\cdot k$ factors between subgraphs are collected as local factors $\epsilon\cdot X$. When $U(1)$-decoupling identity\footnote{The $U(1)$-decoupling identity for scalars can be extended to the subcurrents containing both scalars and gluons because it is essentially determined by the antisymmetry of cubic vertex in the BS current. } and the cyclic sum of $(x_1x_2)$ are considered, the terms where two scalars belong to a same subcurrent must vanish. {\it The sum of $\text{cyc}(a_1a_2a_3a_4)$ is introduced so that the left and right permutations in a BS subcurrent contain the same particles.} The above result can also be obtained by the tensorial PT-factor (\ref{Eq:TensorialPT}) decomposition of CHY left half integrand 
\bea
I^{\text{1-loop}}_{\text{L}}\big(x_1,x_2||\{p,q\}\big)=
&&(\epsilon_p\cdot X'_p(\shuffle))(\epsilon_q\cdot X'_q(\shuffle))\text{PT}_{\text{t}}(x_1,\{x_2\}\shuffle \{p\}\shuffle \{q\})\nn
&&~~+(\epsilon_p\cdot X'_p(\shuffle))(\epsilon_q)_{\mu}\text{PT}_{\text{t}}^{\mu}(x_1,\{x_2\}\shuffle \{p\}\shuffle \{q\})\nn
&&~~+(\epsilon_q\cdot X'_q(\shuffle))(\epsilon_p)_{\mu}\text{PT}_{\text{t}}^{\mu}(x_1,\{x_2\}\shuffle \{p\}\shuffle \{q\})\nn
&&~~+(\epsilon_p)_{\mu}(\epsilon_q)_{\nu}\text{PT}_{\text{t}}^{\mu\nu}(x_1,\{x_2\}\shuffle \{p\}\shuffle \{q\})\,,
\eea
where for a given permutation, ${X'}^{\mu}_p(\shuffle)$ is defined by the total momentum of external particles on the left-hand side of $p$. This tensorial PT factor can be generalized to half integrand with $n$ scalars and $2$ gluons
\bea
I^{\text{1-loop}}_{\text{L}}\big(x_1,x_2,...,x_n||\{p,q\}\big)=
&&(\epsilon_p\cdot X'_p(\shuffle))(\epsilon_q\cdot X'_q(\shuffle))\text{PT}_{\text{t}}(x_1,\{x_2,...,x_n\}\shuffle \{p\}\shuffle \{q\})\nn
&&~~+(\epsilon_p\cdot X'_p(\shuffle))(\epsilon_q)_{\mu}\text{PT}_{\text{t}}^{\mu}(x_1,\{x_2,...,x_n\}\shuffle \{p\}\shuffle \{q\})\nn
&&~~+(\epsilon_q\cdot X'_q(\shuffle))(\epsilon_p)_{\mu}\text{PT}_{\text{t}}^{\mu}(x_1,\{x_2,...,x_n\}\shuffle \{p\}\shuffle \{q\})\nn
&&~~+(\epsilon_p)_{\mu}(\epsilon_q)_{\nu}\text{PT}_{\text{t}}^{\mu\nu}(x_1,\{x_2,...,x_n\}\shuffle \{p\}\shuffle \{q\})\,.
\eea
According to \eqref{Eq:gon1}, the above expression gives the one-loop quadratic-propagator YMS integrand without $\epsilon\cdot\epsilon$ factor as follows
\bea
&&I^{\text{1-loop}}\big(x_1,x_2,...,x_n||\{p,q\}\big|\pmb{\rho}\big)\Label{Eq:ShiftGon}\\
&\cong&
\Sl_{\small\substack{(A_1A_2...A_m)=(x_1\{x_2...x_n\}\shuffle\{p\}\shuffle\{q\})\\(\widetilde{A_1}\widetilde{A_2}...\widetilde{A_m})=\pmb\rho\\A_j=\widetilde{A_j}\\2\leq m\leq n}} \left[(\epsilon_p\cdot (X_p(\shuffle)+k_{A}))(\epsilon_q\cdot (X_q(\shuffle)+k_{A})) gon(A_1,A_2,...,A_m)\prod_{i=1}^m \phi_{A_i|\widetilde{A_i}}\right],\nonumber
%
\eea
where  $X^{\mu}_p(\shuffle)$ denotes the sum of the loop momentum $l^{\mu}$ and the total momentum of the external particles $x_1$, ..., between $l^{\mu}$ and $p$.   We define ${A}$, ${\bar{A}}$  as the two parts separated by $x_1$ in  $A_1=Ax_1\bar{A}$. The $k_A^{\mu}$ is the total momentum of elements in $A$.
%
\begin{figure}
\centering
\includegraphics[width=0.85\textwidth]{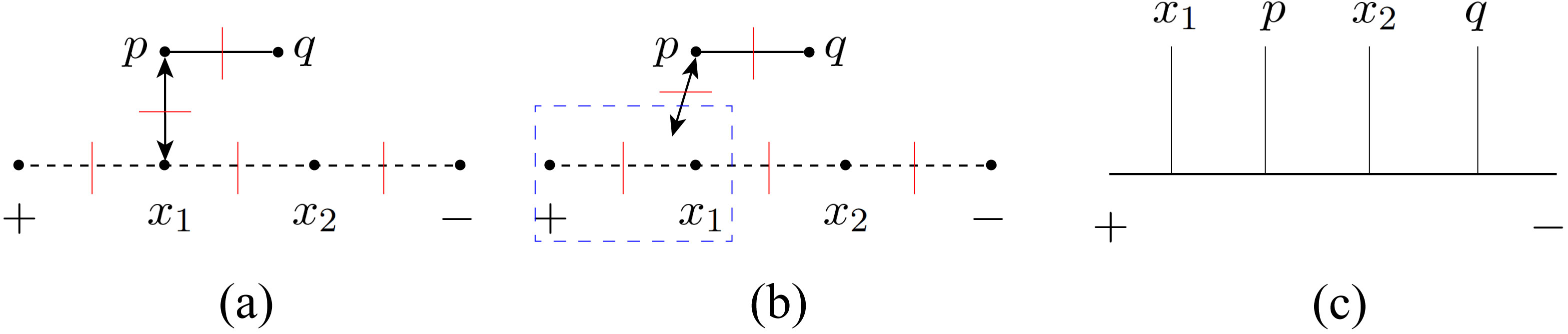}
\caption{In graph (a) associating with the Feynman diagram (c), the subgraph containing node $p$ is contracted with the one containing $q$, but they are separated by linear propagators in the Feynman diagram (c). This nonlocality cannot be treated by approach-1. In fact, when $p$ is connected to $+$, we also have such a nonlocal term with respect to the Feynman diagram (c). The two terms together are presented by (b) which induces an X-pattern by the subgraph $p$, since (b) brings a factor $-k_p\cdot X_p$ to the Feynman diagram (c).  }
\label{Fig:4pt2glu7}
\end{figure}
The expression (\ref{Eq:ShiftGon}) can further be generalized to the case which contains an arbitrary number of gluons and only $\epsilon\cdot k$ lines
\bea
I^{\text{1-loop}}\big(x_1,...,x_n||\mathsf{G}\big|\pmb{\rho}\big)&\cong&\Sl_{\small\substack{(A_1...A_m)=(x_1\{x_2...x_n\}\shuffle\text{perms}\,\mathsf{G})\\(\widetilde{A_1}...\widetilde{A_m})=\pmb\rho\\A_j=\widetilde{A_j}\\2\leq m\leq n}} \left[\,\prod_{g_k\in \mathsf{G}} \epsilon_{g_k}\cdot \left(X_{g_k}(\shuffle)+k_A\right) gon(A_1,...,A_m)\prod_{i=1}^m \phi_{A_i|\widetilde{A_i}}\,\right],\nn
\eea
where $\mathsf{G}$ is the gluon set and $\text{perms}\,\mathsf{G}$ denotes all permutations of elements in $\mathsf{G}$. In the case $n=2$ and $\mathsf{G}=\{p, q\}$, the above expression becomes (\ref{Eq:FourPointResult}).

We conclude that the YMS quadratic-propagator integrands without $\epsilon\cdot\epsilon$ factor can be obtained by dealing with the graphs coming from refined graphic rules or by expanding the CHY half integrand into tensorial PT factors. Although the tensorial PT-factor approach to quadratic propagators seems straightforward, the CHY half integrand with $\epsilon\cdot\epsilon$ factors may not be straightforwardly expanded to the tensorial PT factors. Therefore, in the rest of this paper, we still use the graphic expansion method to convert the linear propagators into quadratic ones.

\subsection{The third class of graphs and approach-2 to canceling the nonlocal terms}\label{sec:TwoGluonEG2}

In the third class of  graphs, some nonlocal contributions cannot be removed by assembling them into  $C\cdot X$ factors. For example, \figref{Fig:4pt2glu7} (a) is a typical graph of this class, due to the nonlocal contraction $\epsilon_p\cdot \epsilon_q$ between subgraphs $p$ and $q$. For this type of nonlocality, we introduce {\it approach-2} to deal with it, which relies on two helpful patterns, {\it BCJ-pattern} and {\it X-pattern}. Note that the nonlocality in \figref{Fig:4pt2glu7} (a) caused by $-k_p\cdot k_{x_1}$ can be treated first, by considering the graph where the subgraph $p$ is connected to $+$ via $-k_p\cdot l$. Thus the factors are summed into a local one $-k_p\cdot X_p$, as shown by \figref{Fig:4pt2glu7} (b). The linear-propagator  contribution of graph  \figref{Fig:4pt2glu7} (b) for Feynman diagram \figref{Fig:4pt2glu7} (c) can be expressed as follows
\bea
&&(-k_p\cdot X_p)(\epsilon_p\cdot\epsilon_q){1\over l^2 }{1\over s_{x_1,l} }{1\over s_{x_{1}p,l} }{1\over s_{x_1px_2,l} }\phi_{x_1|a_1}\phi_{p|a_2}\phi_{x_2|a_3}\phi_{q|a_4}\Label{Eq:02glu}\\
&=&(-{1\over2})(\epsilon_p\cdot\epsilon_q)\bigg[-\bigg({1\over l^2 }{1\over s_{x_{1}p,l} }{1\over s_{x_1px_2,l} }\bigg)+\bigg({1\over l^2 }{1\over s_{x_1,l} }{1\over s_{x_1px_2,l} }\bigg)\bigg]\phi_{x_1|a_1}\phi_{p|a_2}\phi_{x_2|a_3}\phi_{q|a_4}\,,\nonumber
\eea
\begin{figure}
\centering
\includegraphics[width=0.55\textwidth]{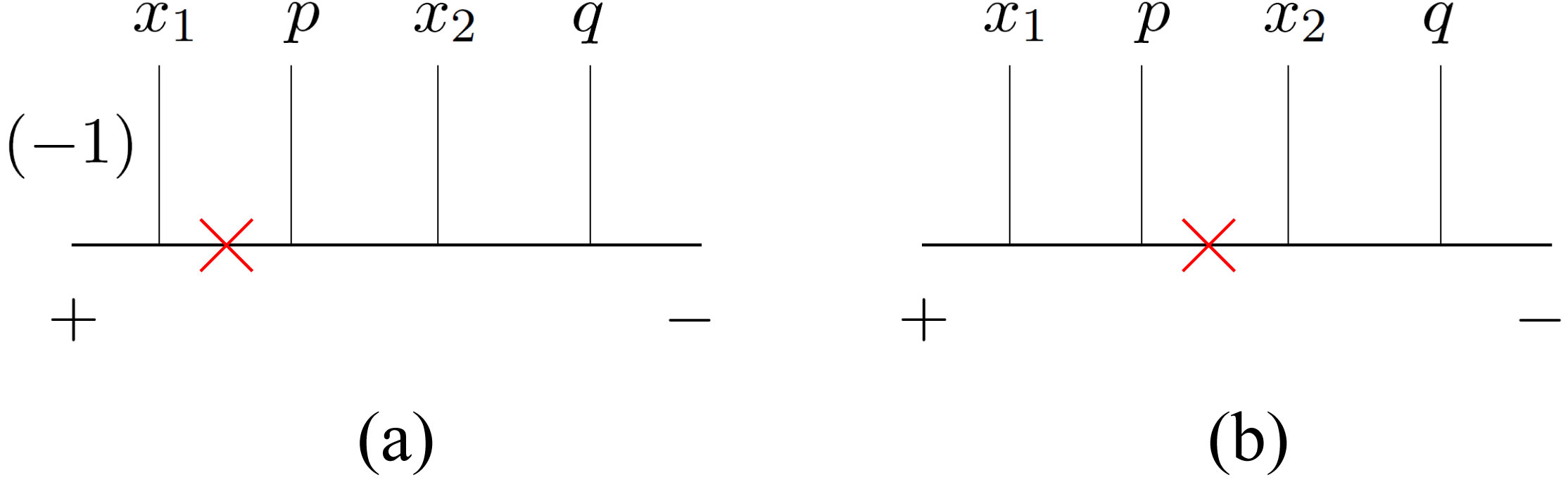}
\caption{The Feynman diagrams (a) and (b) correspond to the two terms on the second line of (\ref{Eq:02glu}).}
\label{Fig:4pt2glu8}
\end{figure}
\begin{figure}
\centering
\includegraphics[width=0.8\textwidth]{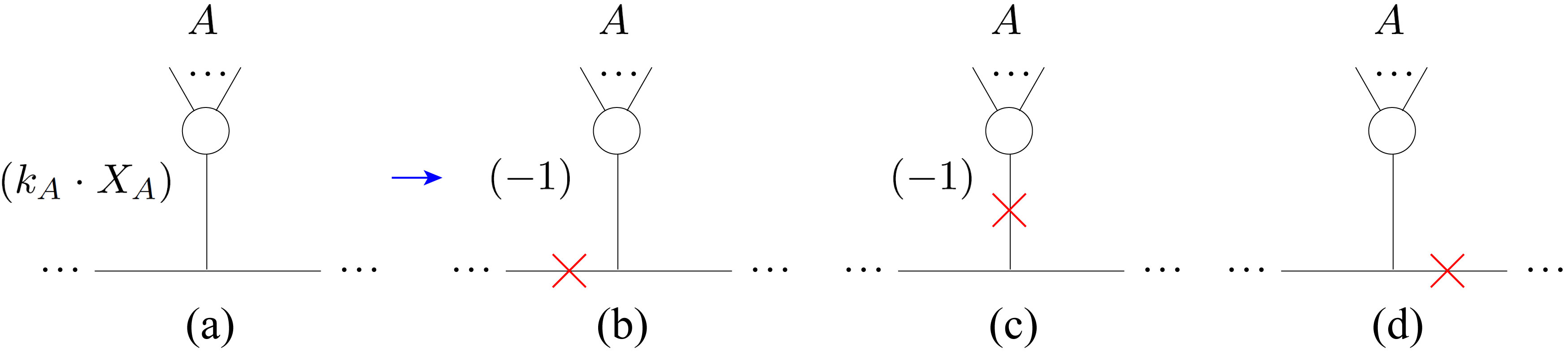}
\caption{When the X-pattern for subgraph $A$ exists, it can be used to delete the linear propagator to either the left (see (b)) or the right (see (d)) of the subcurrent containing $A$, or remove the off-shell line of the subcurrent containing $A$ (see (c)).}
\label{Fig:4pt2glu9}
\end{figure}
where we have used the relation
\bea
k_p\cdot X_p={1\over2}(l^2_{x_1p}-l^2_{x_1}-k^2_p)={1\over2}(s_{x_1p,l}-s_{x_1,l}),
\eea
to cancel out different linear propagators. The on-shell condition $k^2_p=0$ has been used. The two propagator-decreased terms in (\ref{Eq:02glu})  can be characterized by the Feynman diagrams \figref{Fig:4pt2glu8} (a) and (b), respectively \footnote{An overall factor $-{1\over 2}$ is omitted in each diagram.}. We say a graph has an {\it X-pattern} if this graph involves a factor $k_A\cdot X_A$, 
where $k^{\mu}_A$ denotes the total momentum of particles in subgraph\footnote{Here we use the same notation $A$ to denote the subset $A$ and a subgraph corresponding to this subset for convenience. In the general discussions in \secref{Sec:GenYMS} and \secref{sec:FeynDiagrams}, we introduce different notations for a subset and its corresponding subgraphs.} $A$, $X^{\mu}_A$ denotes the momentum of the propagator attached to the subcurrent corresponding to the subset $A$ from left,  in the Feynman diagram. Then the relation
\bea
k_A\cdot X_A={1\over2}\left[(k_A+ X_A)^2-X_A^2-k_A^2\right]={1\over2}\left[\big((k_A+ X_A)^2-l^2\big)-\big(X_A^2-l^2\big)-k_A^2\right]\,,\Label{Eq:XCoefficient}
\eea
can be applied to remove the propagators correspondingly, as shown by Feynman diagrams in \figref{Fig:4pt2glu9}. Apparently, \figref{Fig:4pt2glu7} (b) has an X-pattern. More general discussions of X-pattern will be given in the next section.

%
\begin{figure}
\centering
\includegraphics[width=0.85\textwidth]{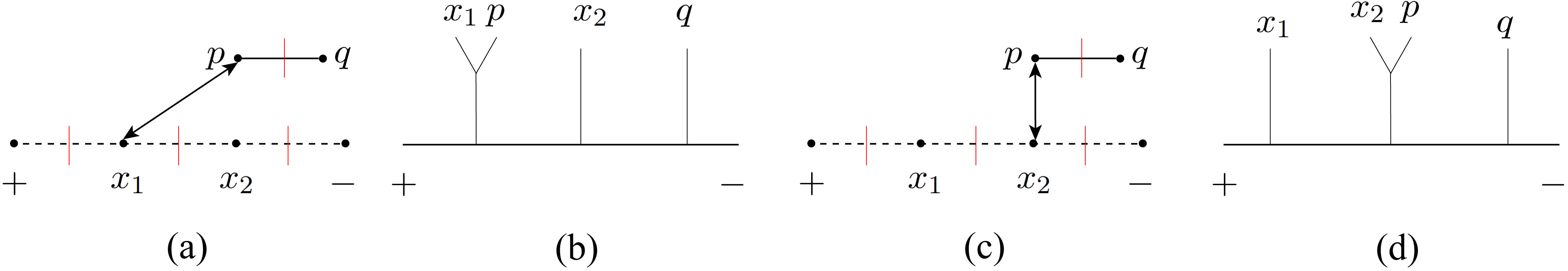}
\caption{Both  graphs (a) and (c) contain a BCJ-pattern since they each has a subgraph with $p$, $x_1$ or $p$, $x_2$, which contributes the LHS of off-shell BCJ relation for the corresponding subcurrent in the related Feynman diagram (b) or (d).  }
\label{Fig:4pt2glu10}
\end{figure}

If there exists a graph with an X-pattern, there also exist some graphs corresponding to it with {\it BCJ-patterns}, which means the graph has a subgraph that can use the off-shell BCJ relation (\ref{Eq:OffBCJ1}). For the graph in \figref{Fig:4pt2glu7} (b), there exists graph \figref{Fig:4pt2glu10} (a) with respect to the Feynman diagram \figref{Fig:4pt2glu10} (b).  This graph contains a BCJ-pattern and contributes
\bea
&&(-k_p\cdot k_{x_1})(\epsilon_p\cdot\epsilon_q){1\over l^2 }{1\over s_{x_1p,l} }{1\over s_{x_1px_2,l} }\phi_{x_1p|a_1a_2}\phi_{x_2|a_{3}}\phi_{q|a_4}\Label{Eq:0Ai-1}\\
&=&(-{1\over2})(\epsilon_p\cdot\epsilon_q)\bigg({1\over l^2 }{1\over s_{x_1p,l} }{1\over s_{x_1px_2,l} }\bigg)\left[-\phi_{p|a_1}\phi_{x_1|a_2}\phi_{x_2|a_{3}}\phi_{q|a_{4}}+\phi_{x_1|a_1}\phi_{p|a_2}\phi_{x_2|a_{3}}\phi_{q|a_4}\right],\nonumber
\eea
where the off-shell BCJ relation (\ref{Eq:OffBCJ1}) has been applied to rewriting $(k_p\cdot k_{x_1})\phi_{x_1p|a_1a_2}$ as ${1\over2}\phi_{x_{1}|a_{1}}\phi_{p|a_2}-{1\over2}\phi_{p|a_{1}}\phi_{x_{1}|a_2}$. A general discussion on BCJ-pattern will be provided in the next section. It can be seen that the second term of the above expression is canceled out by the first term of (\ref{Eq:02glu}). This cancellation between BCJ-pattern and X-pattern is not occasional, we can find another graph with X-pattern to cancel out the first term of (\ref{Eq:0Ai-1}) and another graph with BCJ-pattern to cancel out the second term of (\ref{Eq:02glu}). More precisely, the latter graph containing $x_2$ and $p$ in the same subgraph, as shown by \figref{Fig:4pt2glu10} (c), provides the linear-propagator contribution with respect to the Feynman diagram \figref{Fig:4pt2glu10} (d)
\bea
(-{1\over2})(\epsilon_p\cdot\epsilon_q)\bigg({1\over l^2 }{1\over s_{x_1,l} }{1\over s_{x_{1}x_2p,l} }\bigg)\left[-\phi_{x_1|a_1}\phi_{p|a_2}\phi_{x_2|a_{3}}\phi_{q|a_{4}}+\phi_{x_1|a_1}\phi_{x_2|a_2}\phi_{p|a_{3}}\phi_{q|a_4}\right],
\Label{Eq:0Ai+1}
\eea
where the off-shell BCJ relation (\ref{Eq:OffBCJ1}) has also been applied to $(k_p\cdot k_{x_2})\phi_{x_2p|a_2a_3}$. Thus, the first term of the above expression actually cancels out the second term of (\ref{Eq:02glu}). The complete cancellation map is given by \figref{Fig:4pt2glu11} in \appref{app:CancellationMap}. The BCJ-patterns and the X-patterns cancel out each other  until reaching the boundary case, where no scalar subcurrent is located between subcurrents $p$ and $q$ in the Feynman diagram, as shown by \figref{Fig:4pt2glu11} (c1). The contribution obtained by this boundary case can be written as
\bea
&&(-k_p\cdot X_p)(\epsilon_p\cdot\epsilon_q){1\over l^2 }{1\over s_{x_1,l} }{1\over s_{x_{1}x_2,l} }{1\over s_{x_{1}x_2p,l} }\phi_{x_1|a_1}\phi_{x_2|a_2}\phi_{p|a_3}\phi_{q|a_{4}}\Label{Eq:0Ai2}\\
&&=(-{1\over2})(\epsilon_p\cdot\epsilon_q)\bigg[-\bigg({1\over l^2 }{1\over s_{x_{1},l} }{1\over s_{x_{1}x_2p,l} }\bigg)+\bigg({1\over l^2 }{1\over s_{x_1,l} }{1\over s_{x_{1}x_2,l} }\bigg)
\bigg]\phi_{x_1|a_1}\phi_{x_2|a_2}\phi_{p|a_3}\phi_{q|a_{4}},\nonumber
\eea
with respect to the Feynman diagrams \figref{Fig:4pt2glu11} (c2) and (c3).
Although the first term from the X-pattern also cancels out with the term from a certain BCJ-pattern, the second term is retained with the factor $\epsilon_p\cdot\epsilon_q$ local. After the cyclic summations of $(x_1x_2)$ and $(a_1a_2a_3a_4)$, we finally get the quadratic-propagator contribution 
\bea
&&(-{1\over2})(\epsilon_p\cdot\epsilon_q)\bigg[\,{1\over l^2 }{1\over s_{x_1,l} }{1\over s_{x_{1}x_2,l} }+{1\over l^2 }{1\over s_{pq,l} }{1\over s_{pqx_{1},l} }+{1\over l^2 }{1\over s_{x_{2},l} }{1\over s_{x_{2}pq,l} }\,\bigg]\phi_{x_1|a_1}\phi_{x_2|a_2}\phi_{p|a_3}\phi_{q|a_{4}}\nn
&=&(-{1\over2})(\epsilon_p\cdot\epsilon_q)\left[{1\over l^2 }{1\over l^2_{x_1} }{1\over l^2_{x_{1}x_2,l} }\right]\phi_{x_1|a_1}\phi_{x_2|a_2}\phi_{p|a_3}\phi_{q|a_{4}}\,.
\eea
In general, when other Feynman diagrams are considered, the BCJ-patterns  and the X-patterns where $p$ and $q$ are separated by other subcurrents have to cancel out. The only surviving terms are those Feynman diagrams where $p$, $q$ are adjacent to each other and the linear propagator between them is removed. All such surviving terms are collected as
\bea
&&(-{1\over2})(\epsilon_p\cdot\epsilon_q)\Bigg[{1\over l^2 }{1\over l^2_{x_1} }{1\over l^2_{x_1pq,l} }\phi_{x_1|a_1}\phi_{p|a_2}\phi_{q|a_3}\phi_{x_2|a_{4}}+{1\over l^2 }{1\over l^2_{x_1} }{1\over l^2_{x_1x_2,l} }\phi_{x_1|a_1}\phi_{x_2|a_2}\phi_{p|a_3}\phi_{q|a_{4}}\nn
&+&{1\over l^2 }{1\over l^2_{x_1x_2} }\phi_{x_1x_2|a_1a_2}\phi_{p|a_3}\phi_{q|a_4}+{1\over l^2 }{1\over l^2_{x_2x_1} }\phi_{x_2x_1|a_1a_2}\phi_{p|a_3}\phi_{q|a_4}\Bigg]
+\text{cyc}(a_1a_2a_3a_4).\Label{Eq:0Ai3}
\eea
When the U(1)-decoupling identity (\ref{Eq:GenU1}) is applied, the last two terms cancel out, the above expression gets a further simplification
\bea
&&(-{1\over2})(\epsilon_p\cdot\epsilon_q)\Bigg[{1\over l^2 }{1\over l^2_{x_1} }{1\over l^2_{x_1pq,l} }\phi_{x_1|a_1}\phi_{p|a_2}\phi_{q|a_3}\phi_{x_2|a_{4}}+{1\over l^2 }{1\over l^2_{x_1} }{1\over l^2_{x_1x_2,l} }\phi_{x_1|a_1}\phi_{x_2|a_2}\phi_{p|a_3}\phi_{q|a_{4}}\Bigg]+\text{cyc}(a_1a_2a_3a_4).\Label{Eq:0Ai4}\nn
\eea

\subsection{Compact expression of the four-point integrand with two gluons}\label{sec:TwoGluonEG3}
In the previous discussions, we have shown how to extract the quadratic propagators in the four-point example with two gluons. Our starting point is the  linear-propagator-expressed Feynman diagrams (LPFD), whose coefficients are expressed by the refined graphs. When the nonlocalities are canceled out and the cyclic permutations of scalars are considered, we get the expressions with quadratic propagators. In this subsection, we assemble all results corresponding to any given structure of  quadratic propagators  (i.e., a given  {\it cyclic partition}\footnote{Partitions associating with quadratic propagators in the final result are cyclic partitions which means $\{A_1,A_2,...,A_i\}=\{A_2,...,A_i,A_1\}$, while the partitions accompanied to LPFD are not cyclic ones. } of external particles) and provide a compact formula of the result.

 First, partitions in which each subset contains only one element provide box diagrams. The first term in (\ref{Eq:FourPointResult}) and the first term in the square brackets of (\ref{Eq:FourPointResult}),
\bea
{(\epsilon_p\cdot l_{x_1})(\epsilon_{q}\cdot l_{x_1px_2})\over l^2 l^2_{x_1} l^2_{x_1p}l^2_{x_1px_2}}\phi_{x_1|a_1}\phi_{p|a_2}\phi_{x_2|a_3}\phi_{q|a_4}\,,\,\,\,{(\epsilon_p\cdot l_{x_1x_2})(\epsilon_{q}\cdot l_{x_1x_2p})\over l^2 l^2_{x_1} l^2_{x_1x_2}l^2_{x_1x_2p}}\phi_{x_1|a_1}\phi_{x_2|a_2}\phi_{p|a_3}\phi_{q|a_4}
\eea
correspond to such partitions $\{x_1,p,x_2,q\}$ and $\{x_1,x_2,p,q\}$, respectively.  If the subset has only one scalar, e.g., $x_1$ for partition $\{x_1,x_2,p,q\}$, it supplies a scalar subcurrent $\phi_{x_1|a_1}$. If the subset contains only one gluon, e.g., $p$ for partition $\{x_1,x_2,p,q\}$, it provides $(\epsilon_p\cdot l_{x_1x_2})\phi_{x_1|a_1}=\W J(p)\cdot X_p$. The effective current $\W J^{\mu}(p)=\epsilon_p^{\mu}\phi_{x_1|a_1}$ is  defined by  refined graphic rule when the Lorentz index (associated with off-shell line) is considered as the root. The momentum $X^{\mu}_p=l^{\mu}_{x_1x_2}$ is just the momentum of the (quadratic) propagator to the left of the subcurrent $\W J^{\mu}(p)$. The contributions of all box diagrams can be collected as
\bea
&&{(\epsilon_p\cdot l_{x_1x_2})(\epsilon_{q}\cdot l_{x_1x_2p})\over l^2 l^2_{x_1} l^2_{x_1x_2}l^2_{x_1x_2p}}\phi_{x_1|a_1}\phi_{x_2|a_2}\phi_{p|a_3}\phi_{q|a_4}+{(\epsilon_p\cdot l_{x_1})(\epsilon_{q}\cdot l_{x_1px_2})\over l^2 l^2_{x_1} l^2_{x_1p}l^2_{x_1px_2}}\phi_{x_1|a_1}\phi_{p|a_2}\phi_{x_2|a_3}\phi_{q|a_4}\nn
&&+{(\epsilon_p\cdot l_{x_1})(\epsilon_{q}\cdot l_{x_1p})\over l^2 l^2_{x_1} l^2_{x_1p}l^2_{x_1pq}}\phi_{x_1|a_1}\phi_{p|a_2}\phi_{q|a_3}\phi_{x_2|a_4}+(p\leftrightarrow q)+\text{cyc}(a_1a_2a_3a_4)\nn
&=&{(\W J(p)\cdot X_p)(\W J(q)\cdot X_q)\over l^2 l^2_{x_1} l^2_{x_1x_2}l^2_{x_1x_2p}}\phi_{x_1|a_1}\phi_{x_2|a_2}+{(\W J(p)\cdot X_p)(\W J(q)\cdot X_q)\over l^2 l^2_{x_1} l^2_{x_1p}l^2_{x_1px_2}}\phi_{x_1|a_1}\phi_{x_2|a_3}\nn
&&+{(\W J(p)\cdot X_p)(\W J(q)\cdot X_q)\over l^2 l^2_{x_1} l^2_{x_1p}l^2_{x_1pq}}\phi_{x_1|a_1}\phi_{x_2|a_4}+(p\leftrightarrow q)+\text{cyc}(a_1a_2a_3a_4)
\eea 
where we don't have term $\text{cyc}(x_1x_2)$, since partition $\{x_2,x_1,p,q\}$ is the same as $\{x_1,p,q,x_2\}$ at the loop level.

Second, partitions involving only one two-element subset provide triangle diagrams. There are three types of such partitions which correspond to the two-element subset containing (i). two scalars, (ii). a scalar and a gluon, (iii). two gluons:
\begin{itemize}
\item For the case (i), consider the partitions $\{\{x_1,x_2\},p,q\}$ and $\{\{x_2,x_1\},p,q\}$, whose contribution are given by the second term in the square brackets and $\text{cyc}(x_1x_2)$ term  of (\ref{Eq:FourPointResult})
\bea
{(\epsilon_p\cdot l_{x_1x_2})(\epsilon_{q}\cdot l_{x_1x_2p})\over l^2 l^2_{x_1x_2} l^2_{x_1x_2p}}\phi_{x_1x_2|a_1a_2}\phi_{p|a_3}\phi_{q|a_4}\,,\,\,\,{(\epsilon_p\cdot l_{x_2x_1})(\epsilon_{q}\cdot l_{x_2x_1p})\over l^2 l^2_{x_2x_1} l^2_{x_2x_1p}}\phi_{x_2x_1|a_1a_2}\phi_{p|a_3}\phi_{q|a_4}.
\eea
The numerator in the above expression can be rewritten as $\phi_{x_1x_2|a_1a_2}(\W J(p)\cdot X_p)(\W J(q)\cdot X_q)$ and $\phi_{x_2x_1|a_1a_2}(\W J(p)\cdot X_p)(\W J(q)\cdot X_q)$. Due to the $U(1)$-decoupling identity, the sum of these two partitions vanishes.

\item For the case (ii), consider the partition $\{\{x_1,p\},x_2,q\}$, whose contribution  is the fourth term in the square brackets of (\ref{Eq:FourPointResult})
\bea
{(\epsilon_p\cdot k_{x_1})(\epsilon_{q}\cdot l_{x_1px_2})\over l^2 l^2_{x_1p} l^2_{x_1px_2}}\phi_{x_1p|a_1a_2}\phi_{x_2|a_3}\phi_{q|a_4}.
\eea
The numerator is given by $\W J(x_1,p)\,\phi_{x_2|a_3}\,(\W J(q)\cdot X_q)$, where $\W J(x_1,p)$ is the effective current generated by refined graph about $x_1$ and $p$. We choose the scalar $x_1$ as the root of graph, rather than the off-shell line which is the root for subset with only gluons.

\item The case (iii) reveals a new feature. For example, the partition $\{x_1,\{p,q\},x_2\}$ gets contributions from (\ref{Eq:4pt2glu4}), the fifth term in the square brackets of  (\ref{Eq:FourPointResult}), as well as the $p\leftrightarrow q$ term of  (\ref{Eq:FourPointResult}), and explicitly provides
\bea
\Bigg[(-k_p\cdot X_p)(\epsilon_{q}\cdot \epsilon_p)\phi_{pq|a_2a_3}+(\epsilon_p\cdot X_p)(\epsilon_q\cdot k_p)\phi_{pq|a_2a_3}
+(\epsilon_q\cdot X_q)(\epsilon_p\cdot k_p)\phi_{qp|a_2a_3}\Bigg]{\phi_{x_1|a_1}\phi_{x_2|a_4}\over l^2 l^2_{x_1} l^2_{x_1pq} }.
\eea
The expression inside the square brackets is just $\W J(p,q)\cdot X_{\{p,q\}}$, where the effective current $\W J^{\mu}(p,q)$ corresponds to the sum of refined graphs with the reference order $p\prec q$. 
In addition, when the cancellation between X-pattern and BCJ-pattern has been considered, the graph with $\epsilon_p\cdot\epsilon_q$ line also provides a nontrivial term (the first term of (\ref{Eq:0Ai3})) for the partition $\{x_1,\{p,q\},x_2\}$
\bea
\left(-{1\over2}\right)(\epsilon_p\cdot\epsilon_q){1\over l^2 }{1\over l^2_{x_1} }{1\over l^2_{x_1pq} }\phi_{x_1|a_1}\phi_{p|a_2}\phi_{q|a_3}\phi_{x_2|a_{4}}.
\eea 
This term merged $\{p\}$ and $\{q\}$ into a single subset $\{p,q\}$, and its numerator can be expressed as $\left(-{1\over2}\right)\phi_{x_1|a_1}(\W J(p)\cdot\W J(q))\phi_{x_2|a_{4}}$. 
Thus the final expression for the partition $\{x_1,\{p,q\},x_2\}$ is given by 
\bea
\phi_{x_1|a_1}\left[\W J(p,q)\cdot X_{\{p,q\}}+\left(-{1\over2}\right)(\W J(p)\cdot\W J(q))\right]\phi_{x_2|a_4}{1\over l^2 }{1\over l^2_{x_1} }{1\over l^2_{x_1pq} }.
\eea 
\end{itemize}
The contributions of all triangle diagrams can then be collected as
\bea
&&\left[\W J(p,q)\cdot X_{\{p,q\}}+\left(-{1\over2}\right)(\W J(p)\cdot\W J(q))\right]{\phi_{x_1|a_1}\phi_{x_2|a_4}\over l^2l^2_{x_1} l^2_{x_1pq} }
+\left[\W J(p,q)\cdot X_{\{p,q\}}+\left(-{1\over2}\right)(\W J(p)\cdot\W J(q))\right]{\phi_{x_1|a_1}\phi_{x_2|a_2}\over l^2l^2_{x_1} l^2_{x_1x_2} }\nn
&&~~~+\Bigg[{(\W J(p)\cdot X_p)(\W J(q)\cdot X_q)\over l^2 l^2_{x_1x_2} l^2_{x_1x_2p}}\phi_{x_1x_2|a_1a_2}+{(\W J(p)\cdot X_p)(\W J(q)\cdot X_q)\over l^2 l^2_{x_2x_1} l^2_{x_2x_1p}}\phi_{x_2x_1|a_1a_2}+{\W J(x_1,p)(\W J(q)\cdot X_q)\over l^2 l^2_{x_1p} l^2_{x_1px_2}}\phi_{x_2|a_3}\nn
&&~~~+{\W J(x_1,p)(\W J(q)\cdot X_q)\over l^2 l^2_{x_1p} l^2_{x_1pq}}\phi_{x_2|a_4}+{\W J(x_2,p)(\W J(q)\cdot X_q)\over l^2 l^2_{x_1} l^2_{x_1x_2p}}\phi_{x_1|a_1}+(p\leftrightarrow q)\Bigg]+\text{cyc}(a_1a_2a_3a_4),
\eea
where the $X^{\mu}_{\{p,q\}}$ denote the momentum of the propagator attached to the subset $\{p,q\}$ from left. Particularly, $X^{\mu}_{\{p,q\}}=l^{\mu}_{x_1}$ and  $X^{\mu}_{\{p,q\}}=l^{\mu}_{x_1x_2}$ for the first and the second squarebrackets. The first two terms on the second line cancel out due to the $U(1)$-decoupling identity.

Third, partitions containing two subsets provide bubble diagrams. There are also three possibilities here and the final results can be collected following discussions similar to the above:
\begin{itemize}
\item There exists a three-element subset which contains both gluons and scalars. For example, the contribution of the partition $\{\{x_1,p,q\},x_2\}$ is
\bea
\W J(x_1,p,q)\phi_{x_2|a_4}{1\over l^2l^2_{x_1pq} }=\Bigg[(-k_p\cdot k_{x_1})(\epsilon_{q}\cdot \epsilon_p)\phi_{x_1pq|a_1a_2a_3}+(\epsilon_p\cdot k_{x_1})(\epsilon_q\cdot k_p)\phi_{x_1pq|a_1a_2a_3}\nn
+(\epsilon_q\cdot k_{x_1})(\epsilon_p\cdot k_q)\phi_{x_1qp|a_1a_2a_3}+(\epsilon_p\cdot k_{x_1})(\epsilon_q\cdot k_{x_1})\phi_{x_1\{p\}\shuffle\{q\}|a_1a_2a_3}\Bigg]{\phi_{x_2|a_4}\over l^2l^2_{x_1pq} }.
\eea
The effective current $\W J(x_1,p,q)$ is generated by the refined graphic rule where $x_1$ plays as root. Another example is given by the sum of terms with respect to partitions $\{\{x_1,x_2,p\},q\}$, $\{\{x_2,x_1,p\},q\}$
\bea
\W J(x_1,x_2, p)\phi_{q|a_4}{1\over l^2l^2_{x_1x_2p} }+\W J(x_2,x_1, p)\phi_{q|a_4}{1\over l^2l^2_{x_1x_2p} }
\eea
which vanishes due to $U(1)$-decoupling identity.
 
\item There exist two-element subsets, one containing two scalars and one containing two gluons, i.e., the partition $\{\{x_1,x_2\},\{p,q\}\}$. Although the contribution of this partition has to vanish due to the $U(1)$-decoupling identity, we express it formally as
\bea
\left(\Sl_{\text{cyc}(x_1x_2)}\phi_{x_1x_2|a_1a_2}\right)\,\left[\W J(p,q)\cdot X_{\{p,q\}}+\left(-{1\over2}\right)(\W J(p)\cdot\W J(q))\right]{1\over l^2 }{1\over l^2_{x_1x_2} }.
\eea

\item There exist two-element subsets, each containing a scalar and a gluon.  For instance the partition $\{\{x_1,p\},\{x_2,q\}\}$ gives the quadratic-propagator term 
\bea
\W J(x_1,p)\W J(x_2,q){1\over l^2 }{1\over l^2_{x_1p} }.
\eea
\end{itemize}
The nonvanishing contribution of all bubble diagrams is given by
\bea
\W J(x_1,p,q)\phi_{x_2|a_4}{1\over l^2l^2_{x_1pq} }+\W J(x_2,p,q)\phi_{x_1|a_1}{1\over l^2l^2_{x_1} }+\left[\W J(x_1,p)\W J(x_2,q){1\over l^2 }{1\over l^2_{x_1p} }+(p\leftrightarrow q)\right]+\text{cyc}(a_1a_2a_3a_4).\nn
\eea

Last, the partition with all particles involved in a single subset corresponds to a tadpole diagram. 
Although the tadpole diagram has to vanish due to $U(1)$-decoupling identity, we also display the explicit quadratic-propagator expression along our approach 
\bea
\W J(x_1,x_2,p,q){1\over l^2 }+\text{cyc}(x_1x_2)+\text{cyc}(a_1a_2a_3a_4).
\eea
Here the effective current $\W J(x_1,x_2,p,q)$ is generated by graphs where scalars $x_1$ and $x_2$ are considered as roots. 

\begin{figure}
\centering
\includegraphics[width=0.7\textwidth]{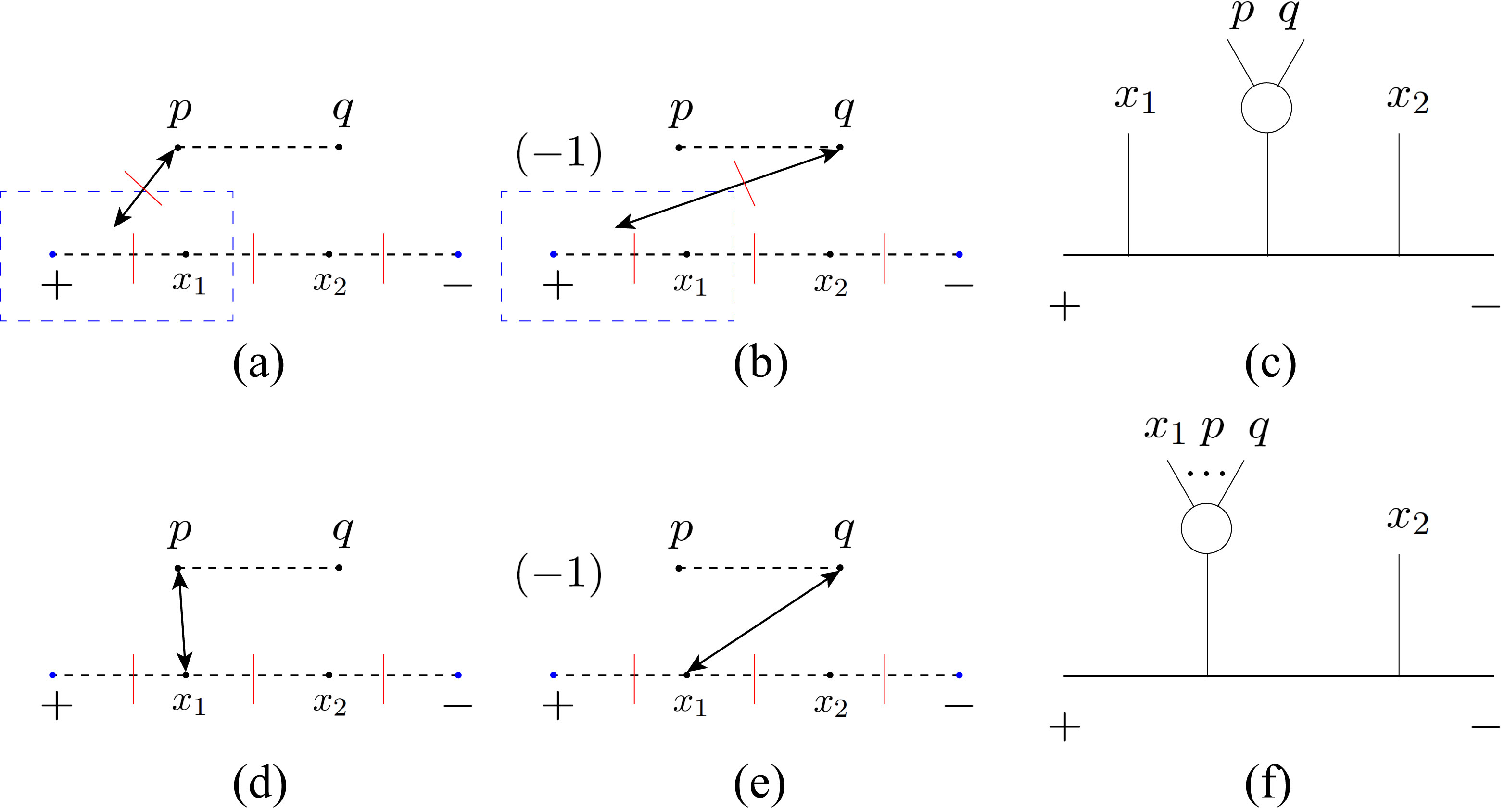}
\caption{The graphs (a) and (b) together provide an X-pattern with respect to the Feynman diagram (c). The graphs (d) and (e) together provide a BCJ-pattern with respect to the Feynman diagram (f).}
\label{Fig:0pattern}
\end{figure}

Now let us summarize the final expression:
\bea
I^{\text{1-loop}}\big(x_1,x_2||\{p,q\}\big)\cong\Sl_{\{A_1,A_2,...,A_I\} }{1\over l^2}\,J[A_1]\, {1\over l^2_{A_1}}\,J[A_2]\cdots\,{1\over l^2_{A_1...A_{I-1}}}\,J[A_I]+\text{cyc}(a_1a_2a_3a_4),
\eea
where we summed over all possible cyclic partitions $\{A_1,A_2,...,A_I\}$ ($I=1,...,4$) of external particles $x_1$, $x_2$, $p$, $q$. The subcurrent $J[A_i]$ is given by
\bea
J[A_i]&=&\W J (A_i),~~~~~~~~~~~~~~~~~~~~~~~~~~~~~~~~~~~~~~~~~~~~~~~~\text{(if $A_i$ contains  scalars)}\nn
J[A_i]&=&\W J (A_i)\cdot X_{A_i}+\left(-{1\over 2}\right)\left[\W J (A_{iL})\cdot\W J (A_{iR})\right],~~~~\text{(if $A_i$ contains only gluons)}
\eea
where $\W J (A_i)$ (or $\W J^{\mu} (A_i)$) denotes the effective current generated by the refined graphic rule, $X^{\mu}_{A_i}$ is the momentum of the loop propagator to the left of the subset $A_i$. When $A_i$ contains only scalars, $\W J (A_i)$  becomes a pure scalar current $\phi_{A_i|\W A_i}$. The effective currents $\W J^{\mu} (A_{iL})$ and $\W J^{\mu} (A_{iR})$ denote the polarizations $\epsilon^{\mu}_p$ and $\epsilon^{\mu}_q$, respectively.


\section{X-pattern  and BCJ-pattern}\label{sec:XBCJ}

In the previous section, we have defined X-pattern and BCJ-pattern, which respectively contain a $k_A\cdot X_A$ factor and a substructure of off-shell BCJ relation (\ref{Eq:OffBCJ1}). In the example, we have  already seen that the subgraph with node $p$ in \figref{Fig:4pt2glu7} (b) (for the Feynman diagram \figref{Fig:4pt2glu7} (c)), and the subgraphs with nodes $p$, $x_1$ in \figref{Fig:4pt2glu10} (a) and $p$, $x_2$ in \figref{Fig:4pt2glu10} (c) (for the Feynman diagrams \figref{Fig:4pt2glu10} (b) and (d)) are respectively X-pattern and BCJ-patterns.  The cancellations between X- and BCJ-patterns in the example play a crucial role for treating the nonlocal terms. In the current  section, we show more examples and summarize the general features of graphs corresponding to X- and BCJ-pattern.

\subsection{Example-1}
The first example is shown by \figref{Fig:0pattern}. In \figref{Fig:0pattern}, both graphs (a) and (b) contain a subgraph with $p$ and $q$, which correspondingly provide factors $(-k_p\cdot X_p)\phi_{pq|a_2a_3}$, $(k_q\cdot X_q)\phi_{qp|a_2a_3}$ for the Feynman diagram \figref{Fig:0pattern} (c). When the graph-based relation (\ref{Eq:GraphBased1}) is applied, the factor for \figref{Fig:0pattern} (b) turns into $(-k_q\cdot X_{\{p,q\}})\phi_{pq|a_2a_3}$, noting that  $X_{\{p,q\}}$ which denotes the momentum of the linear propagator attached to the subset $\{p,q\}$ from left is the same with the $X_p$ and $X_q$ for \figref{Fig:0pattern} (a) and (b). Thus the total factor of \figref{Fig:0pattern} (a) and (b) becomes $[-(k_p+k_q)\cdot X_{\{p,q\}}]\phi_{pq|a_2a_3}$ accompanying to the subcurrent containing $p$, $q$ in the Feynman diagram  \figref{Fig:0pattern} (c). 
\begin{figure}
\centering
\includegraphics[width=0.7\textwidth]{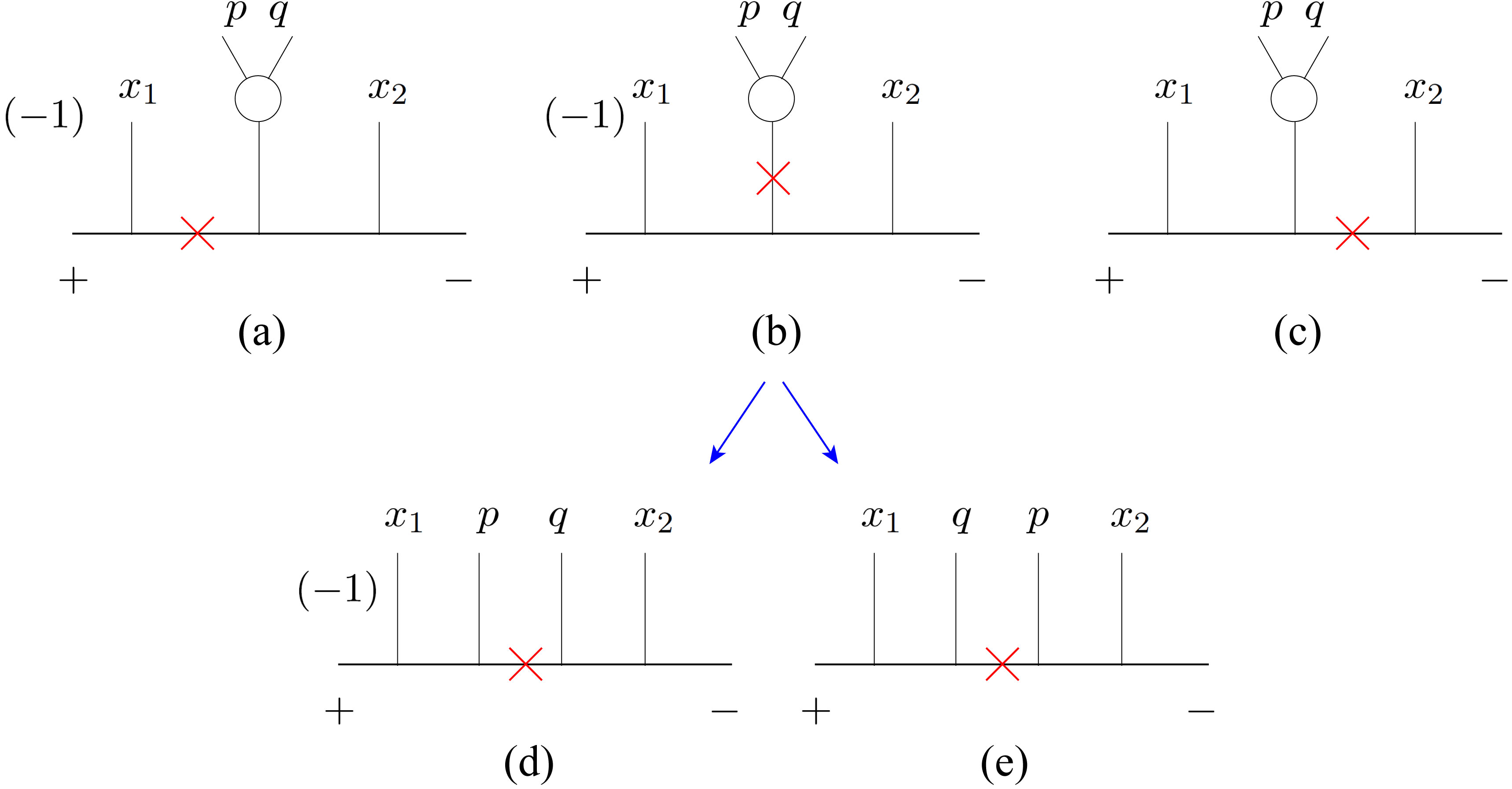}
\caption{The X-pattern \figref{Fig:0pattern} (a), (b) reduces the Feynman diagram \figref{Fig:0pattern} (c) to (a), (b) and (c) by deleting the corresponding propagators. The diagram (b) is further reduced into (d) and (e), when the BG recursion for tree-level BS current is considered.}
\label{Fig:0pattern2}
\end{figure}
This agrees with the definition of X-pattern. According to (\ref{Eq:XCoefficient}) (or equivalently \figref{Fig:4pt2glu9}), \figref{Fig:0pattern} (c) is reduced into the three diagrams \figref{Fig:0pattern2} (a), (b) and (c), in which,  we have correspondingly deleted the linear propagator to the left of $\{p,q\}$, the propagator of the tree structure containing $\{p,q\}$ and the linear propagator to the right of $\{p,q\}$. When the definition of the BG current for BS (\ref{Eq:BScurrent}) is considered, \figref{Fig:0pattern2} (b) is further reduced into \figref{Fig:0pattern2} (d) and (e). The sum of diagrams \figref{Fig:0pattern2} (a), (c), (d), (e) is then given by 
\bea
&&\left(-{1\over 2}\right)\biggl[(-1){1\over l^2}{1\over s_{x_1pq,l}}\phi_{x_1|a_1}\phi_{pq|a_2a_3}\phi_{x_2|a_4}+{1\over l^2}{1\over s_{x_1,l}}\phi_{x_1|a_1}\phi_{pq|a_2a_3}\phi_{x_2|a_4}\nn
&&~~~~~~~~~~~~~~~+(-1){1\over l^2}{1\over s_{x_1,l}}{1\over s_{x_1pq,l}}\phi_{x_1|a_1}\phi_{p|a_2}\phi_{q|a_3}\phi_{x_2|a_4}+{1\over l^2}{1\over s_{x_1,l}}{1\over s_{x_1pq,l}}\phi_{x_1|a_1}\phi_{q|a_2}\phi_{p|a_3}\phi_{x_2|a_4}\biggr].
\eea
\begin{figure}
\centering
\includegraphics[width=0.45\textwidth]{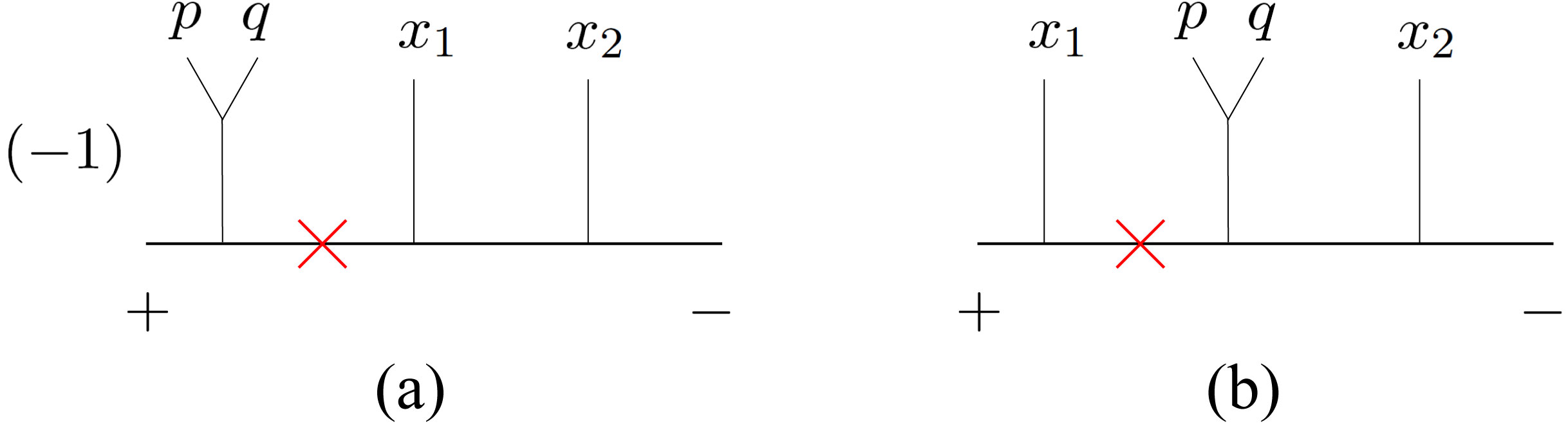}
\caption{ When the off-shell BCJ relation is applied, the diagram \figref{Fig:0pattern} (f) accompanied by the BCJ-pattern \figref{Fig:0pattern} (d), (e) is further reduced into graphs (a) and (b). }
\label{Fig:0pattern1}
\end{figure}
\begin{figure}
\centering
\includegraphics[width=1\textwidth]{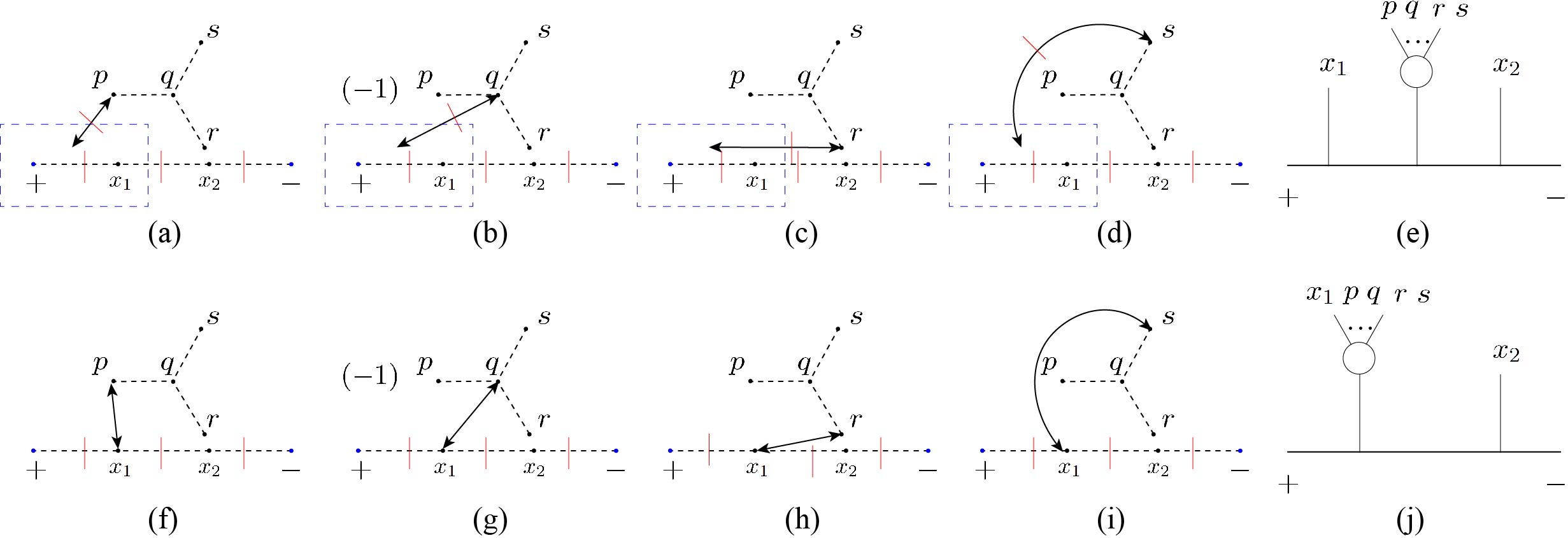}
\caption{Graphs (a)-(d) provide an X-pattern for the Feynman diagram (e), while graphs (f)-(i) provide a BCJ-pattern for the diagram (j). }
\label{Fig:0pattern3}
\end{figure}
\,\,\,\,\,\,\,Associating with the X-pattern shown by \figref{Fig:0pattern} (a), (b), the graphs \figref{Fig:0pattern} (d), (e), (each of which involves a subgraph where $p$ or $q$ is connected to scalar $x_1$) together provide a  BCJ-pattern since the sum of these contributions is the LHS of the off-shell BCJ relation (\ref{Eq:OffBCJ1}):
\bea
 (-k_p\cdot k_{x_1})\phi_{x_1pq|a_1a_2a_3}-(-k_q\cdot k_{x_1})\phi_{x_1qp|a_1a_2a_3}.
\eea
According to the off-shell BCJ relation (\ref{Eq:OffBCJ1}), the above expression becomes
\bea
\left(-{1\over 2}\right)\left[-\phi_{pq|a_1a_2}\phi_{x_1|a_3}+\phi_{x_1|a_1}\phi_{pq|a_2a_3}\right].
\eea
The two terms can be expressed by \figref{Fig:0pattern1} (a) and (b), respectively.

\begin{figure}
\centering
\includegraphics[width=1\textwidth]{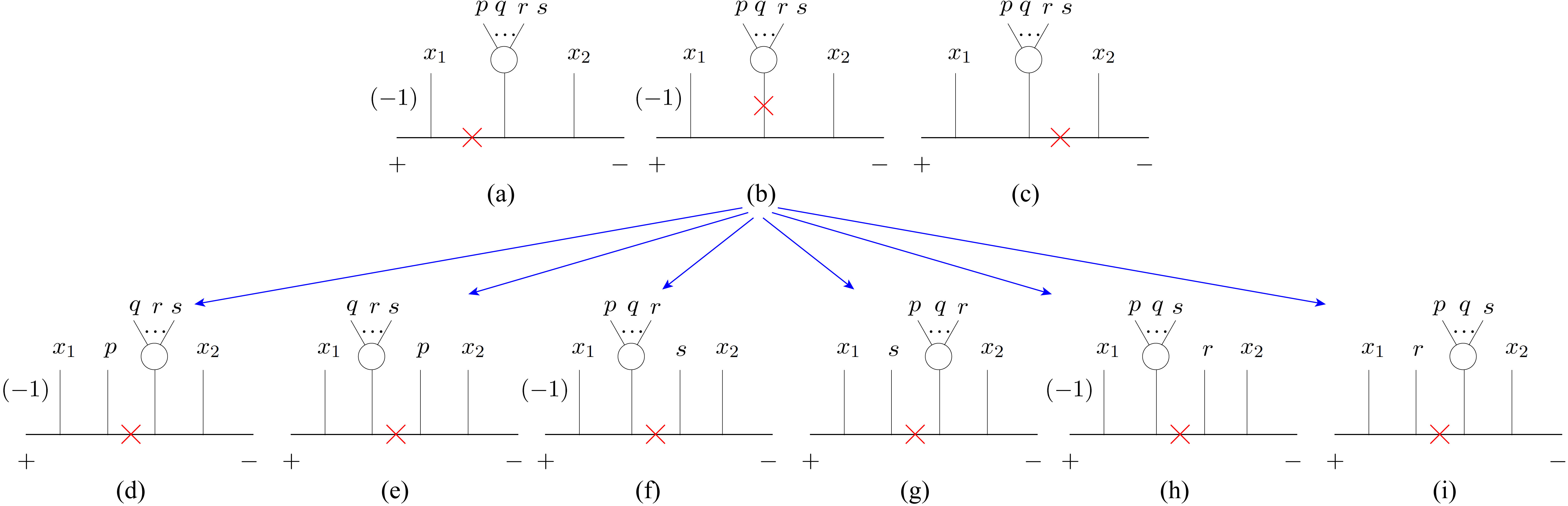}
\caption{ The property \figref{Fig:4pt2glu9}, together with the X-pattern \figref{Fig:0pattern3} (a)-(d) reduces the Feynman diagram \figref{Fig:0pattern3} (e) into (a), (b) and (c). The BG recursion for BS current further reduces the diagram (b) into diagrams (d)-(i). }
\label{Fig:0pattern4}
\end{figure}

\subsection{Example-2}

The second example is presented by \figref{Fig:0pattern3}. The graphs  \figref{Fig:0pattern3} (a), (b), (c) and (d), associating with the subcurrent containing $p$, $q$, $r$, $s$  in the Feynman diagram \figref{Fig:0pattern3} (e) respectively contribute 
\bea
&&(-k_p\cdot X_{\{p,q,r,s\}})\phi_{pq\{s\}\shuffle \{r\}|a_2a_3a_4a_5},\,\,\,-(-k_q\cdot X_{\{p,q,r,s\}})\phi_{q\{p\}\shuffle\{s\}\shuffle \{r\}|a_2a_3a_4a_5},\nn
&&(-k_r\cdot X_{\{p,q,r,s\}})\phi_{rq\{p\}\shuffle\{s\}|a_2a_3a_4a_5},\,\,\,\,~~(-k_s\cdot X_{\{p,q,r,s\}})\phi_{sq\{p\}\shuffle\{r\}|a_2a_3a_4a_5},
\eea 
where $X^{\mu}_{\{p,q,r,s\}}=l^{\mu}+k^{\mu}_{x_1}$ is the momentum of the linear propagator attached to  the subcurrent containing $p$, $q$, $r$, $s$ from left. When the graph-based relation (\ref{Eq:GraphBased1}) is applied to \figref{Fig:0pattern3} (b), (c) and (d), the four terms can be collected into
\bea
-(k_p+k_q+k_r+k_s)\cdot X_{\{p,q,r,s\}}\phi_{pq\{s\}\shuffle \{r\}|a_2a_3a_4a_5},
\eea
therefore agrees with the definition of X-pattern. Once the property of X-pattern (as shown by \figref{Fig:4pt2glu9}) is considered,  the Feynman diagram \figref{Fig:0pattern3} (e) turns into the three diagrams \figref{Fig:0pattern4} (a), (b), (c). In the diagram \figref{Fig:0pattern4} (b), the off-shell line of the subcurrent is deleted, hence it further splits into products of two subcurrents. This can be represented by the diagrams \figref{Fig:0pattern4} (d)-(i), whose expressions respectively contain 
\bea
{1\over 2}\phi_{p|a_2}\phi_{q\{r\}\shuffle\{s\}|a_3a_4a_5},&&\,{\left(-{1\over 2}\right)}\phi_{q\{r\}\shuffle\{s\}|a_2a_3a_4}\phi_{p|a_5},\nn
{1\over 2}\phi_{pqr|a_2a_3a_4}\phi_{s|a_5},~~~~~~&&\,{\left(-{1\over 2}\right)}\phi_{s|a_2}\phi_{pqr|a_3a_4a_5},\nn
{1\over 2}\phi_{pqs|a_2a_3a_4}\phi_{r|a_5},~~~~~~&&\,{\left(-{1\over 2}\right)}\phi_{r|a_2}\phi_{pqs|a_3a_4a_5}.
\eea
Note that there also exist terms ${1\over 2}\phi_{pq|a_2a_3}\phi_{\{r\}\shuffle\{s\}|a_4a_5},$ and ${\left(-{1\over 2}\right)}\phi_{\{r\}\shuffle\{s\}|a_2a_3}\phi_{pq|a_4a_5}$, but these two terms have to vanish due to $U(1)$-decoupling identity. The graphs \figref{Fig:0pattern3} (f)-(i) provide the LHS of off-shell BCJ relation for the subcurrent involving $x_1,p,q,r,s$ in \figref{Fig:0pattern3} (j):
\bea
&&(-k_p\cdot k_{x_1})\phi_{x_1pq\{r\}\shuffle\{s\}|a_1a_2a_3a_4a_5}-(-k_q\cdot k_{x_1})\phi_{x_1q\{p\}\shuffle\{r\}\shuffle\{s\}|a_1a_2a_3a_4a_5}\nn
&+&(-k_r\cdot k_{x_1})\phi_{x_1rq\{p\}\shuffle\{s\}|a_1a_2a_3a_4a_5}+(-k_s\cdot k_{x_1})\phi_{x_1sq\{p\}\shuffle\{r\}|a_1a_2a_3a_4a_5}.
\eea
According to the off-shell BCJ relation (\ref{Eq:OffBCJ1}), the above expression is given by 
\bea
{\left(-{1\over 2}\right)}\left[\phi_{x_1|a_2}\phi_{pq\{r\}\shuffle\{s\}|a_2a_3a_4a_5}-\phi_{pq\{r\}\shuffle\{s\}|a_1a_2a_3a_4}\phi_{x_1|a_5}\right].
\eea
\begin{figure}
\centering
\includegraphics[width=0.9\textwidth]{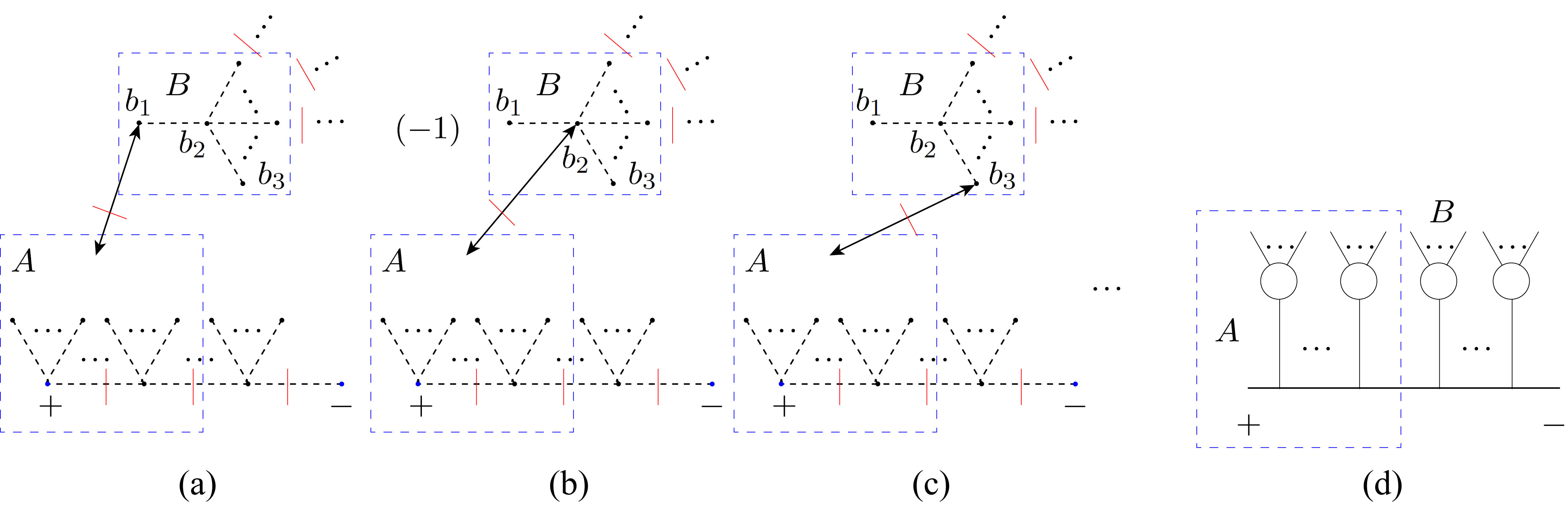}
\caption{Graphs (a), (b), (c), ... together provide an X-pattern, associating with the Feynman diagram (d) where all elements in $B$ are contained in a single subcurrent.}
\label{Fig:GeneralX}
\end{figure}
\subsection{General X- and BCJ-pattern}

Now we are ready for extracting  the general features of X- and BCJ-patterns from the two examples. 

In general, an {\it $X$-pattern} occurs in a set of graphs $\{\mathcal{F}_i\}$ if the following conditions are satisfied:
\begin{itemize}
\item (i). All graphs in $\{\mathcal{F}_i\}$ contain two {\it connected subtree structures} $A$ and $B$ which are interconnected via a type-3  (i.e., {$k_a\cdot k_b$}) line for all $(a,b)$ ($a\in A$, $b\in B$) pairs, as shown by \figref{Fig:GeneralX} (a), (b), (c), ... The node $+$ belongs to the subtree structure $A$. 

\item (ii). Each graph is associated with a sign which relies on the choice of node $b\in B$. Particularly, the relative signs for two graphs $\mathcal{F}_i$ and $\mathcal{F}_j$ is given by $(-1)^{d}$, where $d$ is the distance between nodes $b_i$ and $b_j$ (which are the choices of nodes $b\in B$ in graphs  $\mathcal{F}_i$ and $\mathcal{F}_j$, respectively).
\begin{figure}
\centering
\includegraphics[width=0.7\textwidth]{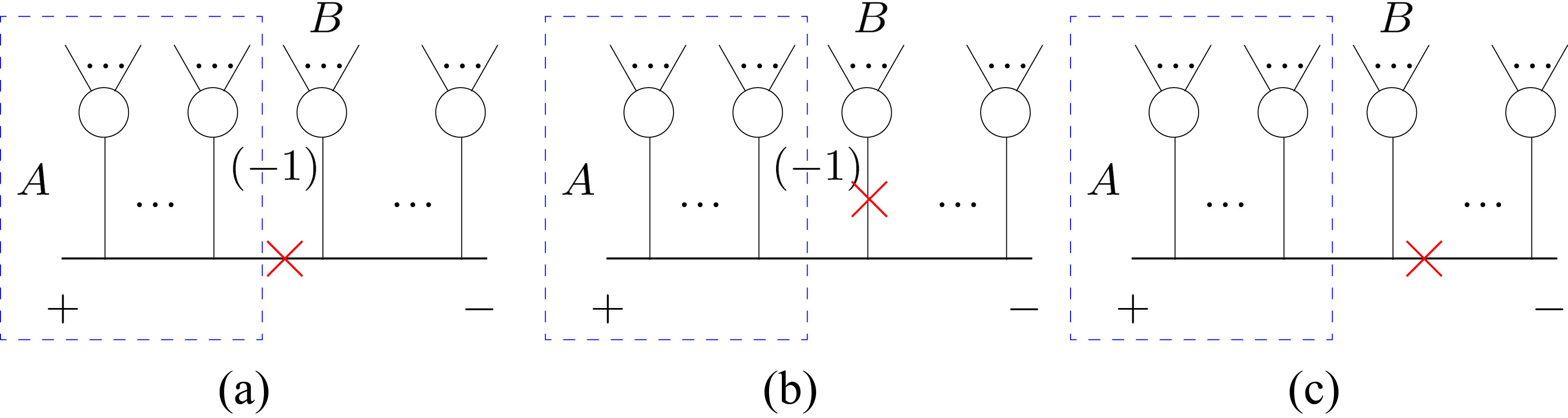}
\caption{ When the property of X-pattern shown by \figref{Fig:4pt2glu9} is applied, \figref{Fig:GeneralX} (d) splits into three diagrams (a), (b) and (c). }
\label{Fig:GeneralX1}
\end{figure}
\item (iii). All graphs in $\{\mathcal{F}_i\}$ are identical when the line between $A$, $B$ in each graph is removed. Each graph corresponds to a choice of $(a,b)$ pair, and graphs for all choices of  $(a,b)$ pairs form the full set $\{\mathcal{F}_i\}$.

\begin{figure}
\centering
\includegraphics[width=0.9\textwidth]{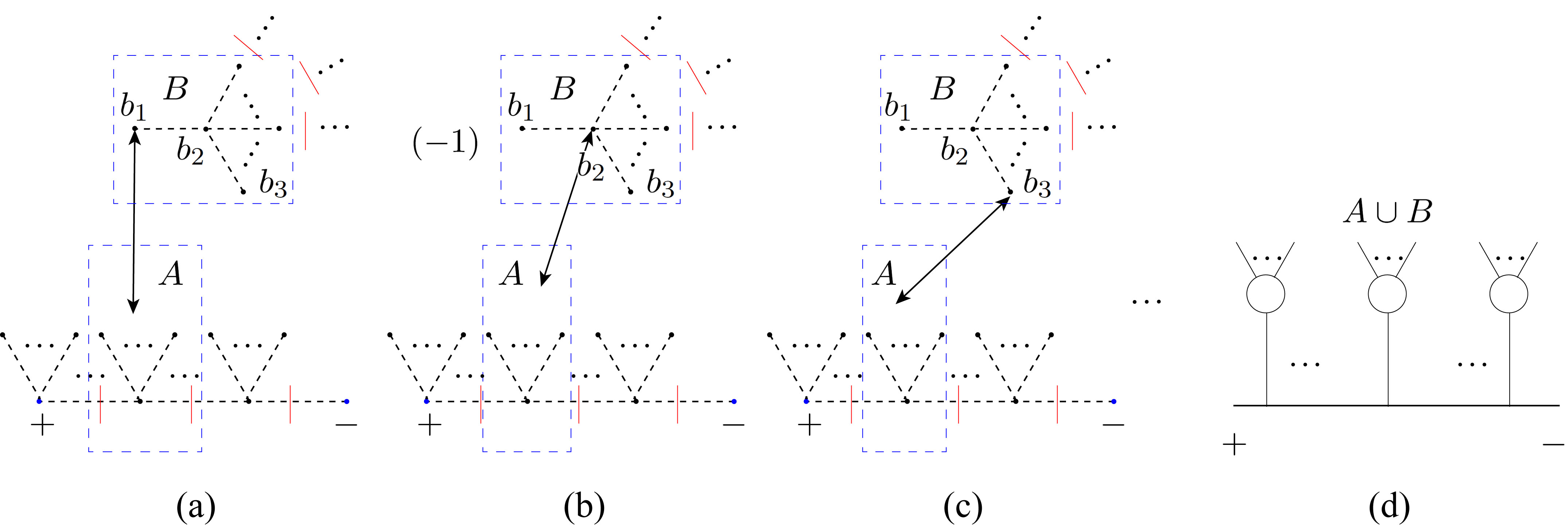}
\caption{The graphs (a), (b), (c)... provide a BCJ-pattern for the Feynman diagram (d), in which elements in $A\cup B$ are contained by a single subcurrent.}
\label{Fig:GeneralBCJ}
\end{figure}
 \end{itemize}
Now we associate each graph $\mathcal{F}_i$  with the Feynman diagram \figref{Fig:GeneralX} (d), where all elements in $B$ are contained by a subcurrent $\phi_{(\pmb\sigma^{B}|_{b_i})|\W B}$ ($\pmb\sigma^{B}|_{b_i}$ denotes the permutations established by graph ${B}$ when the node $b_i\in B$ is chosen as the leftmost element in $B$), and all elements in $A$ are contained by the full subtree structure to the left of $B$. For a given choice of $b=b_i\in B$, the graphs with different choices of $a\in A$ can be collected together to produce a factor $-k_{b_i}\cdot X_{B}$ (where $X^{\mu}_B$ is the total momentum of nodes in the region A, in each of \figref{Fig:GeneralX} (a), (b), (c)..., or in other words, the momentum of the linear propagator attached to $B$ in the Feynman diagram  \figref{Fig:GeneralX} (d) from left). When the graph-based relation (\ref{Eq:GraphBased1}) is further applied, the sum of these terms (up to an overall sign) can be transformed as follows
\bea
&&(-k_{b_1}\cdot X_{B})\phi_{(\pmb\sigma^{B}|_{b_1})|\W B}+
(-1)^{d_2}(-k_{b_2}\cdot X_{B})\phi_{(\pmb\sigma^{B}|_{b_2})|\W B}+(-1)^{d_3}(-k_{b_3}\cdot X_{B})\phi_{(\pmb\sigma^{B}|_{b_3})|\W B}+\dots\nn
&=&\big[-(k_{b_1}+k_{b_2}+k_{b_3}+\dots)\cdot X_{B}\big]\phi_{(\pmb\sigma^{B}|_{b_1})|\W B},
\eea
where $d_2,d_3,...$ are distances between $b_2,b_3,...$ and $b_1$, respectively. On the second line, all subcurrents with permutations $\pmb\sigma^{B}|_{b_i}$ are transformed into the subcurrent with permutation $\pmb\sigma^{B}|_{b_1}$, and the coefficients are precisely collected as $-k_B\cdot X_B$. This  agrees with the definition of X-pattern. Hence, the Feynman diagram further splits into \figref{Fig:GeneralX1} (a), (b) and (c).

\begin{figure}
\centering
\includegraphics[width=0.55\textwidth]{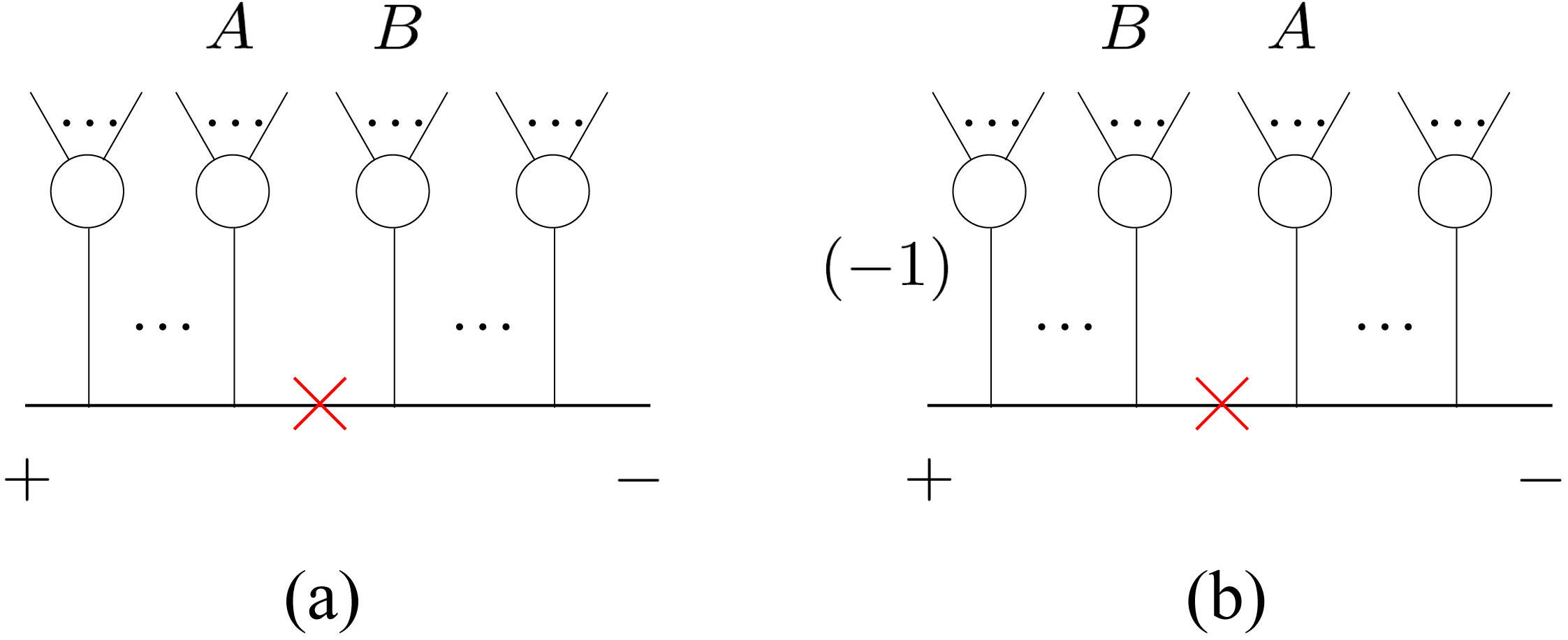}
\caption{Off-shell BCJ relation further reduces \figref{Fig:GeneralBCJ} (d) into the two diagrams (a) and (b).}
\label{Fig:GeneralBCJ1}
\end{figure}

The general BCJ-pattern is characterized by \figref{Fig:GeneralBCJ}. Concretely, each graph \figref{Fig:GeneralBCJ} (a), (b), (c)... contains a subgraph where $A$ is connected with $B$ via a type-3 line. If $b_i$ and $b_j$ are two adjacent nodes (e.g. the $b_1$ and $b_2$), the signs assocating with graphs $b=b_i$ and $b=b_j$ must be opposite. When the kinematic coefficient $-k_a\cdot k_b$  and the relative sign are dressed, the sum of all such graphs produce the LHS of the off-shell BCJ relation (\ref{Eq:OffBCJ1}). Then we arrive at
\bea
{\left(-{1\over 2}\right)}\left[\phi_{\pmb\sigma^A\big| \W C_L}\phi_{(\pmb\sigma^B|_{b_1})\big| \W C_R}-\phi_{(\pmb\sigma^B|_{b_1})\big| \W C_L}\phi_{\pmb\sigma^A\big| \W C_R}\right],
\eea
 where $\W C=\W C_L\W C_R$ denotes a fixed permutation of elements in $A\cup B$ for the right side. The above two terms are respectively characterized by the diagrams \figref{Fig:GeneralBCJ1} (a) and (b).

%

\section{Spurious graphs and five-point integrand with three gluons}\label{sec:5ptExample}

\begin{figure}
\centering
\includegraphics[width=0.9\textwidth]{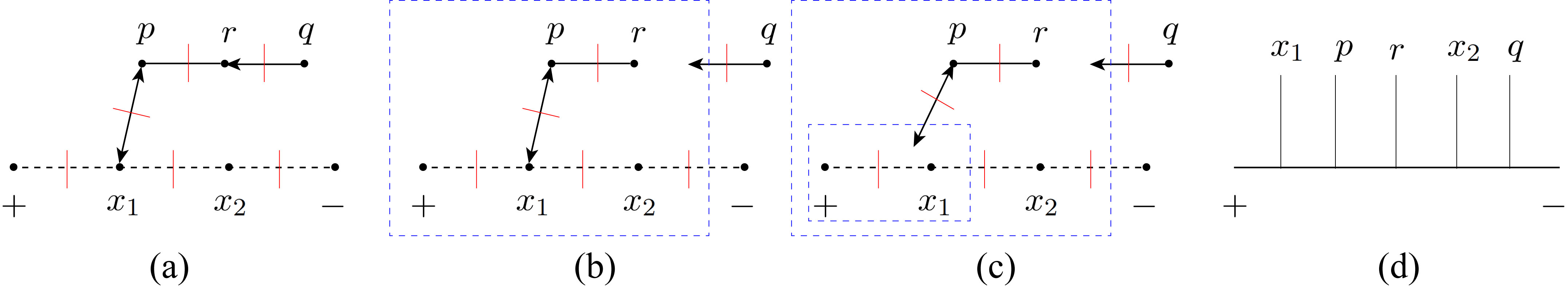}
\caption{Graph (a) induces nonlocalities by the line between nodes $q$, $r$, nodes $p$, $x_1$ as well as the nodes $p$, $r$, for the Feynman diagram (d). The nonlocality caused by the line between $q$, $r$ can be overcomed by collecting all related contributions as shown by (b), while the nonlocality caused by the line between $p$ and $x_1$ is further treated in a similar way, as shown by (c). These treatments follow from approach-1. }
\label{Fig:3GluonEg1}
\end{figure}

In the YMS amplitudes with more than two gluons, the two approaches in \secref{sec:TwoGluonEG1} and \secref{sec:TwoGluonEG2} can be combined to cancel nonlocal terms. However, these approaches are not sufficient for treating all nonlocal terms produced by graphic rule in the cases with more than two gluons. To complete the cancellation of nonlocalities, we have to introduce  pairs of {\it spurious graphs} with opposite signs. Once the spurious graphs are included, the nonlocal terms can always be arranged into  forms that can further be treated via approach-1 and -2. In this section, we use  five-point example with three gluons $p$, $q$, $r$ (with the reference order $p\prec q\prec r$) and two scalars $x_1$, $x_2$ to demonstrate this new mechanism. We further collect terms corresponding to a given partition of external particles, and propose a compact formula expressed by effective currents.

\begin{figure}
\centering
\includegraphics[width=0.8\textwidth]{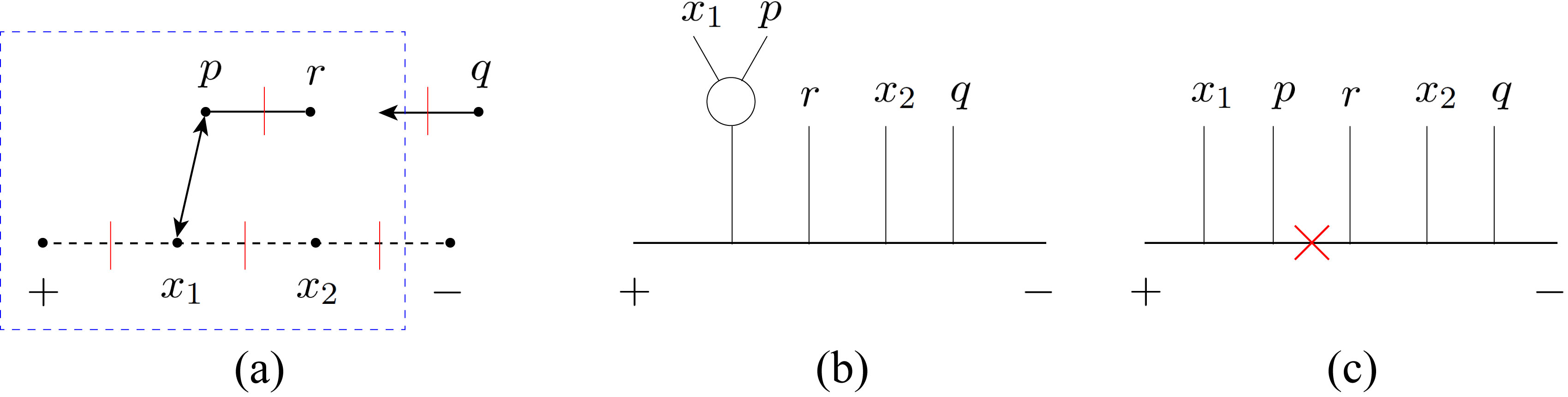}
\caption{Graph (a) contains a BCJ-pattern, the subgraph with $x_1$ and $p$, for the diagram (b). When the cancellation between the X-pattern \figref{Fig:3GluonEg1} (c) and the BCJ-pattern (a) is performed, we get the diagram (c) which comes from the X-pattern in \figref{Fig:3GluonEg1} (c).}
\label{Fig:3GluonEg2}
\end{figure}

\subsection{ Five-point example and spurious graphs}

The first example for combining the two approachs is the graph  \figref{Fig:3GluonEg1} (a) accompanied to the Feynman diagram \figref{Fig:3GluonEg1} (d) with the partition $\{x_1, p, r, x_2, q\}$. The subgraphs $q$ and $r$ in \figref{Fig:3GluonEg1} (a) are contracted via a line $\epsilon_q\cdot k_r$, but they are separated by the linear propagators between them in \figref{Fig:3GluonEg1} (d), thus this contribution is exactly a nonlocal one. Nevertheless, one can follow {\it approach-1}, by collecting other corresponding contributions with $\epsilon_q\cdot k$, all graphs together (as shown by \figref{Fig:3GluonEg1} (b)) produce a total factor $\epsilon_q\cdot X_q=\epsilon_q\cdot (l+k_{x_1}+k_p+k_r+k_{x_2})$. This factor is expressed by the contraction of $\epsilon^{\mu}_q$ with the momentum of the linear propagator ${1\over s_{x_1prx_2,l}}$ attached to the subcurrent $q$, hence this nonlocality has been canceled. Similarly, for the graph \figref{Fig:3GluonEg1} (b), the nonlocality caused by $-k_p\cdot k_{x_1}$ can also be treated by  considering the graph where $p$ is connected to $+$ via $-k_p\cdot l$. Thus the factors are summed into  $-k_p\cdot X_p$, as shown by  \figref{Fig:3GluonEg1} (c).

Though the nonlocalities caused by $-k_p\cdot k$ and $\epsilon_q\cdot k$ have already been rearranged into local contributions, the nonlocality caused by the type-1 line between $p$ and $q$ cannot be dealt with in this way. We have to follow {\it approach-2}. To see this, we find that the factor $-k_p\cdot X_p$ in \figref{Fig:3GluonEg1} (c) also indicates an X-pattern for the gluon $p$. The corresponding BCJ-pattern is shown by \figref{Fig:3GluonEg2} (a), associated with the diagram \figref{Fig:3GluonEg2} (b). When the cancellation between X- and BCJ-pattern has been carried out, the surviving term of X-pattern reads
\bea
{\left(-{1\over 2}\right)}{1\over l^2}{1\over s_{x_1,l}}{1\over s_{x_1pr,l}}{1\over s_{x_1prx_2,l}}(\epsilon_p\cdot\epsilon_r)(\epsilon_q\cdot X_q)\phi_{x_1|a_1}\left[\,\phi_{p|a_2}\phi_{r|a_3}\,\right]\phi_{x_2|a_4}\phi_{q|a_5},
\eea
which is expressed by the diagram \figref{Fig:3GluonEg2} (c) and does not have any nonlocality. The cyclic permutations of $(a_1a_2a_3a_4a_5)$ and $(x_1x_2)$ together with graphic rule allow the cyclic summation of the subcurrents separated by the linear propagators in the above expression. Using (\ref{Eq:partial}), we get the final expression with quadratic propagators
\bea
\left(-{1\over 2}\right){1\over l^2}{1\over l^2_{x_1}}{1\over l^2_{x_1pr}}{1\over l^2_{x_1prx_2}}(\epsilon_p\cdot\epsilon_r)(\epsilon_q\cdot X_q)\phi_{x_1|a_1}\left[\,\phi_{p|a_2}\phi_{r|a_3}\,\right]\phi_{x_2|a_4}\phi_{q|a_5}.\Label{Eq:FivePointResult1}
\eea
\begin{figure}
\centering
\includegraphics[width=0.9\textwidth]{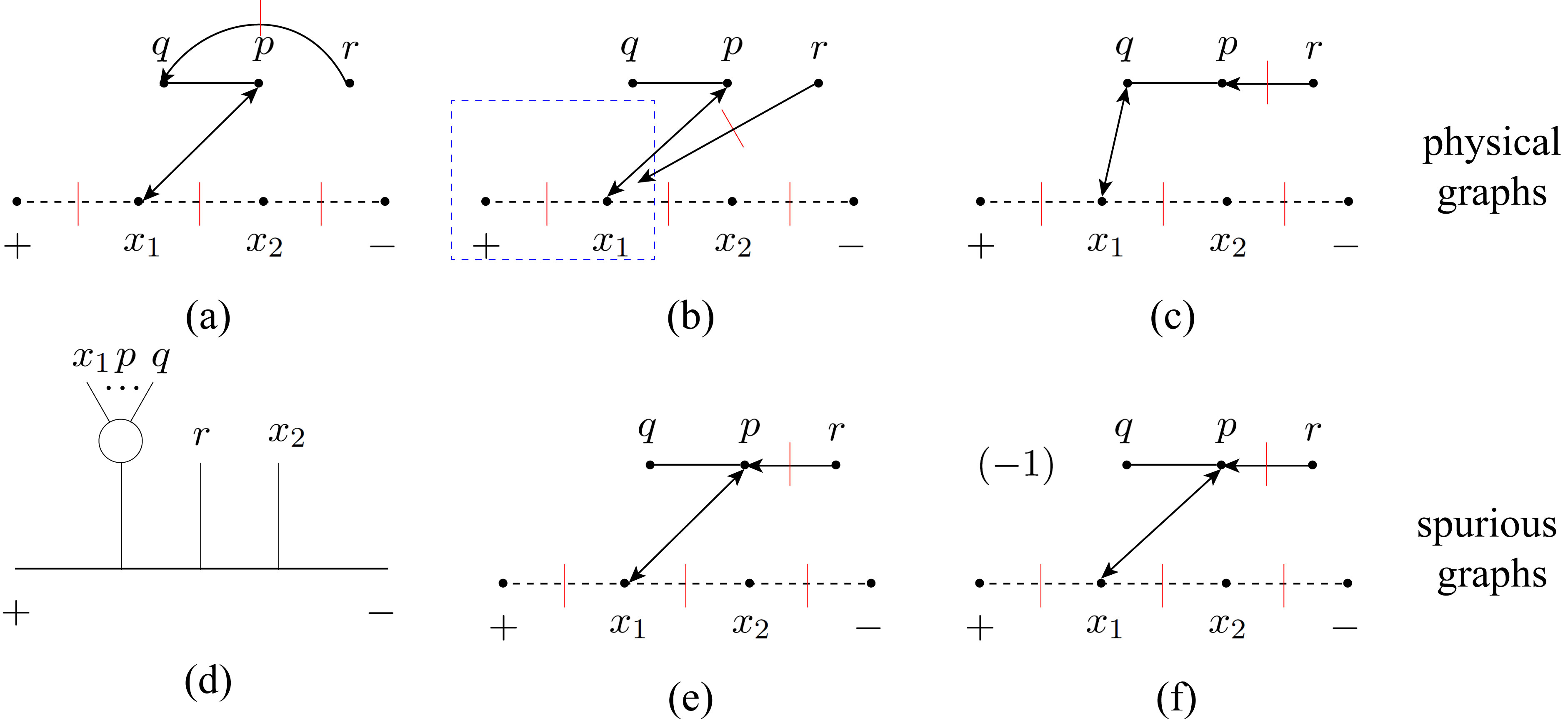}
\caption{The nonlocalities of graphs (a), (b) and (c), accompanied to the diagram (d), cannot be treated via approach-1 or -2. This is a consequence  of the graphic rule with reference order. To remedy this shortcoming, we introduce a pair of spurious graphs (e) and (f) with opposite signs.}
\label{Fig:3GluonEg3}
\end{figure}

The above approach cannot always be applied straightforwardly. In some situations, the physical graphs (graphs defined by the graphic rule) may neither be collected together to cancel the nonlocality, nor contain complete BCJ- or X-pattern. Thus their nonlocal contributions cannot be canceled by directly combining the two approaches. To treat this new situation, we should introduce {\it spurious graphs}. Let us begin with the graphs \figref{Fig:3GluonEg3} (a), (b) and (c) which associate to the Feynman diagram with linear propagators \figref{Fig:3GluonEg3} (d).  \figref{Fig:3GluonEg3} (a) and (b) provide nonlocal terms since each includes a factor $\epsilon_r\cdot k_i$ but $r$ and $i$ ($q$ for (a), $+,x_1$ for (b)) are separated by the linear propagator ${1\over s_{x_1pq,l}}$. One may try to collect such nonlocal contributions to get an $\epsilon_r\cdot X_r$, but the graphic rule does not allow \figref{Fig:3GluonEg3} (e) which contains the factor $\epsilon_r\cdot k_p$. On the other hand, one may try to cancel the nonlocality accompanied to \figref{Fig:3GluonEg3} (c) following the  approach-2: the cancellation between BCJ-pattern and X-pattern (apparently approach-1 is ineffective for this graph because the subgraph containing $x_1$, $p$, $q$ is not a physical graph defined by graphic rule).
\begin{figure}
\centering
\includegraphics[width=0.8\textwidth]{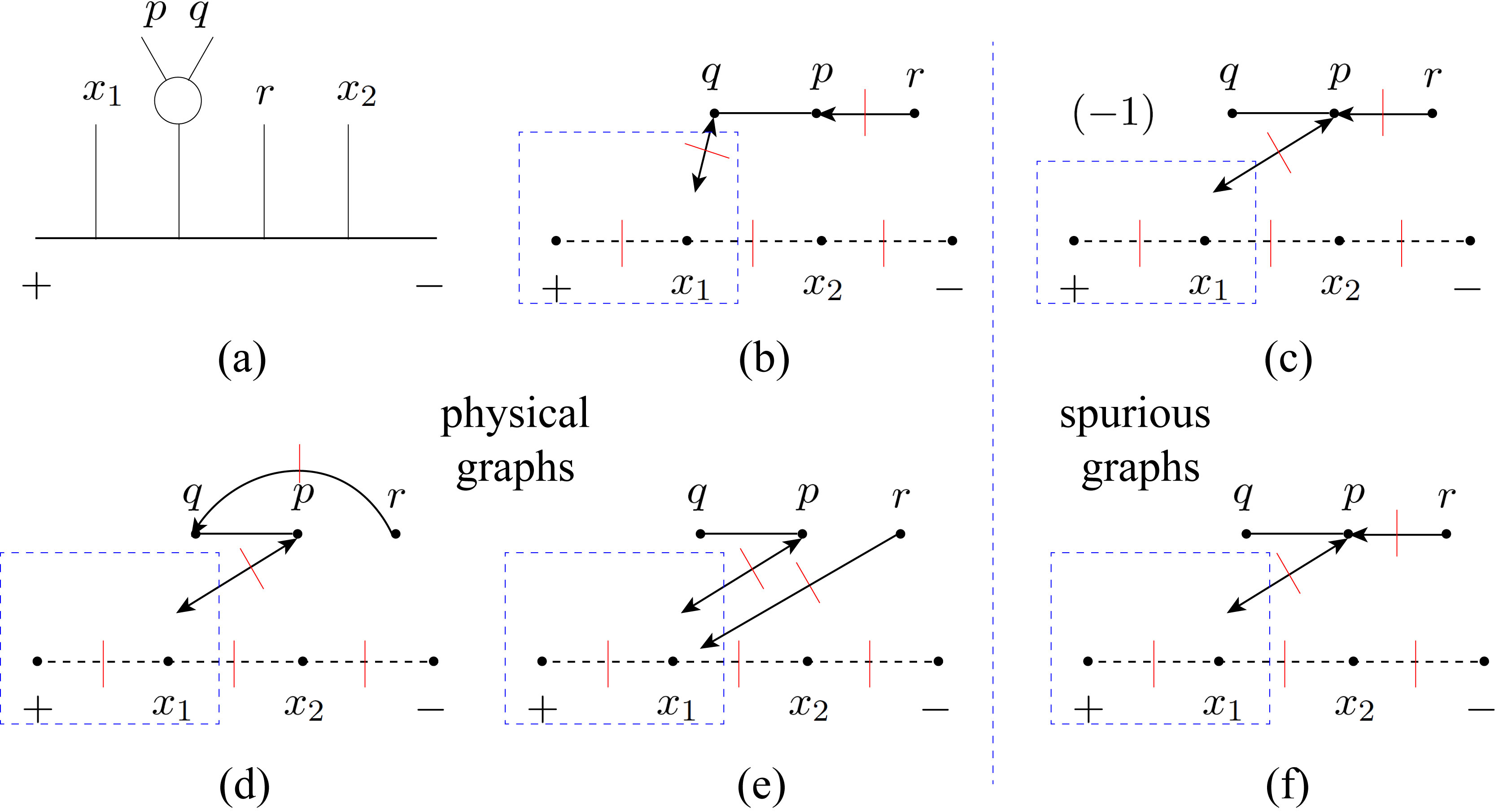}
\caption{Graphs (b), (d) and (e) are physical graphs for diagram (a). A pair of spurious graphs (c) and (f) are introduced. Graphs (b) and (c) provide an X-pattern for (a), thus the nonlocality can be treated according to approach-2. Graphs (d), (e) and (f) can be collected together so that a factor $\epsilon_r\cdot X_r$ arises and cancels the related nonlocality according to approach-1.}
\label{Fig:3GluonEg4}
\end{figure}
 However, it is still lack of a graph  \figref{Fig:3GluonEg3} (f) for a complete BCJ-pattern. To overcome the obstacles mentioned above, we add a pair of graphs \figref{Fig:3GluonEg3} (e) and (f), which are the same spurious graphs but have opposite signs. Though these graphs are not allowed by graphic rule, they cancel each other. Now the sum of graphs \figref{Fig:3GluonEg3} (a), (b) and (e) provides 
\bea
{1\over l^2}{1\over s_{x_1pq,l}}{1\over s_{x_1pqr,l}}(\epsilon_p\cdot\epsilon_q)(-k_p\cdot k_{x_1})(\epsilon_r\cdot X_r)\phi_{x_1pq|a_1a_2a_3}\phi_{r|a_4}\phi_{x_2|a_5},
\eea
where $X^{\mu}_r=l^{\mu}+k^{\mu}_{x_1}+k^{\mu}_p+k^{\mu}_q$ which is the momentum of the linear propagator ${1\over s_{x_1pq,l}}$ to the left of $r$. Hence we have obtained a local expression. The cyclic permutations of $(x_1x_2)$ and $(a_1a_2a_3a_4a_5)$ allow those terms obtained by acting cyclic permuations of the subsets. Altogether provides the quadratic propagator expression 
\bea
{1\over l^2}{1\over l^2_{x_1pq}}{1\over l^2_{x_1pqr}}(\epsilon_p\cdot\epsilon_q)(-k_p\cdot k_{x_1})(\epsilon_r\cdot X_r)\phi_{x_1pq|a_1a_2a_3}\phi_{r|a_4}\phi_{x_2|a_5}.\Label{Eq:FivePointResult2a}
\eea
\begin{figure}
\centering
\includegraphics[width=0.75\textwidth]{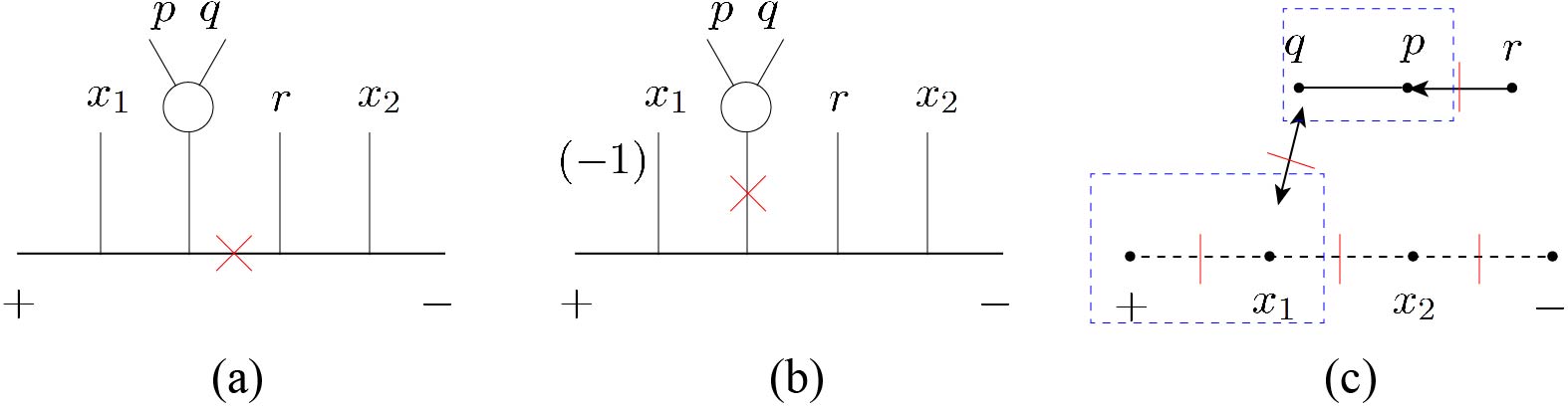}
\caption{When the property \figref{Fig:4pt2glu9} of X-pattern is applied, the X-pattern \figref{Fig:3GluonEg4} (b) and (c) reduces the diagram \figref{Fig:3GluonEg4} (a) into three terms. One of them is canceled with the BCJ-pattern  \figref{Fig:3GluonEg3} (c) and (f) (with respect to diagram \figref{Fig:3GluonEg3} (d)), the remaining two terms are given by diagrams (a) and (b), with the coefficient $(\epsilon_q\cdot \epsilon_p)(\epsilon_r\cdot k_p)$ coming from (c).}
\label{Fig:3GluonEg5}
\end{figure}
\begin{figure}
\centering
\includegraphics[width=0.85\textwidth]{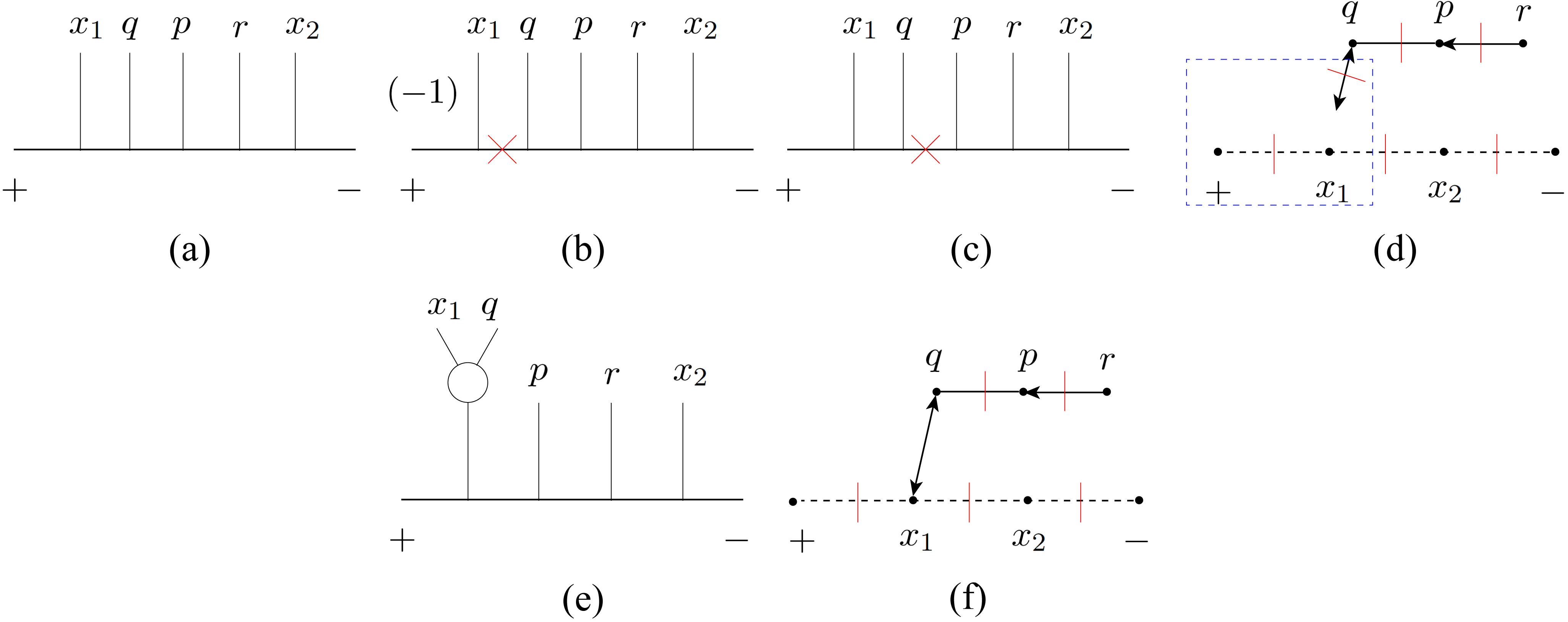}
\caption{The X-pattern for the partition $\{x_1,q,p,r,x_2\}$ is given by graph (d) for the Feynman diagram (a). This diagram splits into diagrams (b) and (c). Diagram (b) cancels with (e) accompanied by the BCJ-pattern (f).}
\label{Fig:3GluonEg6}
\end{figure}
\begin{figure}
\centering
\includegraphics[width=0.87\textwidth]{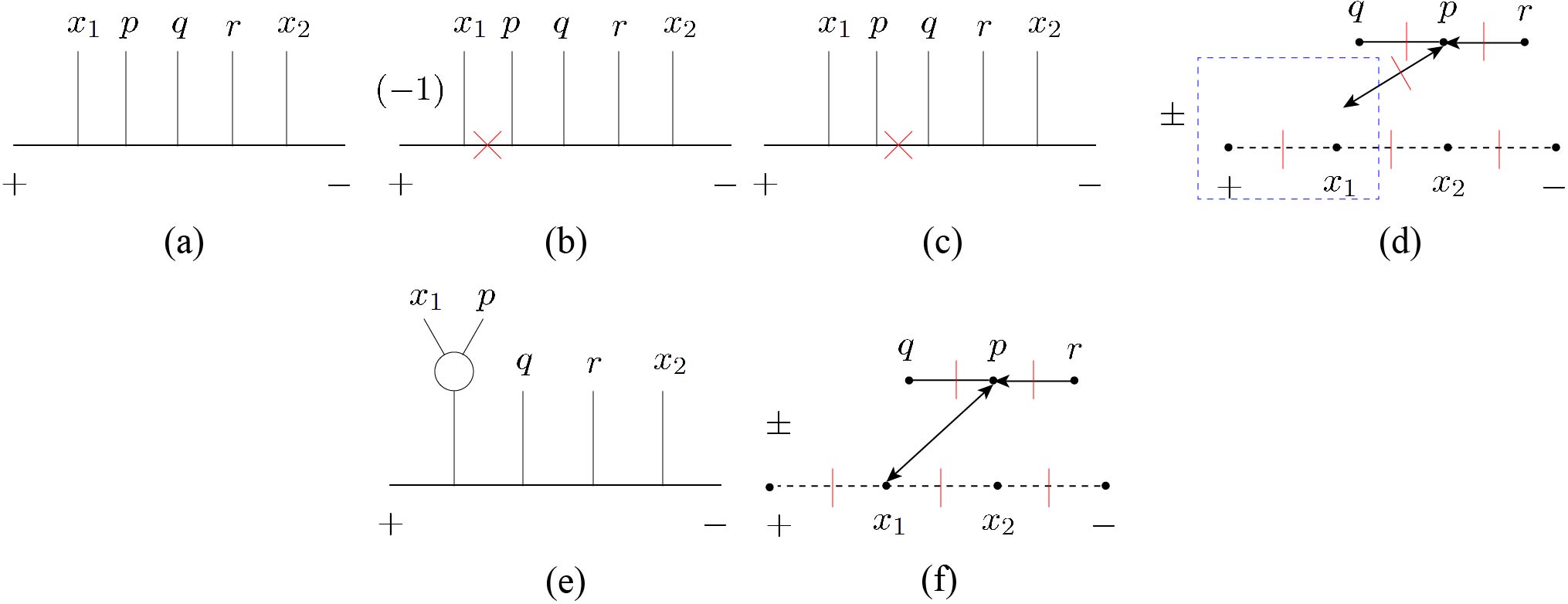}
\caption{The X-pattern for the partition $\{x_1,p,q,r,x_2\}$ is given by the spurious graph (d) (with a minus sign) for the Feynman diagram (a). This diagram splits into diagrams (b) and (c). Diagram (b)  cancels with (e) accompanied by the BCJ-pattern in the spurious graph (f) (with a minus sign).}
\label{Fig:3GluonEg6a}
\end{figure}

The sum of graphs \figref{Fig:3GluonEg3} (c) and (f) involves a BCJ-pattern. We therefore can find the corresponding $X$-pattern, the subgraphs consisting of $q$ and $p$ in the graphs \figref{Fig:3GluonEg4} (b) and (c) accompanied to the Feynman diagram \figref{Fig:3GluonEg4} (a). The graph \figref{Fig:3GluonEg4} (b) is a physical graph, but \figref{Fig:3GluonEg4} (c) is a spurious one. In fact, the spurious graph \figref{Fig:3GluonEg4} (c) with opposite sign (see \figref{Fig:3GluonEg4} (f)) is also introduced for treating the nonlocalities in \figref{Fig:3GluonEg4} (d) and (e). 
According to the approach-2, the Feynman diagram  \figref{Fig:3GluonEg3} (d) with the BCJ-pattern from graphs \figref{Fig:3GluonEg3} (c) and (f), has to cancel with a part of \figref{Fig:3GluonEg4} (a) accompanied by the X-pattern in \figref{Fig:3GluonEg4} (b) and (c), when \figref{Fig:GeneralBCJ1} and \figref{Fig:GeneralX1} are considered. After this step, the remaining part of the diagram \figref{Fig:3GluonEg4} (a) becomes the sum of diagrams \figref{Fig:3GluonEg5} (a) and (b) with the coefficient defined by graph \figref{Fig:3GluonEg5} (c) where the type-3 line is removed. \figref{Fig:3GluonEg5} (a) produces
\bea
\left(-{1\over 2}\right){1\over l^2}{1\over s_{x_1,l}}{1\over s_{x_1pqr,l}}(\epsilon_q\cdot\epsilon_p)(\epsilon_r\cdot k_p)\phi_{x_1|a_1}\Big[\phi_{qp|a_2a_3}\phi_{r|a_4}\Big]\phi_{x_2|a_5},
\eea
which has no nonlocal contribution now. When the cyclic permutations of $(x_1x_2)$ and $(a_1a_2a_3a_4a_5)$ are considered, we get a quadratic-propagator form 
\bea
\left(-{1\over 2}\right){1\over l^2}{1\over l^2_{x_1}}{1\over l^2_{x_1pqr}}(\epsilon_q\cdot\epsilon_p)(\epsilon_r\cdot k_p)\phi_{x_1|a_1}\Big[\phi_{qp|a_2a_3}\phi_{r|a_4}\Big]\phi_{x_2|a_5}.\Label{Eq:FivePointResult2b}
\eea

One may notice that there remains a diagram \figref{Fig:3GluonEg5} (b) which still has nonlocality. This diagram, actually cancels with \figref{Fig:3GluonEg6} (c), \figref{Fig:3GluonEg6a} (c) which come from other two Feynman diagrams \figref{Fig:3GluonEg6} (a), \figref{Fig:3GluonEg6a} (a) with respect to the partitions $\{x_1,q,p,r,x_2\}$, $\{x_1,p,q,r,x_2\}$ accompanied by the X-patterns \figref{Fig:3GluonEg6} (d) and \figref{Fig:3GluonEg6a} (d). The remaining parts of \figref{Fig:3GluonEg6} (a),  \figref{Fig:3GluonEg6a} (a) (i.e., \figref{Fig:3GluonEg6} (b),  \figref{Fig:3GluonEg6a} (b)), finally cancel against the diagrams \figref{Fig:3GluonEg6} (e), \figref{Fig:3GluonEg6a} (e) with the BCJ-patterns \figref{Fig:3GluonEg6} (f), \figref{Fig:3GluonEg6a} (f), respectively. Noting that graphs \figref{Fig:3GluonEg6a} (d) and (f) are spurious ones, in the above cancellations, we have used \figref{Fig:3GluonEg6a} (d) and (f) with minus signs. The remaining two spurious graphs with positive signs are used to cancel the nonlocalities with other graphs.

To sum up, the nonlocalities can be canceled by introducing spurious graphs and combining approach-1 and -2.

\begin{figure}
\centering
\includegraphics[width=0.8\textwidth]{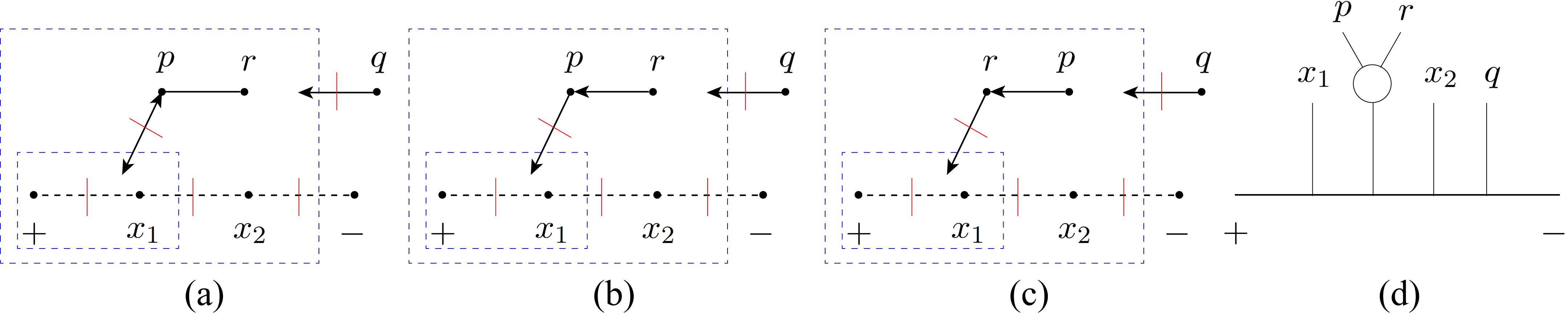}
\caption{Graphs (a), (b) and (c) with respect to the Feynman diagram (d) and the graph \figref{Fig:3GluonEg1} (c) with respect to diagram \figref{Fig:3GluonEg2} (c) together provide all contributions to the partition $\{x_1,\{p,r\},x_2,q\}$.}
\label{Fig:3GluonEg7}
\end{figure}

\subsection{Compact expression of the five-point integrand with three gluons}
In the previous subsection, we began with  graphs associated with LPFD for a given partition of external particles, and arrived at quadratic-propagator Feynman diagrams, without nonlocality. 
We can always collect all contributions corresponding to a same cyclic partition in the final results. Now we demonstrate this by explicit examples.

{\it The first example} is the partition $\{x_1,\{p,r\},x_2,q\}$, which gets contributions from  (\ref{Eq:FivePointResult1}) and \figref{Fig:3GluonEg7} (a), (b), (c) with the Feynman diagram \figref{Fig:3GluonEg7} (d). The local expression  (\ref{Eq:FivePointResult1}) which comes from the cancellations between X- and BCJ-patterns  have already been discussed.  \figref{Fig:3GluonEg7} (a), (b) and (c) with the diagram \figref{Fig:3GluonEg7} (d) are collected as
\bea
&&{1\over l^2}{1\over l^2_{x_1}}{1\over l^2_{x_1pr}}{1\over l^2_{x_1prx_2}}\Bigl[\big((-k_p\cdot X_p)(\epsilon_p\cdot\epsilon_r)+(\epsilon_r\cdot k_p)(\epsilon_p\cdot X_p)\big)\phi_{pr|a_2a_3}+(\epsilon_p\cdot k_r)(\epsilon_r\cdot X_r)\phi_{rp|a_2a_3}\Bigr](\epsilon_q\cdot X_q)\nn
&&~~~~~~~~~~~~~~~~~~~~~~~~~~~~~~~~~~~~~~~~~~~~~~~~~~\times\phi_{x_1|a_1}\phi_{x_2|a_4}\phi_{q|a_5}.\Label{Eq:FivePointResult1b}
\eea
%
%
%
%
In the above expression, the terms inside the square brackets can be collected as $\W J(p,r)\cdot X_{\{p,r\}}$, where $\W J^{\mu}(p,r)$ is the effective current \cite{Wu:2021exa} constructed by the graphic rule, under the reference order $p\prec r$. The $X^{\mu}_{\{p,r\}}$ is the momentum of the quadratic propagator attached to $\W J^{\mu}(p,r)$ from left. The sum of (\ref{Eq:FivePointResult1}) and (\ref{Eq:FivePointResult1b}) together provides 
\bea
&&{1\over l^2}{1\over l^2_{x_1}}{1\over l^2_{x_1pr}}{1\over l^2_{x_1prx_2}}\Bigl[\W J(p,r)\cdot X_{\{p,r\}}+\big(-{1\over 2}\W J(p)\cdot\W J(r)\big)\Bigr](\W J(q)\cdot X_q)\phi_{x_1|a_1}\phi_{x_2|a_4},\Label{Eq:FivePointResult1c}
\eea
where $\epsilon^{\mu}_p\phi(p|a_2)$, $\epsilon^{\mu}_r\phi(r|a_3)$ and $\epsilon^{\mu}_q\phi(q|a_5)$ are also absorbed into the corresponding subcurrents  $\W J^{\mu}(p)$, $\W J^{\mu}(q)$ and $\W J^{\mu}(r)$, respectively. 

%
\begin{figure}
\centering
\includegraphics[width=0.9\textwidth]{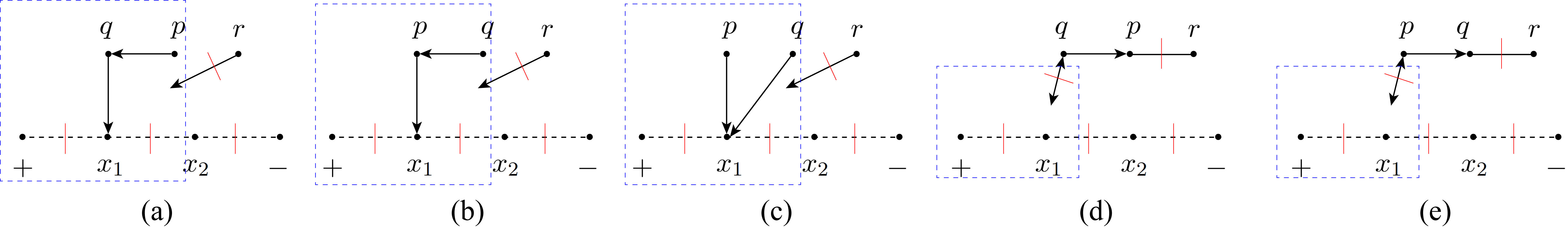}
\caption{Graphs (a), (b) and (c), together with \figref{Fig:3GluonEg3} (a), (b) and (e), provide all contributions to the partition $\{\{x_1,p,q\},r,x_2\}$. Graphs (d) and (e) together with  \figref{Fig:3GluonEg5} (c) provide a contribution to the partition $\{x_1,\{p,q,r\},x_2\}$, corresponding to the case with contraction $\W J(p,q)\cdot\W J(r)$.}
\label{Fig:3GluonEg9}
\end{figure}

{\it The second example} is given by the partition $\{\{x_1,p,q\},r,x_2\}$.  The expression (\ref{Eq:FivePointResult2a})  makes a contribution to this partition, while other contributions are collected as 
\bea
&&{1\over l^2}{1\over l^2_{x_1pq}}{1\over l^2_{x_1pqr}}\Bigl[(\epsilon_p\cdot k_q)(\epsilon_q\cdot k_{x_1})\phi_{x_1qp|a_1a_2a_3}+(\epsilon_q\cdot k_p)(\epsilon_p\cdot k_{x_1})\phi_{x_1pq|a_1a_2a_3}\nn
&&~~~~~~~~~~~~~~~~~~~~~~~~~~~~~+(\epsilon_p\cdot k_{x_1})(\epsilon_q\cdot k_{x_1})\phi_{x_1\{q\}\shuffle\{p\}|a_1a_2a_3}\Bigr](\epsilon_r\cdot X_r)\phi_{r|a_4}\phi_{x_2|a_5},\Label{Eq:FivePointResult2c}
\eea
in which, the three terms come from \figref{Fig:3GluonEg9} (a), (b) and (c), respectively. The expressions (\ref{Eq:FivePointResult2a}) and (\ref{Eq:FivePointResult2c}) sum into
\bea
{1\over l^2}{1\over l^2_{x_1pq}}{1\over l^2_{x_1pqr}}\W J(x_1,p,q)(\W J(r)\cdot X_r)\phi_{x_2|a_5},\Label{Eq:FivePointResult2d}
\eea
which is given by the effective subcurrent $\W J(x_1,p,q)$ with the scalar $x_1$ and gluons $p$, $q$, and the subcurrent $\W J^{\mu}(r)$.

{\it The last example} is the partition  $\{x_1,\{p,q,r\},x_2\}$, which contains the term  (\ref{Eq:FivePointResult2b}) already induced by the graph \figref{Fig:3GluonEg5} (c) with corresponding diagram \figref{Fig:3GluonEg5} (a). The terms induced by graphs \figref{Fig:3GluonEg9} (d) and (e) further provide
\bea
\left(-{1\over 2}\right){1\over l^2}{1\over l^2_{x_1}}{1\over l^2_{x_1pqr}}\phi_{x_1|a_1}\Big[(-\epsilon_q\cdot k_p)(\epsilon_p\cdot \epsilon_r)\phi_{qp|a_2a_3}\phi_{r|a_4}+(-\epsilon_p\cdot k_q)(\epsilon_q\cdot \epsilon_r)\phi_{pq|a_2a_3}\phi_{r|a_4}\Big]\phi_{x_2|a_5}.\Label{Eq:FivePointResult2e}
\eea
The total contribution of (\ref{Eq:FivePointResult2b}) and (\ref{Eq:FivePointResult2e}) is given by
\bea
\left(-{1\over 2}\right){1\over l^2}{1\over l^2_{x_1}}{1\over l^2_{x_1pqr}}\left[\W J(p,q)\cdot\W J(r)\right]\phi_{x_2|a_5}.\Label{Eq:FivePointResult2f}
\eea
Other contributions for this partition can also be obtained as
\bea
&&\left(-{1\over 2}\right){1\over l^2}{1\over l^2_{x_1}}{1\over l^2_{x_1pqr}}\left[\W J(p)\cdot\W J(q,r)\right]\phi_{x_2|a_5},\Label{Eq:FivePointResult2f1}\\
&&\left(-{1\over 2}\right){1\over l^2}{1\over l^2_{x_1}}{1\over l^2_{x_1pqr}}\left[\W J(q)\cdot\W J(p,r)\right]\phi_{x_2|a_5},\Label{Eq:FivePointResult2f2}\\
&&{1\over l^2}{1\over l^2_{x_1}}{1\over l^2_{x_1pqr}}\left[\W J(p,q,r)\cdot X_{\{p,q,r\}}\right]\phi_{x_2|a_5}.\Label{Eq:FivePointResult2f3}
\eea
 The sum of (\ref{Eq:FivePointResult2f}), (\ref{Eq:FivePointResult2f1}), (\ref{Eq:FivePointResult2f2}) and (\ref{Eq:FivePointResult2f3}) gives rise to
\bea
&&{1\over l^2}{1\over l^2_{x_1pq}}{1\over l^2_{x_1pqr}}\biggl[\W J(p,q,r)\cdot X_{\{p,q,r\}}+\left(-{1\over 2}\right)\left(\W J(p,q)\cdot\W J(r)+\W J(p)\cdot\W J(q, r)+\W J(q)\cdot\W J(p, r)\right)\biggr]\phi_{x_2|a_5}.\Label{Eq:FivePointResult2f4}\nn
\eea
The $\W J^{\mu}(a,b)$ and $\W J^{\mu}(p,q,r)$ are the effective currents \cite{Wu:2021exa} constructed via the graphic rule (see \appref{app:GraphsThreeGluons}), where the off-shell node (the Lorentz index $\mu$) is considered as the root. The reference order of gluons in each effective current inherits the relative reference order of the full graph. 
%
%

The contributions of other partitions in the final result can be assembled in an analogue way. When all possible partitions are considered, we get a compact expression for the final reduction result of the five-point integrand with three gluons
\bea
I^{\text{1-loop}}\big(x_1,x_2||\{p,q,r\}\big)\cong\,\Sl_{\{A_1,A_2,...,A_I\}}{1\over l^2}\,J[A_1]\, {1\over l^2_{A_1}}\,J[A_2]\cdots\,{1\over l^2_{A_1...A_{I-1}}}\,J[A_I]+\text{cyc}(a_1a_2a_3a_4a_5),
\eea
in which, all possible cyclic partitions $\{A_1,A_2,...,A_I\}$ of external particles $x_1$, $x_2$, $p$, $q$, $r$, where the relative cyclic order of  $x_1$, $x_2$ is always kept, have been summed over.  The $J[A_i]$ are defined as 
\bea
J[A_i]&=&\W J (A_i),~~~~~~~~~~~~~~~~~~~~~~~~~~~~~~~~~~~~~~~~~~~~~~~~~~~~~~~~~~~~~~~~\text{(if $A_i$ contains  scalars)}\nn
J[A_i]&=&\W J (A_i)\cdot X_{A_i}+\left(-{1\over 2}\right)\left[\Sl_{A_i\to A_{iL},A_{iR}}\W J (A_{iL})\cdot\W J (A_{iR})\right],~~~~\text{(if $A_i$ contains only gluons)}
\eea
where $A_i\to A_{iL},A_{iR}$ stands for splitting $A_i$ into two nonempty subsets such that the highest-weight element in $A_i$  belongs to $A_{iR}$. Concretely, if $A_i$ contains only two elements $a$ and $b$ with reference order $a\prec b$, the splitting is $\{a,b\}\to\{a\}\,\{b\}$.
If $A_i$ contains three elements $p$, $q$ and $r$, the possible splittings are 
\bea
\{p,q\}\,\{r\},~~~~~\{p\}\,\{q,r\},~~~~~\{q\}\,\{p,r\},
\eea
corresponding to (\ref{Eq:FivePointResult2f}), (\ref{Eq:FivePointResult2f1}), (\ref{Eq:FivePointResult2f2}).


\section{General discussions and the final result for single-trace YMS}\label{Sec:GenYMS}
In \secref{sec:SimpleExample} and \secref{sec:5ptExample}, explicit examples for extracting quadratic propagators have been provided. We have seen that the cancellation of nonlocalities played a crucial way and the final result had compact expressions in these examples. In this section, we provide a general approach to the cancellation of nonlocal terms and the compact formula of the final quadratic-propagator result.  We begin with the following expression of YMS integrand 
\bea
\frac{1}{l^2}\int \text{d}\mu_{\,n+2}^{\,\text{tree}}I_{\,\text{L}}^{\,\text{1-loop}}I_{\,\text{R}}^{\,\text{1-loop}}&=&\Sl_{\mathcal{F}}\mathcal{C}^{\mathcal{F}}\Bigg[\,\Sl_{\footnotesize\substack{(A_1A_2...A_i)={\pmb{\sigma}^{\mathcal{F}}}\\(\W A_1\W A_2...\W A_i)={\pmb{\gamma}}\\ A_j=\W A_j}}\,{1\over l^2}\,{1\over s_{A_1,l}}\,{1\over s_{A_1A_2,l}}\,\cdots\,{1\over s_{A_1A_2\cdots A_{i-1},l}}\phi_{A_1|\W A_1}\phi_{A_2|\W A_2}\cdots \phi_{A_i|\W A_i}\nn
&&~~~~~~~~~~~~~~~~+\text{cyc}(x_1x_2...x_r)\Bigg]+\text{cyc}(\pmb\gamma),\Label{Eq:LoopCHYRefinedGraphsNew}
\eea
which is obtained by substituting (\ref{Eq:Expansion1}) and (\ref{Eq:One-loopBasis-1}) into (\ref{Eq:LoopCHY}), directly. Here, $\pmb{\sigma}^{\mathcal{F}}$ denotes permutations established by the graph $\mathcal{F}$, $x_1,...,x_r$ are the scalars, $\pmb\gamma$ is a permutation of all scalars and gluons in the right half integrand. This expression turns back to \eqref{Eq:LoopCHYRefinedGraphs2} if the integrand involves two scalars and two gluons.

To collect the coefficients corresponding to a Feynman diagram, we arrange the two summations in the above expression as follows
\bea
\Sl_{\mathcal{F}}\mathcal{C}^{\mathcal{F}}\,\Sl_{\footnotesize\substack{(A_1A_2...A_i)={\pmb{\sigma}\shuffle\pmb{\rho}^{\mathcal{F}}}\\(\W A_1\W A_2...\W A_i)={\pmb{\gamma}}\\ A_j=\W A_j}}\,\to\,\Sl_{\footnotesize\substack{\text{partitions}\\ \{A_1, A_2, ..., A_i\}}}\,\left[\,\Sl_{\footnotesize\substack{\text{topologies of subsets}\\ \text{for a given partition}}}
\,\,\Sl_{\footnotesize\substack{\text{configurations of}\\ \text{subgraphs for a topology}}} \text{Contractions}\,\left[\mathcal{C}^{\mathcal{F}_j}\right]\,\right].
\eea
This means we can collect contributions of {\it all partitions} (or in other words LPFD) instead of contributions of {\it all graphs}. For a given partition $\{A_1,...,A_i\}$ of external particles, one can establish the possible topologies by connecting lines (with ignoring the types of these lines) between these subsets and sum over all such topologies. Fixing a partition and a topology, one determines the possible configurations of subgraphs by graphic rule when the tree structures attached to this subset are regarded as off-shell nodes. When we contract the Lorentz indices between coefficients $\mathcal{C}^{{\mathcal{F}}_j}$ for subgraphs ${\mathcal{F}}_j$ corresponding to the subset $A_j$, the full coefficient for this LPFD is obtained. Therefore, the integrand turns into
\bea
&&~~\frac{1}{l^2}\int \text{d}\mu_{\,n+2}^{\,\text{tree}}I_{\,\text{L}}^{\,\text{1-loop}}I_{\,\text{R}}^{\,\text{1-loop}}\nn
&=&\Sl_{\footnotesize\substack{\text{partitions}\\ \{A_1, A_2, ..., A_i\}}}\,\Biggl[\,\Sl_{\footnotesize\substack{\text{topologies of subsets}\\ \text{for a partition}}}
\Sl_{\footnotesize\substack{\text{configurations of}\\ \text{subgraphs for a topology}}} \text{Contractions}\,\left[\mathcal{C}^{{\mathcal{F}}_j}\right]\nn
&&\times{1\over l^2}\,{1\over s_{A_1,l}}\,{1\over s_{A_1A_2,l}}\,\cdots\,{1\over s_{A_1A_2\cdots A_{i-1},l}}\phi_{A_1|\W A_1}\phi_{A_2|\W A_2}\cdots \phi_{A_i|\W A_i}+\text{cyc}(x_1x_2...x_r)\,\Biggr]+\text{cyc}(\pmb\gamma).\Label{Eq:LoopCHYRefinedGraphsNew1}
\eea
\begin{figure}
\centering
\includegraphics[width=0.97\textwidth]{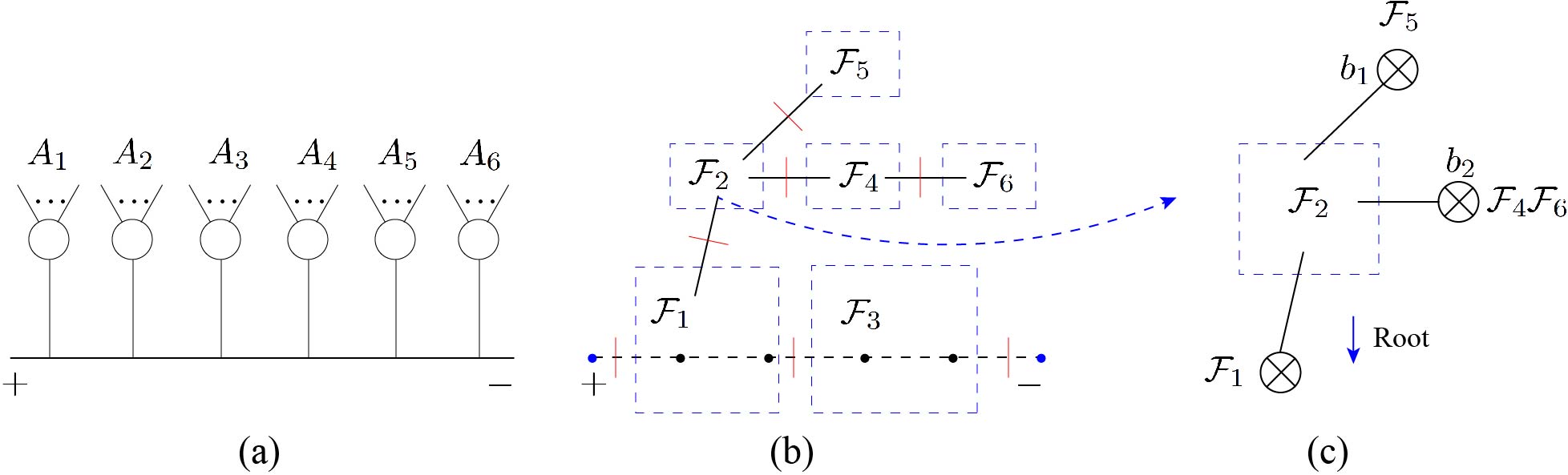}
\caption{The LPFD for partition $\{A_1,A_2,A_3,A_4,A_5,A_6\}$ is given by diagram (a). Assume that $A_1$ and $A_3$ contain scalars, while $A_2$, $A_4$, $A_5$, $A_6$ are pure gluon subsets. Graph (b), where $\mathcal{F}_1$, ...,  $\mathcal{F}_6$ stand for the subgraphs corresponding to subsets $A_1$, ..., $A_6$, is a possible topology for this partition. The subgraphs containing scalars are interconnected via type-4  lines to record the relative order between scalars. Other lines between subgraphs can be of other three types and are expressed by solid lines with no arrow.   When other subgraphs connected to a given subgraph are considered as off-shell nodes, the subgraph under consideration can be regarded as a graph with both on-shell  and off-shell nodes, e.g., (c). In (c), the off-shell node below $\mathcal{F}_2$ stands for the subgraph $\mathcal{F}_1$, while the off-shell nodes $b_1$ and $b_2$ (which locate above $\mathcal{F}_2$) respectively represent the subtree $\mathcal{F}_5$ and the subtree consisting of $\mathcal{F}_4$, $\mathcal{F}_6$.}
\label{Fig:GeneralPatternPatition}
\end{figure}

Generally speaking, in (\ref{Eq:LoopCHYRefinedGraphsNew1}), {\it there exist terms containing contractions between subgraphs corresponding to distinct subsets. Since these subsets are separated by linear propagators in the Feynman diagram, such terms are nonlocal ones}. These nonlocalities have to be canceled. Actually, we can reconstruct the subgraphs by adjusting the refined graphic rule in a proper way so that (i). the X- and BCJ- patterns arise and/or (ii). the $C\cdot X$ factor can be collected. Hence the cancellations of nonlocalities can also be achieved, following the same approaches in \secref{sec:SimpleExample} and \secref{sec:5ptExample}. When the cyclic permutations of  $(x_1x_2...x_r)$ and $\pmb{\gamma}$ are considered, we finally get a compact formula with quadratic propagators.

In the remaining part of this section, we demonstrate the above general approach in more detail, including generating topologies for a partition, the construction rule for a subgraph and the cancellation between X- and BCJ-patterns. After that, we provide the final compact expression for single-trace YMS  integrand (which contains a pure scalar loop) with quadratic propagators.

\subsection{Topologies for a given partition}





When a partition of external particles has been given, e.g., the partition $\{A_1,A_2,A_3,A_4,A_5,A_6\}$ which corresponds to the LPFD \figref{Fig:GeneralPatternPatition} (a), one can easily find out all possible topologies for these subsets that agree with the graphic rule:
\begin{itemize}

\item Subsets containing scalars are interconnected according to their relative order in the partition, via type-4 lines.  The leftmost (rightmost) subset containing scalars is further connected to $+$ ($-$) via a type-4 line.

\item If a subset $A_i$ contains only gluons, one can  connect (i). a line between $A_i$ and any subset, say $A_j$, which locates on the left of $A_i$ in the partition, or (ii). a line between $A_i$ and the node $+$. The line style is not fixed.

\end{itemize}
A typical topology for the Feynman diagram \figref{Fig:GeneralPatternPatition} (a), is presented by \figref{Fig:GeneralPatternPatition} (b) which encodes relative positions of subgraphs.

In \appref{sec:Subgraph1} and  \appref{sec:Subgraph2}, we provide systematic rules (which are equivalent to each other as pointed in \cite{Hou:2018bwm}) for constructing subgraphs with proper spurious graphs in a given topolopy.  The key result following the construction in \appref{sec:Subgraph2} is that a subgraph consists of an upper and a lower block $\mathcal{U}$ and $\mathcal{L}$, which are (i). connected via a type-3 line if both blocks has on-shell nodes (see \figref{Fig:SubgraphSituations0} (a)), or (ii). connected via a half-arrow line if one of the block has no on-shell node (see \figref{Fig:SubgraphSituations0} (b) and (c)). According to \appref{sec:Subgraph2}, a subgraph may (a). contain a BCJ-pattern (whose two subgraphs are $\mathcal{U}$ and $\mathcal{L}$) or an X-pattern (only occurs in the case where the lower block has no on-shell node), (b). play as a physical subgraph when cross nodes are extracted out (in the case where the upper block has no on-shell node), (c). be neither the case (a) nor (b) and relies on the subgraph above it (in the case where the lower block has no on-shell node). In the coming subsections, we provide a general discussion on the cancellations between X- and BCJ-patterns and summarize the final expression with no nonlocal term.



\subsection{General cancellation between X- and BCJ-patterns}\label{sec:CancellationXBCJ}

\begin{figure}
\centering
\includegraphics[width=1\textwidth]{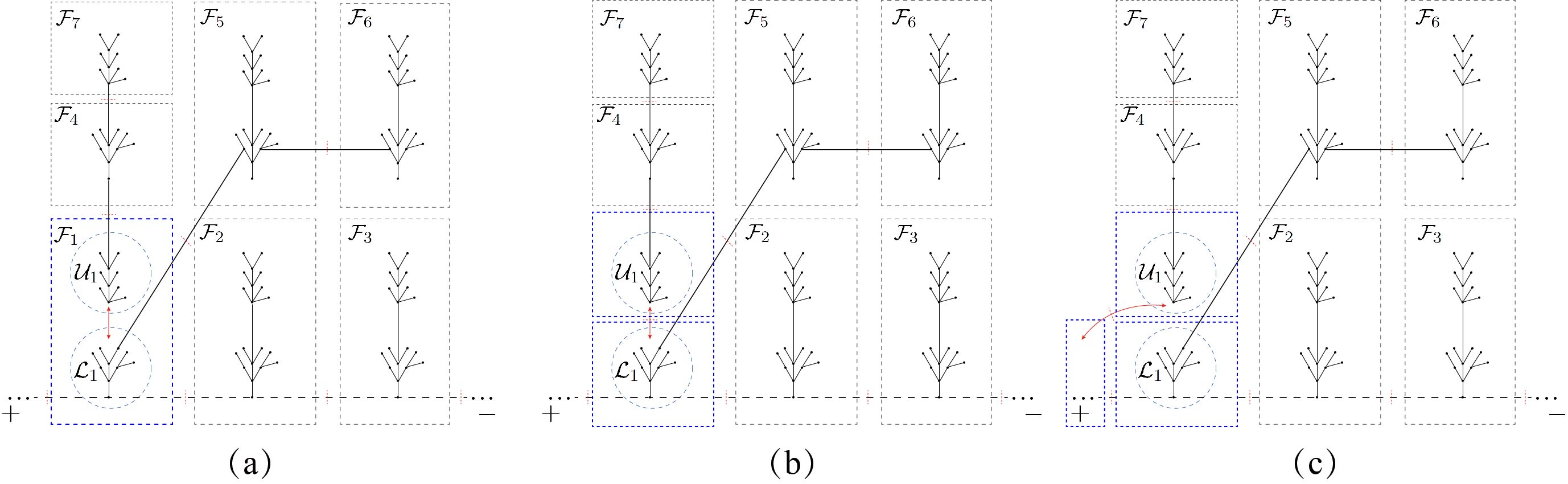}
\caption{Three typical graphs which are decomposed into subgraphs. The subgraph $\mathcal{F}_1$ in graph (a) is a BCJ-pattern whose upper and lower blocks are $\mathcal{U}_1$ and $\mathcal{L}_1$. Graph (b) is obtained by decomposing $\mathcal{F}_1$ in (a) into two subgraphs $\mathcal{U}_1$ and ${\mathcal{L}_1}$. When the  $\mathcal{U}_1$ in (b) is connected to the substructure below $\mathcal{L}_1$ via a type-3 line, we get graph (c). Graph (b) and (c) provide an X-pattern with $\mathcal{U}_1$.}
\label{Fig:StructureOfCuts}
\end{figure}
 Now, let us demonstrate the general cancellation between X- and BCJ-patterns. We begin with the typical graph \figref{Fig:StructureOfCuts} (a) (which may contribute to different partitions), where each subgraph is considered as any graph constructed by combining version-1 and -2 rules when the partition and topology have been fixed. We will show that the cancellations occur among graphs with respect to different topologies and different partitions.

\noindent{\bf Cancellations of BCJ-patterns}~~Without loss of generality, we concentrate on the subgraph $\mathcal{F}_1$ and suppose that it is a subgraph with BCJ-pattern. In other words, it consists of upper and lower blocks $\mathcal{U}_1$, $\mathcal{L}_1$, which are connected via a type-3 line, as shown by \figref{Fig:StructureOfCuts} (a). We further assume that the partition under consideration for this graph has the form $\{...,A_1,...\}$. This graph then provides
\bea
\left(-{1\over 2}\right)\Big[\dots\left(\mathcal{C}^{\mathcal{L}_1}\mathcal{C}^{\mathcal{U}_1}\phi_{\left[\mathcal{L}_1,\mathcal{U}_1\right]}\right)\dots\Big]\dots {1\over s_{A_0,l} }{1\over s_{A_0A_1,l} }\dots,\Label{Eq:Diagram1}
\eea
where the off-shell BCJ relation has been applied.  In the above, the $A_0$ denotes the set of all elements on the left of $A_1$ in the partition. The right permutation in a BS current was omitted for short. The $\phi_{ [A_{L},A_R]}$ stands for 
\bea
\phi_{ A_{L}|\W A_{L}}\phi_{ A_{R}|\W A_{R}}-\phi_{ A_{R}|\W A'_{L}}\phi_{ A_{L}|\W A'_{R}}\Label{Eq:phiLR}
\eea
 where $\W A_{L}\W A_{R}=\W A'_{L}\W A'_{R}$. The subscripts $\mathcal{U}_1$ and $\mathcal{L}_1$ in $\phi_{\mathcal{U}_1}$ and $\phi_{\mathcal{L}_1}$ stand for permutations established by the corresponding subgraphs. 

For \figref{Fig:StructureOfCuts} (a) with the partition $\{...,A_1,...\}$, one can always find other graphs (with other topologies) \figref{Fig:StructureOfCuts} (b) and (c).  \figref{Fig:StructureOfCuts} (b) can be considered as the graph  obtained by dividing the upper and lower blocks of $\mathcal{F}_1$ in \figref{Fig:StructureOfCuts} (a) into two subgraphs. \figref{Fig:StructureOfCuts} (c) can be obtained by connecting a type-3 line between $\mathcal{U}_1$ and the subsets to the left of $\mathcal{U}_1$ instead of $\mathcal{L}_1$ in \figref{Fig:StructureOfCuts} (b).  These two graphs contribute X-patterns, i.e., the subgraph with $\mathcal{U}_1$. Particularly, for partition $\{...,\mathcal{L}_1,\mathcal{U}_1,...\}$, both \figref{Fig:StructureOfCuts} (b) and (c) produce an X-pattern, while for partition $\{...,\mathcal{U}_1,\mathcal{L}_1,...\}$, the graph \figref{Fig:StructureOfCuts} (c) has an X-pattern. The explicit expression for partition $\{...,\mathcal{L}_1,\mathcal{U}_1,...\}$ (with graphs \figref{Fig:StructureOfCuts} (b), (c)) is given by
\bea
%
&&\Big[\dots\left(\mathcal{C}^{\,\mathcal{L}_1}\phi_{\mathcal{L}_1}\right)\left(\mathcal{C}^{\,\mathcal{U}_1}\phi_{\mathcal{U}_1}\right)\dots\Big]\dots[-k_{\mathcal{U}_1}\cdot \left(l+k_{A_0\mathcal{L}_1}\right)]{1\over s_{A_0\mathcal{L}_1,l}} {1\over s_{A_0A_1,l} }\dots~\Label{Eq:Diagram2}\\
%
%
&=&\Big[\dots\left(\mathcal{C}^{\,\mathcal{L}_1}\phi_{\mathcal{L}_1}\right)\left(\mathcal{C}^{\,\mathcal{U}_1}\phi_{\mathcal{U}_1}\right)\dots\Big]\dots\left(-{1\over 2}\right)\left[-{1\over s_{A_0A_1,l} }+{1\over s_{A_0\mathcal{L}_1,l}}-{k^2_{\mathcal{U}_1}\over s_{A_0\mathcal{L}_1,l}s_{A_0A_1,l}}\right]\dots\nonumber
\eea
The definition of X-pattern and the property (\ref{Eq:XCoefficient})  have been used. Analogously, the contribution of \figref{Fig:StructureOfCuts} (c) for the partition $\{...,\mathcal{U}_1,\mathcal{L}_1,...\}$ is given by
\bea
%
&&\Big[\dots\left(\mathcal{C}^{\,\mathcal{L}_1}\phi_{\mathcal{L}_1}\right)\left(\mathcal{C}^{\,\mathcal{U}_1}\phi_{\mathcal{U}_1}\right)\dots\Big]\dots[-k_{\mathcal{U}_1}\cdot (l+k_{A_0})]{1\over s_{A_0,l}}{1\over s_{A_0\mathcal{U}_1,l} }\dots\Label{Eq:Diagram2c}\\
%
%
&=&\Big[\dots\left(\mathcal{C}^{\,\mathcal{L}_1}\phi_{\mathcal{L}_1}\right)\left(\mathcal{C}^{\,\mathcal{U}_1}\phi_{\mathcal{U}_1}\right)\dots\Big]\dots\left(-{1\over 2}\right)\left[-{1\over s_{A_0\,\mathcal{U}_1,l} }+{1\over s_{A_0,l}}-{k^2_{\mathcal{U}_1}\over s_{A_0,l}s_{A_0\,\mathcal{U}_1,l}}\right]\dots\nonumber
\eea
When we expand the `commutator' in (\ref{Eq:Diagram1}), the first and the second terms of (\ref{Eq:Diagram1}) respectively cancel against the first term of (\ref{Eq:Diagram2}) and the second term of (\ref{Eq:Diagram2c}). Hence the full BCJ-pattern has been canceled.

\begin{figure}
\centering
\includegraphics[width=0.95\textwidth]{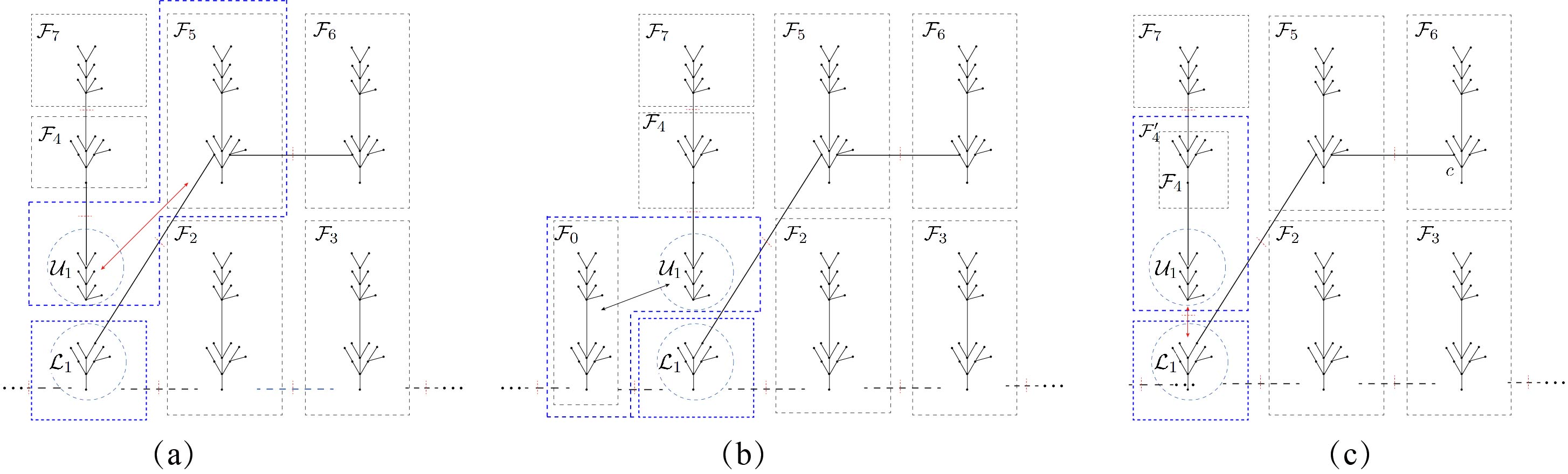}
\caption{Graphs (a), (b) and (c)  provide further cancellations with the remaining contributions of  the X-pattern in \figref{Fig:StructureOfCuts} (b) and (c). Graphs (a) and (b) contain BCJ-patterns whose upper block is $\mathcal{U}_1$ and lower blocks are $\mathcal{F}_5$ and $\mathcal{F}_0$, respectively. In graph (c), $\mathcal{F}_4$ and $\mathcal{U}_1$ are merged into a larger X-pattern. }
\label{Fig:StructureOfCutsNew2}
\end{figure}

{\bf Further cancellations of X-patterns}~~When the BCJ-pattern in \figref{Fig:StructureOfCuts} (a) has been canceled out, there are remaining contributions of X-patterns in \figref{Fig:StructureOfCuts} (b) and (c), i.e., the second and the third terms in (\ref{Eq:Diagram2}), the first and the third terms in (\ref{Eq:Diagram2c}). Now we concentrate on these terms and further classify our discussions by distinct  partitions  for  \figref{Fig:StructureOfCuts} (b) and (c):
\begin{itemize}
\item {\bf
The subset  next to  $\mathcal{U}_1$ in the partition for \figref{Fig:StructureOfCuts} (b) is not involved by a cross node above $\mathcal{U}_1$}, e.g., the partition $\{...,\mathcal{L}_1,\mathcal{U}_1,A_5,...\}$. For this case, one can always find a BCJ-pattern \figref{Fig:StructureOfCutsNew2} (a), whose upper and lower blocks are  $\mathcal{U}_1$ and $\mathcal{F}_5$, respectively. When the off-shell BCJ relation is applied, this graph provides
\bea
%
&&\left(-{1\over 2}\right)\Big[\dots\left(\mathcal{C}^{\,\mathcal{L}_1}\phi_{\mathcal{L}_1}\right)\left(\mathcal{C}^{\,\mathcal{U}_1}\mathcal{C}^{\mathcal{F}_5}\phi_{\left[\,\mathcal{U}_1,\mathcal{F}_5\right]}\right)\dots\Big]\dots{1\over s_{A_0\mathcal{L}_1,l} }{1\over s_{A_0A_1A_5,l} }\dots,\Label{Eq:Diagram1c}
\eea
whose first term cancel out the second term of  (\ref{Eq:Diagram2}) (where $\left(\mathcal{C}^{\mathcal{F}_5}\phi_{\mathcal{F}_5}\right){1\over s_{A_0A_1A_5,l} }$ is absorbed by the dots), according to the cancellation of BCJ-pattern with the partition $\{...,\mathcal{L}_1,\{\mathcal{U}_1,A_5\},...\}$. Similarly, the first term of (\ref{Eq:Diagram2c}) (for the partition $\{...,\mathcal{F}_0,\mathcal{U}_1,\mathcal{L}_1,...\}$ where $\mathcal{F}_0$ refers to the subset near to $\mathcal{U}_1$ from left and the corresponding subgraph) is canceled by one term of the BCJ-pattern in \figref{Fig:StructureOfCutsNew2} (b) with the partition $\{...,\{\mathcal{F}_0,\mathcal{U}_1\},\mathcal{L}_1,...\}$.

\item  {\bf The subset  next to  $\mathcal{U}_1$ in the partition for \figref{Fig:StructureOfCuts} (b) is contained by a cross node above $\mathcal{U}_1$}, e.g., the partition $\{...,\mathcal{L}_1,\mathcal{U}_1,A_4,...\}$\footnote{In this example, there is only one cross node above $\mathcal{U}_1$. For cases with more cross nodes above $\mathcal{U}_1$, the subset next to $\mathcal{U}_1$ may belong to any one of these cross nodes. However, the discussion in this part is also effective. }. If {\it the lower block of $\mathcal{F}_4$ does not contain on-shell node}, one can always find a graph with a larger X-pattern $\mathcal{F}'_4$, which is obtained from \figref{Fig:StructureOfCuts} (b) by merging $\mathcal{U}_1$ and $A_4$ into a single X-pattern, as shown by \figref{Fig:StructureOfCutsNew2} (c). Explicitly, this graph provides
\bea
%
%
&&\left[\dots\left(\mathcal{C}^{\mathcal{L}_1}\phi_{\mathcal{L}_1}\right) \left(\mathcal{C}^{\mathcal{F}'_4}\phi_{^{\mathcal{F}'_4}}\right)\dots\right]\left[-k_{A_4\mathcal{U}_1}\cdot \left(l+k_{A_0}+k_{\mathcal{L}_1}\right)\right]{1\over s_{A_0\mathcal{L}_1,l}}{1\over s_{A_0A_1A_4,l} }\dots\Label{Eq:Diagram3}\\
&=&\left(-{1\over 2}\right)\left[\dots\left(\mathcal{C}^{\mathcal{L}_1}\phi_{\mathcal{L}_1}\right) \left(\mathcal{C}^{\mathcal{F}'_4}\phi_{\mathcal{F}'_4}\right)\dots\right]\left[-{1\over s_{A_0A_1A_4,l} }+{1\over s_{A_0\mathcal{L}_1,l}}-{\left(k_{\mathcal{U}_1}+k_{A_4}\right)^2\over s_{A_0A_1A_4,l}s_{A_0\mathcal{L}_1,l}}\right]\dots\nonumber
\eea
The third term on the last line in the above expression involves
\bea
-\left(-{1\over 2}\right)\mathcal{C}^{\mathcal{F}'_4}{\left(k_{\mathcal{U}_1}+k_{A_4}\right)^2\over s_{A_0A_1A_4,l}s_{A_0\mathcal{L}_1,l}}\phi_{\mathcal{F}'_4}.
\eea
According to the BG recursion (\ref{Eq:BScurrent}) of the BS currents, the numerator together with the current $\phi$ produces a combination of product of two subcurrents, which contains the following term
\bea
{1\over 2}{1\over s_{A_0A_1A_4,l}s_{A_0\mathcal{L}_1,l}}\left(\mathcal{C}^{\mathcal{F}'_4}\phi_{\left[\mathcal{U}_1,\mathcal{F}_4\right]}\right)= {1\over 2}{1\over s_{A_0A_1A_4,l}s_{A_0\mathcal{L}_1,l}}\left(\mathcal{C}^{\mathcal{U}_1}\mathcal{C}^{\mathcal{F}_4}\phi_{\left[\mathcal{U}_1,\mathcal{F}_4\right]}\right),
\eea
where $\mathcal{C}^{\mathcal{F}'_4}$ has been rewritten as the contraction between $\mathcal{C}^{\mathcal{U}_1}$ and $\mathcal{C}^{\mathcal{F}_4}$.  Noting that the expression $\mathcal{C}^{\mathcal{F}_4}\phi_{\mathcal{F}_4}{1\over s_{A_0A_1A_4,l}}$ is absorded by dots in  (\ref{Eq:Diagram2}), once the commutator is expanded, we find the first term cancels with the second term on the last line of (\ref{Eq:Diagram2}). 
 Following a similar discussion, the last terms of (\ref{Eq:Diagram2}), (\ref{Eq:Diagram2c}) cancel with the X-patterns which come from subdivisions of $\mathcal{U}_1$.  If {\it the lower block of $\mathcal{F}_4$ in \figref{Fig:StructureOfCuts} (b) involves on-shell nodes}, the second term of (\ref{Eq:Diagram2}) survives because one cannot find a larger X-pattern to cancel it. In this case, when the first and the last terms are canceled out, (\ref{Eq:Diagram2}) becomes
\bea
\Big[\dots\left(\mathcal{C}^{\,\mathcal{L}_1}\phi_{\mathcal{L}_1}\right)\left(\mathcal{C}^{\,\mathcal{U}_1}\phi_{\mathcal{U}_1}\right)\left(\mathcal{C}^{\mathcal{F}_4}\phi_{\mathcal{F}_4}\right)\dots\Big]\dots\left(-{1\over 2}\right){1\over s_{A_0\mathcal{L}_1,l}}{1\over s_{A_0A_1A_4,l}}\dots
\eea
If the $\mathcal{F}_4$ further involves a BCJ-pattern, it has to cancel out. {\it If the upper block of $\mathcal{F}_4$ has no on-shell node, the above expression finally survives and provides a contraction between $\left(\mathcal{C}^{\,\mathcal{U}_1}\phi_{\mathcal{U}_1}\right)$ and $\left(\mathcal{C}^{\mathcal{F}_4}\phi_{\mathcal{F}_4}\right)$ ($\mathcal{F}_4$ is a physical graph constructed by graphic rule), with the linear propagator between them deleted.}
\end{itemize}
There is a boundary case that $\mathcal{U}_1$ in (\ref{Eq:Diagram2}) contains only one node. In this case, the on-shell condition gives $k_{\mathcal{U}_1}^2=0$, and $\mathcal{U}_1$ cannot further be  divided. Therefore, the third term in (\ref{Eq:Diagram2})  itself vanishes but do not cancel with a further division.

\subsection{From linear to quadratic propagators}

 We now summarize what we have learnt in the previous subsections and then provide the general expression with quadratic propagators.
\begin{itemize}
\item For a given partition and a topology, if a subgraph contains a BCJ-pattern, one can always find the corresponding graph with an X-pattern to cancel it. Therefore, the surviving graphs cannot involve any subgraph with BCJ-pattern.

\item If a subgraph contains an X-pattern (the on-shell nodes in this subgraph can only be gluons), the corresponding Feynman diagram further splits into three terms, as shown by \figref{Fig:GeneralX1}, in which, (a) cancels with a BCJ-pattern, (c) may cancel with either a BCJ-pattern or a larger X-pattern, while (b) may cancel with an X-pattern (which can be considered as a subdivision of the original X-pattern). The only possible nonvanishing term from this subgraph (which corresponds to a subset say, $A_{iL}$) is given by the contraction between $\mathcal{F}_{iL}$ and another subgraph, say {$\mathcal{F}_{iR}$} whose upper block has no on-shell node {(in other words, a subgraph constructed by version-1 rule)}. On the LPFD side, this structure exists in the diagram where $\mathcal{F}_{iR}$ is adjacent to the $\mathcal{F}_{iL}$ from right. The linear propagator between them is deleted. Such contraction provides a factor 
$\left(-{1\over 2}\right)\left[\,\W J (A_{iL})\cdot\W J (A_{iR})\right]$, where $\W J_{\mu}(A_{iL})=\Sl_{\mathcal{F}_{iL}} \mathcal{C}_{\mu}^{\mathcal{F}_{iL}} \phi_{\mathcal{F}_{iL}}$ and $\W J_{\mu}(A_{iR})=\Sl_{\mathcal{F}_{iR}} \mathcal{C}_{\mu}^{\mathcal{F}_{iR}} \phi_{\mathcal{F}_{iR}}$  are the effective currents defined by graphic rule with an off-shell node.

\item For a given partition and a given topology, if a cross node, say $b_i$, of a subgraph $\mathcal{F}_j$ itself plays as the upper block (in other words, the upper block of this subgraph contains no on-shell node), one can find out other related topologies for the same partition, where the full substructure involved by  $b_i$ plays as the upper block of other subgraphs to the left of $A_j$ in the partition. The leftmost subset $A_k$ involved in $b_i$ for the partition then produces a factor $(\mathcal{C}^{\mathcal{F}_k}\cdot X_{A_k})\,\phi_{\mathcal{F}_k}=\W J(A_k)\cdot X_{A_k}$.

\item  For a subset which contains scalars (and maybe gluons), the only surviving subgraphs are the ones whose upper block does not contain on-shell node. In this case, all on-shell nodes (scalars and possible gluons) are used to construct the lower blocks. These subgraphs are just physical graphs constructed by version-1 rule and all together provide a $\W J (A_{j})$.


\end{itemize}
When all possibilities are considered, we get a local expression of integrand
\bea
I^{\text{1-loop}}(x_1,x_2,\dots,x_r||G|\pmb{\rho})&\cong&\,\left[\Sl_{\{A_1,A_2,...,A_I\}}{1\over l^2}\,J[A_1]\, {1\over s_{A_1,l}}\,J[A_2]\cdots\,{1\over s_{A_1...A_{I-1},l}}\,J[A_I]+\text{cyc}(x_1x_2\dots x_{r})\right]\nn
&&~~~~~~~~~~~~~~~~~~~~~~~~~~~~~~~~~~~~~~~~~~~~~~~~~~~~~~~~+\text{cyc}(a_1a_2\dots a_{r+s}).\Label{Eq:GenResultNew0}
\eea
\begin{figure}
\centering
\includegraphics[width=0.8\textwidth]{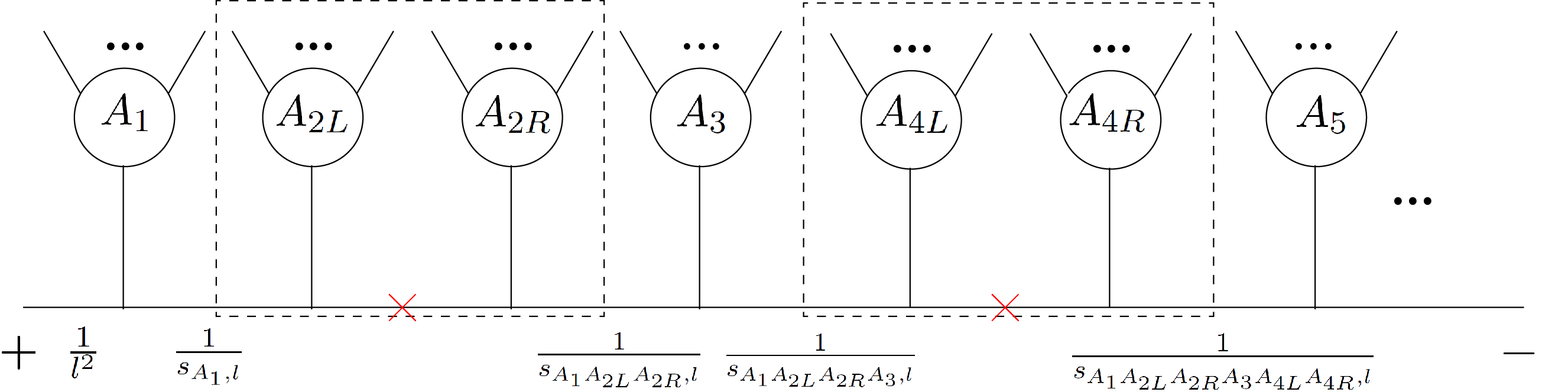}
\caption{ A typical diagram with surviving terms, where $A_1$, $A_3$, $A_5$ are the (YMS or pure gluon) subcurrents corresponding to the physical subgraphs. The substructures containing $A_{2L}$, $A_{2R}$ and $A_{4L}$, $A_{4R}$ provide contractions between pure gluon effective subcurrents $\W J(A_{2L})\cdot \W J(A_{2R})$ and  $\W J(A_{4L})\cdot \W J(A_{4R})$.}
\label{Fig:SurvivingGraphs}
\end{figure}
In the first term, we summed over all possible partitions $\{A_1,A_2,...,A_I\}$ of external particles (including $r$ scalars $x_1$, $x_2$,..., $x_r$ and $s$ gluons). The relative  order of  $x_1$, $x_2$, ..., $x_r$  in the partition $\{A_1,A_2,...,A_I\}$ is always kept. Terms related by cyclic permutations of scalars and terms related by the cyclic permutations of all elements on the right side $\pmb{\rho}=(a_1a_2\dots a_{r+s})$ are also taken into account. The $J[A_i]$ are defined as 
\bea
J[A_i]&=&\W J (A_i),~~~~~~~~~~~~~~~~~~~~~~~~~~~~~~~~~~~~~~~~~~~~~~~~~~~~~~~~~~~~~~~~\text{(if $A_i$ contains  scalars)}\Label{Eq:GenResultNewJ1}\\
J[A_i]&=&\W J (A_i)\cdot X_{A_i}+\left(-{1\over 2}\right)\left[\Sl_{A_i\to A_{iL},A_{iR}}\W J (A_{iL})\cdot\W J (A_{iR})\right].~~~~\text{(if $A_i$ contains only gluons)}\Label{Eq:GenResultNewJ2}
\eea
A typical diagram of (\ref{Eq:GenResultNew0}) is shown by \figref{Fig:SurvivingGraphs}.

The cyclic permutations of $(x_1x_2...x_r)$ and $(a_1a_2...a_{r+s})$ in (\ref{Eq:GenResultNew0}) together with graphic rule allow the cyclic summation of the subcurrents $J[A_1]$, ...., $J[A_I]$. Using (\ref{Eq:partial}), we finally conclude that the integrand with quadratic propagators is expressed by
\bea
I^{\text{1-loop}}(x_1,x_2,\dots,x_r||G|\pmb{\rho})\cong\,\Sl_{\substack{\{A_1,A_2,...,A_I\}\\ (\W A_1\,\W A_2\,...\,\W A_I)\\ A_i=\W A_i}}{1\over l^2}\,J[A_1]\, {1\over l^2_{A_1}}\,J[A_2]\cdots\,{1\over l^2_{A_1...A_{I-1}}}\,J[A_I],\Label{Eq:GenResultNew}
\eea
where $\{A_1,A_2,...,A_I\}$ are possible cyclic partitions of external particles, while $(\W A_1\,\W A_2\,...\,\W A_I)$ are possible divisions of the cyclic permutation of $(a_1a_2...a_{r+s})$ such that the elements in $\W A_i$ ($i=1,...,I$), which are hidden inside $\W J[A_i]$, are identical with those in  $A_i$. Each subset $\W A_i$ cannot contain all scalars, because this case has to vanish due to the $U(1)$-decoupling identity under the cyclic sum of scalars in (\ref{Eq:GenResultNew0}). The four-point and the five-point examples precisely agree with this general expression.  More detail about (\ref{Eq:GenResultNew0}) are demonstrated by YMS integrand with four gluons in \appref{app:ExplicitFourGluons}.

\begin{figure}
\centering
\includegraphics[width=0.94\textwidth]{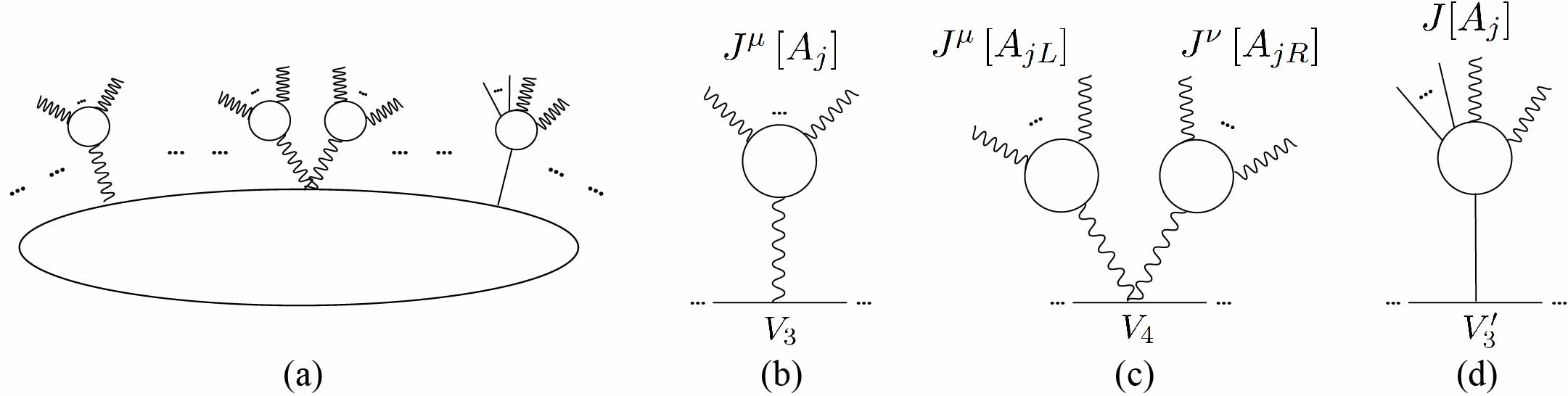}
\caption{Traditional Feynman diagrams for single-trace YMS in general can be packaged into (a), where tree structures are collected as BG currents which are further attached to the loop with quadratic propagators via the vertices \figref{Fig:Feynman0} (c), (d) and/or (e). Thus there are three types of such substructures, as shown by (b), (c) and (d) (the vertices \figref{Fig:Feynman0} (c), (d) and (e) are respectively denoted by $V_3$, $V'_3$ and $V_4$), where $J^{\mu}[A_j]$'s  in (b) and (c) are pure gluon BG currents, $J[A_j]$ in (d) is a YMS BG current with both scalar and gluon external lines. }
\label{Fig:Feynman}
\end{figure}

\section{Traditional Feynman diagrams and the extension to  multi-trace YMS and YM}\label{sec:FeynDiagrams}

In this section, we reveal the relationship between the main result (\ref{Eq:GenResultNew}) that was derived from one-loop CHY formula, and the direct analysis from Feynman diagrams of single-trace YMS. We further extend these results straightforwardly to multi-trace YMS amplitudes (with a pure scalar loop) and pure YM amplitudes.

\subsection{The relationship with traditional Feynman diagrams}

Now we demonstrate the explicit relationship between the quadratic propagator form (\ref{Eq:GenResultNew}) and the Feynman diagrams. To achieve this, we display the expressions of vertices in traditional Feynman rule for single-trace YMS  in \figref{Fig:Feynman0} and then package the  Feynman diagrams by attaching tree-level BG currents to the loop with quadratic propagators, via vertices in \figref{Fig:Feynman0}. The sum of all Feynman diagrams becomes the sum of diagrams with structures like \figref{Fig:Feynman} (a).
%
%
%
%
%
%
The BG currents in general can be planted to the quadratic-propagator scalar loop via three types of vertices, as shown by \figref{Fig:Feynman} (b), (c), (d). Each $J^{\mu}$ in  \figref{Fig:Feynman} (b) and (c) is a BG current for gluons. Upto an overall factor,  this current has been proven to satisfy the off-shell decomposition formula \cite{Wu:2021exa} (also see \cite{Lee:2015upy, Bridges:2019siz})\footnote{Here we make some clarifications about the result in \cite{Wu:2021exa}: (i). In \cite{Wu:2021exa}, the reference order was fixed as the inverse of the right permutation of the BS amplitude. However, this result can be generalized to the cases with other reference orders by changing the particular expressions of the $K^{\mu}$ term and $L^{\mu}$ term. (ii). In \cite{Wu:2021exa}, the expansion of $\W J^{\mu}$ is expressed by summing over permutations, but in this paper, we sum over graphs. In fact they are equivalent to each other as pointed out in \cite{Hou:2018bwm}. (iii). In \cite{Wu:2021exa}, there are two equivalent decompositions of  $\W J^{\mu}$: (a) The first is choosing  the off-shell line as the last element, and an on-shell line as the first. Assuming that there are $m$ external on-shell particles, this is an $(m-1)!$-decomposition and the coefficients were mentioned as type-A numerators. (b). The second is to choose the off-shell element as the first one but the last element is not fixed. This is an $m!$-decomposition and the coefficients are mentioned as type-B numerators in \cite{Wu:2021exa}. In this work, the $\W J$ in  \eqref{Eq:JA1} is considered as the second decomposition.  (iv). When extracting a factor ${i\over \sqrt{2}}$, ${i\over 2}$ and $-i$ from each three-point vertex, four-point vertex and gluon propagator in standard Feynman diagrams, we get an overall factor $\left({1/\sqrt{2}}\right)^{N_g-1}$, where $N_g$ denotes the total number of external gluons in the current. This factor does not affect our discussions and is not included in (\ref{Eq:JA1}).
}:
\bea
J^{\mu}=\W J^{\mu}+K^{\mu}+L^{\mu}, \Label{Eq:JA1}
\eea
in which the first term $\W J^{\mu}$ can be decomposed into a combination of BS currents, according to the refined graphic rule while all external momenta in the numerator are rescaled as $k^{\mu}\to 2k^{\mu}$. The second term, $K^{\mu}$ is proportional to the total momentum of external gluons in the current. The third term $L^{\mu}$ contains substructures which are obtained via replacing subcurrents by objects proportional to the total momentum of these subcurrents.

Note that the {\it three-point vertex} \figref{Fig:Feynman0} (c) with two scalar lines and one gluon line is proportional to $\delta^{ab}2k_1^{\mu}$\footnote{For each two-scalar one-gluon vertex, two-scalar two-gluon vertex and each scalar propagator, we extract a factor $-i/\sqrt{2}$, $-i/2$ and $i$, respectively. These factors only contribute to an overall factor and do not affect our discussions. }, due to the Ward identity $k_{A_j}\cdot J(A_j)=0$ where $J^{\mu}(A_j)$ is the subcurrent contracted with this vertex. When the color factor is extracted out, the substructure \figref{Fig:Feynman} (b), together with $\W J^{\mu}$ produces 
$2\W J\left[A_j\right]\cdot X_{A_j}$, where $X^{\mu}_{A_j}$ denotes the momentum of the propagator attached to $J^{\mu}(A_j)$ from left. In \figref{Fig:Feynman} (c), {\it the four-point vertex}, together with the $\W J^{\mu}$ contributes a $-\W J[A_{jL}]\cdot\W J[A_{jR}]$.

The {\it remaining terms} resulted from the $K^{\mu}$ and $L^{\mu}$ can be collected as 
\bea
V_3&\sim&\left[(X_{A_j}-(-X_{A_j}-k_{A_j}))-k_{A_j}\right]\cdot(K[A_j]+L[A_j]),\Label{Eq:FeynKL}\\
V_4&\sim&-(L[A_{jL}]+K[A_{jL}])\cdot J[A_{jR}]-J[A_{jL}]\cdot(L[A_{jR}]+K[A_{jR}])+(L[A_{jL}]+K[A_{jL}])\cdot (L[A_{jR}]+K[A_{jR}]).\nonumber
\eea
In the above expression, the first line comes from the three-point vertex with the new expression $\delta^{ab}2k_1^{\mu}$, where $2k_1^{\mu}$ (from the vertex  \figref{Fig:Feynman0} (c)) in this case is $2X_{A_j}^{\mu}$. Terms in (\ref{Eq:FeynKL}) can be treated as follows. (i). The term $(X_{A_j}-(-X_{A_j}-k_{A_j}))\cdot(K[A_j]+L[A_j])$  on the first line and all terms on the second line can be collected into a sum of Feynman diagrams, where some subcurrents are replaced by their total momenta (This discussion follows from section 5.1 in \cite{Wu:2021exa}). This part, together with the corresponding part in YMS currents (see \figref{Fig:Feynman0} (d)), has to vanish due to Ward identity. (ii). The term $-k_{A_j}\cdot L[A_j]$ in (\ref{Eq:FeynKL}) is proportional to the sum of tree-level Feynman diagrams where some of the subcurrents are replaced by their total momenta, hence also vanishes. (iii). Following discussions parallel with \cite{Wu:2021exa}, we find that the $K^{\mu}$ has the form
\bea
K^{\mu}={1\over k^2_{A_j}}k^{\mu}_{A_j}\Sl_{A_j\to A_{jL},A_{jR}}\W J[A_{jL}]\cdot\W J[A_{jR}],
\eea
when the reference order for expanding $\W J^{\mu}$ in (\ref{Eq:JA1}) is chosen as the same with the right permutation in the BS currents.
Consequently, the term $-k_{A_j}\cdot K[A_j]$ can further be rewritten as
\bea
-k_{A_j}\cdot K[A_j]=-\Sl_{A_j\to A_{jL},A_{jR}}\W J[A_{jL}]\cdot\W J[A_{jR}].
\eea

To sum up, the total contribution of \figref{Fig:Feynman} (b) and (c) is
\bea
2\W J\left[A_j\right]\cdot X_{A_j}-2\Sl_{A_j\to A_{jL},A_{jR}}\W J[A_{jL}]\cdot\W J[A_{jR}].\Label{Eq:FeynmanThreeFourVertex}
\eea
The prefactors of the two terms above are  $2$ and $-2$, respectively. On another hand, we choose the reference order in  (\ref{Eq:LoopCHYRefinedGraphsNew}) as the permutation in the right half integrand, then multiply each external momentum and the loop momentum in the coefficients in (\ref{Eq:LoopCHYRefinedGraphsNew}) by $2$ (thus all the momenta in numerators of $\W J^{\mu}$ are rescaled by a factor $2$). The prefactors of the two terms in (\ref{Eq:GenResultNewJ2}) respectively become $2$ and $-2$. Thus,  the two terms in (\ref{Eq:GenResultNewJ2}) precisely agree with the two terms in (\ref{Eq:FeynmanThreeFourVertex}), which are derived from Feynman diagrams.

Although the decomposition formula of the YMS BG current has not been explicitly given, it is natural to expect that the current mixing scalars and gluons can also split into the part decomposing according to the graphic rule, and another part which has to vanish in the on-shell limit. When the current is attached to the loop via three-scalar vertex as shown by \figref{Fig:Feynman0} (d), the former part matches with (\ref{Eq:GenResultNewJ1}), while the latter part, in general can be expanded into terms, in which some YM subcurrents are replaced by objects proportional to their total momenta. The sum of this part and the (i) part of (\ref{Eq:FeynKL}) has to vanish due to Ward identity \footnote{The detail of this statement will be presented in a coming work.}. Hence, the Feynman diagrams match with the expression (\ref{Eq:GenResultNew}) obtained from one-loop CHY.

\subsection{The extensions to multi-trace YMS and YM}
Although all the previous discussions are carried around single-trace YMS amplitudes, it is straightforward to generalize them to multi-trace YMS amplitudes with a pure scalar loop. This is because the multi-trace YMS amplitudes that are derived from tree-level CHY formula can be obtained via introducing another type of components: components with a gluon trace as stated in \appref{Multitrace}. Such a component has an equal status with the components containing a type-1 line. Therefore, one just includes these new components in the previous discussions, and finally also arrives at the conclusion (\ref{Eq:GenResultNew}). But in this case, the graphs $\mathcal{F}_{jL}$, $\mathcal{F}_{jR}$ referred in (\ref{Eq:GenResultNewJ2}) can be those with both gluons and scalar traces \footnote{ On the Feynman diagram side, this feature naturally implies the four-scalar vertex for multi-trace YMS amplitudes.}.  Note that the multi-trace case discussed above is only the case that all loop propagators are scalar ones and the two particles $\pm$ with the forward limit are adjacent to each other in the $(n+2)$-point tree-level half integrand. As classified in \cite{Porkert:2022efy}, according to distinct orders of coupling constants, there still exist (i). half integrands which contribute only scalar loop propagators but the particles $\pm$  are not adjacent to each other,  and (ii). the half integrands which contribute gluon propagators on the loop. The former case is related to the results in the current paper via a KK relation at one-loop level (see \appref{App:KKOneLoop}), while we leave a detailed study of the latter case in future work.

For YM amplitudes, we have to expand the reduced Pfaffian  (see \tabref{table:LoopIntegrands}) with $n+2$ external particles in terms of tree-level PT factors and then take the forward limit. This can be achieved via \eqref{Eq:ExpandReducedPfaffian2} where each term is accompanied by a Pfaffian for YMS. More specifically, apart from the prefactor $\text{Tr}[F_{i_1}\cdot F_{i_2}\cdot...\cdot F_{i_l}]$, each term in \eqref{Eq:ExpandReducedPfaffian2} is a YMS half integrand with the scalar trace $+,i_1,i_2,...,i_l,-$. Hence, together with the other half integrand, which is expressed as an $(n+2)$-point PT factor where the $\pm$ are considered as the two ends and the cyclic permutations over other $n$ elements are summed over, such a term will produce 
\bea
(-1)^{l}\,\text{Tr}[F_{i_1}\cdot F_{i_2}\cdot...\cdot F_{i_l}]\,\Sl_{\{A_1,A_2,...,A_I\}}\,gon\left(A_1,A_2,...,A_I\right)\prod\limits_{j=1}^{I}J\left[A_j\right],
\eea
where the cyclic symmetry of $\text{Tr}[F_{i_1}\cdot F_{i_2}\cdot...\cdot F_{i_l}]$ has been applied. We summed over all possible cyclic partitions $\{A_1,A_2,...,A_I\}$ of external particles such that the relative cyclic ordering of $i_1, i_2, ..., i_l$ is preserved. The $J\left[A_j\right]$ has been defined in (\ref{Eq:GenResultNewJ1}) and (\ref{Eq:GenResultNewJ2}), where elements of $\{i_1,...,i_l\}$ are considered as scalars. If the set $\{i_1,...,i_l\}$ is empty, we do not have the $\text{Tr}[F_{i_1}\cdot F_{i_2}\cdot...\cdot F_{i_l}]$ and the extra factor $(D-2)$ in \eqref{Eq:ExpandReducedPfaffian2} should be dressed. For this special case, the $J\left[A_j\right]$'s are those defined in (\ref{Eq:GenResultNewJ2}) which contains only external gluons. When all terms in \eqref{Eq:ExpandReducedPfaffian2} are collected, one expresses the one-loop YM amplitudes in terms of expressions with quadratic propagators
\bea
 I_{\,\text{YM}}^{\,\text{1-loop}}&=&(D-2)\Sl_{\{A_1,A_2,...,A_I\}} gon\left(A_1,A_2,...,A_I\right)\prod\limits_{j=1}^{I}J\left[A_j\right]\Label{Eq:YMQuadraticForm1}\\
&+&\Sl_{l=2}^n(-1)^l\Sl_{\{i_1,i_2,...,i_l\}\in \text{S}_{l}\setminus\text{Z}_{l}}\text{Tr}[F_{i_1}\cdot F_{i_2}\cdot...\cdot F_{i_l}]\Sl_{\{A_1,A_2,...,A_I\}} gon\left(A_1,A_2,...,A_I\right)\prod\limits_{j=1}^{I}J\left[A_j\right],\nonumber
\eea
where each $A_j$  can at least contain one element. Hence the $I$ in each term runs from $1$ to $n$ (the total number of external particles). Terms with all $i_1,...,i_l$ belonging to a same subset have to vanish due to the $U(1)$-decoupling identity of YMS current.

\section{Conclusions and further discussions}\label{sec:Conclusions}

In this paper, we provided a general approach to converting the linear propagators which exist in the one-loop Cachazo-He-Yuan (CHY) formula \cite{Cachazo:2013gna,Cachazo:2013hca,Cachazo:2013iea,Cachazo:2014nsa,Cachazo:2014xea} into quadratic propagators which exist in traditional Feynman diagrams. We expanded $n$-point one-loop CHY half integrands for Yang-Mills-scalar (YMS) in terms of $(n+2)$-point Parke-Taylor (PT) factors with forward limits,  according to the tree-level refined graphic rule. Then decomposed a graph into subgraphs and  established correspondence between subgraphs and subcurrents in the linear-propagator Feynman diagrams (LPFD). Nonlocalities occur when two subgraphs are contracted with each other but are separated by linear propagators in the LPFD. We treated the nonlocalities by combining two approaches. Once the nonlocalities are canceled, the expressions can always be arranged into a cyclic sum which implies a quadratic-propagator expression.

The formula with quadratic propagators has further been shown to match the traditional Feynman diagrams in YMS. The quadratic form reduced from CHY formula of multi-trace YMS amplitudes and pure Yang-Mills (YM) amplitudes were obtained via the single-trace YMS.  Although, the starting point of this work is the one-loop CHY formula, the full discussion essentially relies only on the forward limit. Therefore, the discussion in this paper may be directly extended to other approaches which are based on forward limit. Moreover, since the quadratic form of the pure YM amplitudes has been obtained, it is straightforward to generalize the results to super-Yang-Mills theories and the gluon amplitudes with fermions circulating in the loop.

There are still some related problems that deserve further study. The first one is to implement the conversion from linear to quadratic propagators  for YMS with gluon propagators on the loop, Einstein-Yang-Mills (EYM) and gravity (GR) integrands, which have more complicated CHY half integrands. Being different from the situation studied in this paper, they do not have a  half integrand that can be directly expressed as a scalar PT factor with a cyclic symmetry \cite{Feng:2022wee}, which means that the contribution of the tadpole diagrams needs to be revisited and cannot be straightforwardly canceled out by U(1)-decoupling. The second is how to expand the CHY half integrands into combinations of tensorial PT factors in a generic way. The third is, we can analyze the algebraic structures of results in this paper and further study the construction of BCJ numerators \cite{Bern:2008qj,Bern:2010ue} at one-loop (see e.g. \cite{Bern:2013yya,Chiodaroli:2013upa,Johansson:2014zca,Berg:2016fui,Edison:2022jln}). The relations between BCJ numerators in various theories at tree-level \cite{Dong:2021qai,Cao:2022vou} may further help one to extend the discussions to one-loop amplitudes in different theories. Last but not least, although we have already analyzed the relationship between the result in the current paper and Feynman diagrams for YMS, it is still worth working out the full connection between the YM results and the YM Feynman diagrams.

\section*{Acknowledgements}
We would like to thank Bo Feng, Song He, Yong Zhang and Kang Zhou for helpful suggestions and comments on the draft. We also thank Chih-Hao Fu and Yihong Wang for helpful discussions on kinematic algebra. We thank Yixiao Tao, Xinhan Tong and Konglong Wu for helpful discussions on the decomposition of Berends-Giele currents in YMS. We acknowledge the refrees of this paper for helpful suggestions. The authors acknowledge the organizers of ``3rd seminar on quantum field theory and its applications'' (2023, Bejing, China). YD would like to thank the organizers of ``4th seminar on field theory and string theory in China'' (2023, Nanjing, China) as well as Fudan University, Tianjin University for kind invitation. CX would like to thank the organizers of ``2023 Frontier Summer School on String Theory, Field Theory, and Holography'' (2023, Nanjing, China). This work is supported by NSFC under Grant No. 11875206.

\appendix

\section{Expansion formula of  $\text{PT}(+,x_1,...,x_r,-)\,\text{Pf}\,[\,\Psi_{\mathsf{G}}\,]$}\label{sec:ExpansionNew}

The expansion formula (\ref{Eq:Expansion1}) can be equivalently expressed by the strength tensors $F_i^{\mu\nu}$ as follows:
\begin{itemize}
\item Define reference order for gluons by an ordered set $\mathsf{R}=\{\gamma_1,...,\gamma_s\}$ and define the root set by $\mathcal{R}=\{+,x_1,x_2,...,x_{r}\}$.

\item Pick out the highest-weight element $a=\gamma_s$ and possible other elements $i_1,...,i_l$ from the ordered set $\mathsf{R}$. Construct a chain towards an element $b$ in the root set $\mathcal{R}$
\bea
\epsilon_a\cdot F_{i_1}\cdot F_{i_2}\cdot...\cdot F_{i_l}\cdot k_b.
\eea
Redefine the ordered set and the root set by moving $a, i_1, ..., i_l$ from $\mathsf{R}$ to $\mathcal{R}$.

\item Repeating the previous step by the new defined $\mathsf{R}$ and $\mathcal{R}$ until the ordered set $\mathsf{R}$ has been cleared up, we finally get a fully connected tree graph $\mathcal{F}$. This graph is accompanied by a coefficient $\mathcal{C}^{\mathcal{F}}$ which is a product of Lorentz contractions between $\epsilon^{\mu}$, $F^{\rho\sigma}$ and $k^{\nu}$. Permutations established by $\mathcal{F}$ are determined in the same way with those in \secref{sec:RefinedGraph}.

\item When all graphs are summed over, we get an expansion formula of  $\text{PT}(+,x_1,...,x_r,-)\,\text{Pf}\,[\,\Psi_{\mathsf{G}}\,]$.
\end{itemize}
When all strength tensors $F_i^{\mu\nu}$ are expanded into $k_i^{\mu} \epsilon_i^{\nu}-\epsilon_i^{\mu} k_i^{\nu}$, the result turns into that was generated in \secref{sec:RefinedGraph}. This has been proven in \cite{Hou:2018bwm}.

\section{Expansion of $\text{PT}(+,\pmb{\sigma}_1,-)\text{PT}(\pmb{\sigma}_2)\cdots\text{PT}(\pmb{\sigma}_m) \,\mathcal{P}$}\label{Multitrace}
\begin{figure}
\centering
\includegraphics[width=0.4\textwidth]{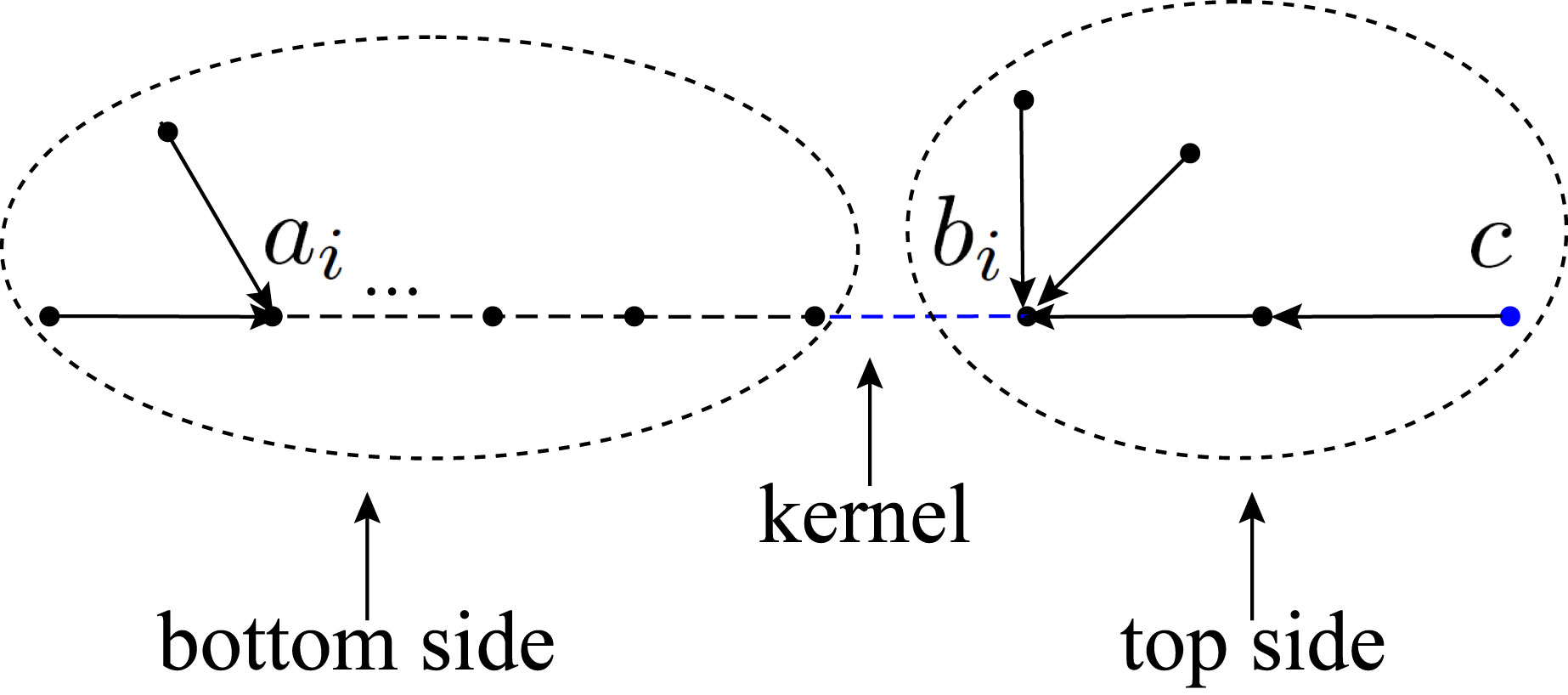}
\caption{In the multitrace case, there exists a new type of component.  In such a component, a trace is characterized by a simple chain where nodes are connected via type-4 lines, possible type-2 lines are connected to the trace. If  the highest-weight node in this component is a gluon $c$, the contribution of the graph can always be expanded in terms of those graphs where $c$ is connected to an end node $b_i$ of the trace $\mathsf{KK}[\,i,a_i,b_i\,]$ via type-2 lines. The path from this gluon to root must pass through the full trace $\mathsf{KK}[\,i,a_i,b_i\,]$ where $a_i$ is the nearest-to-root scalar. If the weight of the trace $\pmb{\text{tr}}_i$ is higher than all gluons in that component, the node $b_i$ carries the weight of the trace. In both cases, the kernel is defined by the type-4 line attached to $b_i$. The top side is  the side containing $b_i$. }
\label{Fig:TraceComponent}
\end{figure}

We provide the graphic rule for the half integrand  $\text{PT}(+,\pmb{\sigma}_1,-)\text{PT}(\pmb{\sigma}_2)\cdots\text{PT}(\pmb{\sigma}_{m}) \,\mathcal{P}$ which exists in a multi-trace amplitude with $m$  traces $\pmb{\text{tr}}_1=\{+,\pmb{\sigma}_1,-\},\pmb{\text{tr}}_2=\{\pmb{\sigma}_2\},...,\pmb{\text{tr}}_m=\{\pmb{\sigma}_m\}$. We define a KK basis for a trace $\pmb{\text{tr}}_i$ by
\bea
\mathsf{KK}[\,i,a_i,b_i\,]\equiv (-1)^{|\pmb{\beta}|}\left\{a_i, \pmb{\alpha}\shuffle\pmb{\beta}^{\text{T}}, b_i\right\},
\eea
where the trace $\pmb{\text{tr}}_i$ can be expressed as  $\pmb{\text{tr}}_i=\{a_i,\pmb{\alpha},b_i,\pmb{\beta}\}$. As proven in \cite{Du:2019vzf}, the refined graphic rule for $\text{PT}(+,\pmb{\sigma}_1,-)\text{PT}(\pmb{\sigma}_2)\cdots\text{PT}(\pmb{\sigma}_{m}) \,\mathcal{P}$ can be obtained by adjusting the rules in \secref{sec:RefinedGraph1} as follows:
\begin{itemize}
\item Enlarge the reference order set $\mathsf{R}=\{\gamma_1,...,\gamma_s\}$ in step-1 by including the traces $\{\pmb{\text{tr}}_2,...,\pmb{\text{tr}}_m\}$, the reference order then becomes $\mathsf{R}=\{\gamma_1,...,\gamma_{s+m-1}\}$.

\item Consider the $C_0$ (defined in \secref{sec:RefinedGraph}) as the component involving $\pmb{\text{tr}}_1$. Enlarge the kernel set by including traces $\{\pmb{\text{tr}}_2,...,\pmb{\text{tr}}_m\}$. For each trace $\pmb{\text{tr}}_i$ ($i=2,...,m$), we choose a pair $a_i,b_i\in \pmb{\text{tr}}_i$ and then arrange the elements of $\pmb{\text{tr}}_i$ as $\mathsf{KK}[\,i,a_i,b_i\,]$. For a given permutation in $\mathsf{KK}[\,i,a_i,b_i\,]$, two adjacent nodes are connected by a type-4 line. As demonstrated by \figref{Fig:TraceComponent}, components containing the trace $\pmb{\text{tr}}_i$ are constructed as follows: (i). If the trace has the highest weight in the component, we plant gluons to scalars of the trace via type-2 lines. (ii). If the highest-weight element is a gluon say $c$, we further require that $c$ is connected to $b_i$ via type-2 lines. {\it The kernel of such a component is defined as the type-4 line attached to $b_i$. The top side of this component is defined by the side containing $b_i$}.

\end{itemize}
With the above adjustment, one finally gets all possible graphs $\mathcal{F}$ which include graphs with $k$ pairs of type-2 lines and $m$ traces. The summation over $\mathcal{F}$ in \eqref{Eq:Expansion1} means summing over all possible values of $k$ ($k=1,...,\lfloor{s\over 2}\rfloor$), all possible graphs corresponding to a given $k$. Given trace $\pmb{\text{tr}}_i$, we sum over all possible $a_i\in \pmb{\text{tr}}_i (a_i\neq b_i)$ for a fixed $b_i\in \pmb{\text{tr}}_i$ when the following two conditions are satisfied simultaneously: (i). the trace $\pmb{\text{tr}}_i$ is involved in a starting component of a chain and (ii). the trace $\pmb{\text{tr}}_i$ has the highest weight in this component. For other cases, all ordered pairs of $\{a_i,b_i\}$ in $\pmb{\text{tr}}_i$  are summed over. The kinematic coefficient $\mathcal{C}^{\mathcal{F}}$ is also given by the product of the kinematic factors corresponding to the lines, reminding that a type-4 line does not bring any kinematic factor. In addition to the signs counted in the single-trace case, each trace is also accompanied by a sign $(-1)^{|\pmb{\beta}_i|}$. 



\newpage

\section{All graphs with three gluons and all graphs for the effective current $\W J^{\mu}(p,q,r)$}\label{app:GraphsThreeGluons}
\begin{figure}
\centering
\includegraphics[width=0.8\textwidth]{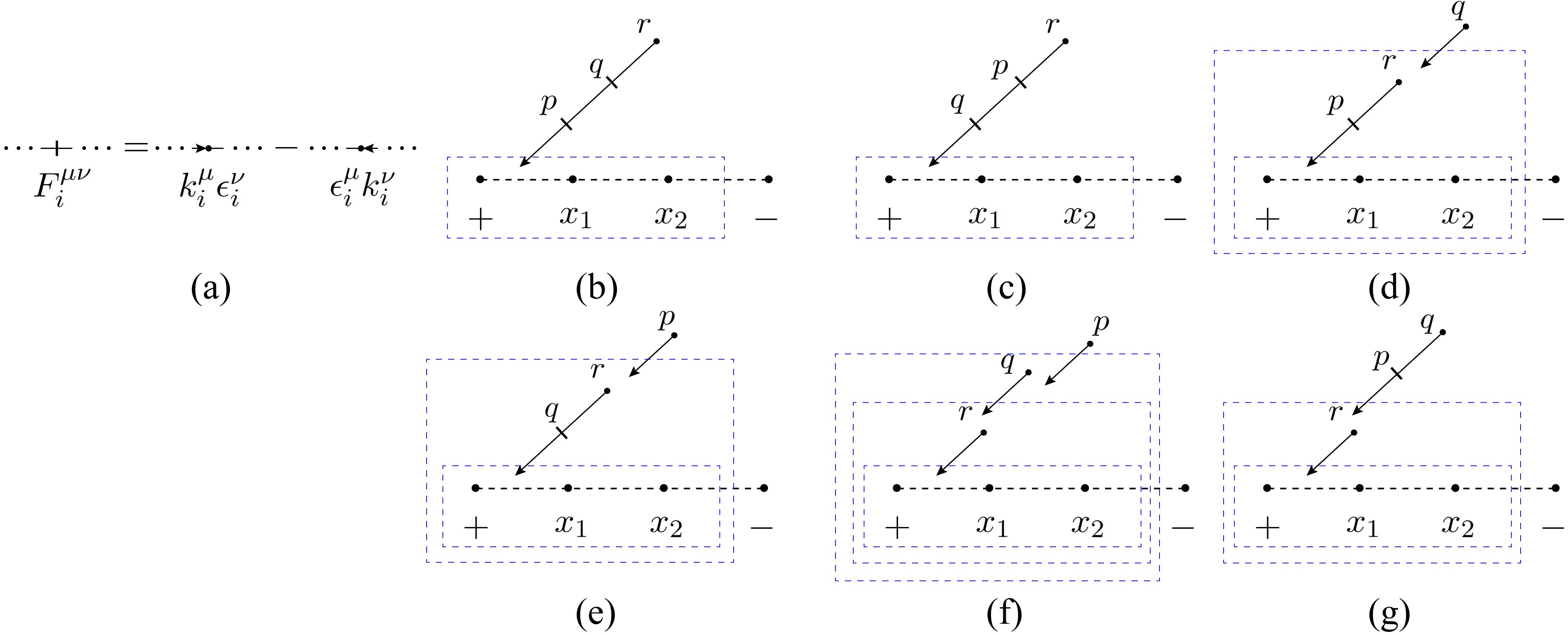}
\caption{When we introduce the compact expression (a) for the strength tensor, all graphs for three-gluon two-scalar integrands are presented by graphs (b)-(g). The reference order is fixed as $p\prec q\prec r$.}
\label{Fig:3GluonAllGraphs0}
\end{figure}
\begin{figure}
\centering
\includegraphics[width=0.55\textwidth]{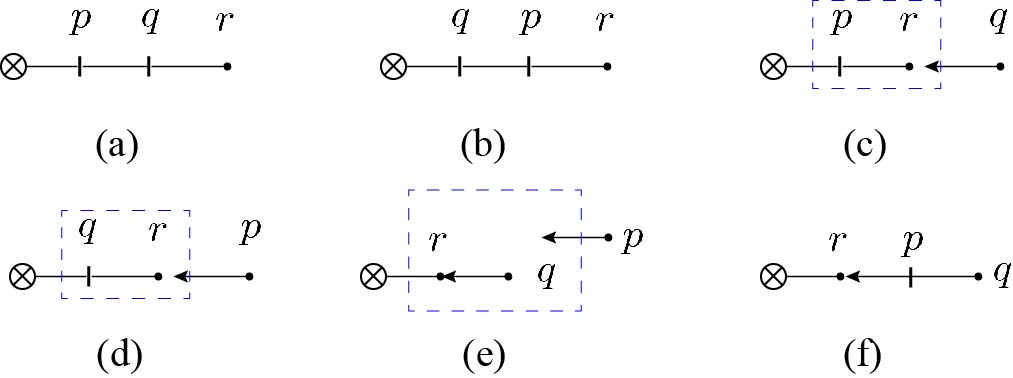}
\caption{All graphs for the effective current $\W J^{\mu}(p,q,r)$ with the reference order $p\prec q\prec r$. }
\label{Fig:3GluonAllGraphs}
\end{figure}

We present all graphs with three gluons and two scalars in  \figref{Fig:3GluonAllGraphs0} and all graphs for the effective current $\W J^{\mu}(p,q,r)$ in \figref{Fig:3GluonAllGraphs}. The reference order is supposed to be $p\prec q\prec r$. The compact form, which is expressed by strength tensors $F_i^{\mu\nu}=k_i^{\mu}\epsilon_i^{\nu}-\epsilon_i^{\mu}k_i^{\nu}$ in \figref{Fig:3GluonAllGraphs0} (a), is directly generated by the rule in \appref{sec:ExpansionNew}. The graphs in \figref{Fig:3GluonAllGraphs} can be obtained from \figref{Fig:3GluonAllGraphs0} (b)-(g), via replacing the root set by a cross node which stands for other structure contracted with $\W J^{\mu}(p,q,r)$, and deleting the disconnected graphs which have to vanish due to generalized $U(1)$-decoupling identity. The explicit expression of $\W J^{\mu}(p,q,r)$ is 
\bea
\W J^{\mu}(p,q,r)&=&(\epsilon_r\cdot F_q\cdot F_p)^{\mu}\phi_{pqr|a_1a_2a_3}+(\epsilon_r\cdot F_p\cdot F_q)^{\mu}\phi_{qpr|a_1a_2a_3}\Label{Eq:EffCurrentpqr}\\
&&+\Big[(\epsilon_q\cdot k_r)\phi_{prq|a_1a_2a_3}+(\epsilon_q\cdot k_p)\phi_{p\{r\shuffle q\}|a_1a_2a_3}\Big](\epsilon_r\cdot F_p)^{\mu}\nn
&&+\Big[(\epsilon_p\cdot k_r)\phi_{qrp|a_1a_2a_3}+(\epsilon_p\cdot k_q)\phi_{q\{r\shuffle p\}|a_1a_2a_3}\Big](\epsilon_r\cdot F_q)^{\mu}\nn
&&+\Big[(\epsilon_q\cdot k_r)(\epsilon_p\cdot k_q)\phi_{rqp|a_1a_2a_3}+(\epsilon_q\cdot k_r)(\epsilon_p\cdot k_r)\phi_{r\{q\shuffle p\}|a_1a_2a_3}\Big]\epsilon^{\mu}_r+(\epsilon_q\cdot F_p\cdot k_r)\epsilon^{\mu}_r\phi_{rpq|a_1a_2a_3}.\nonumber
\eea

\newpage

\begin{figure}[h]
\centering
\includegraphics[width=0.93\textwidth]{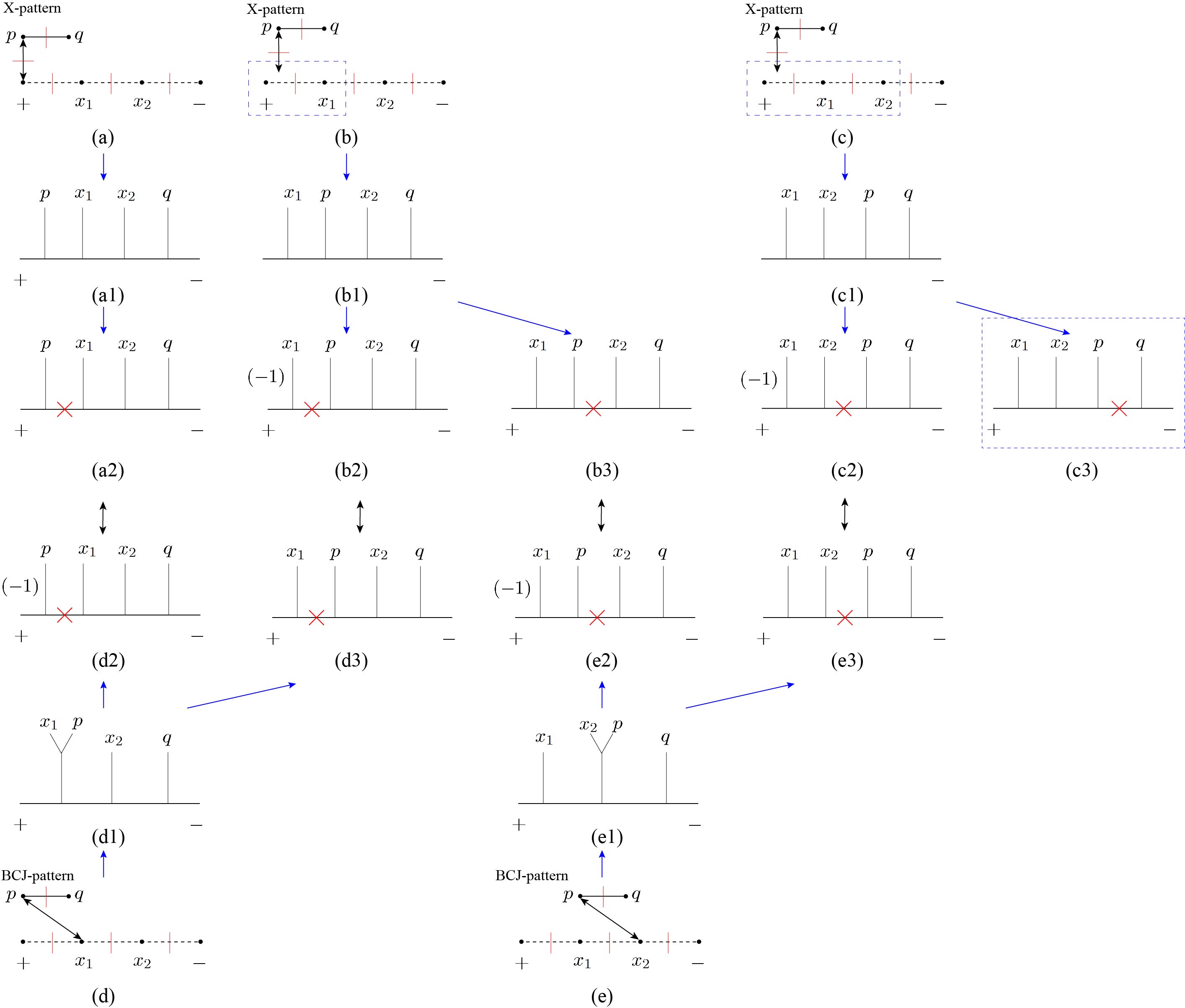}
\caption{The cancellation map for the example in \secref{sec:TwoGluonEG2} }
\label{Fig:4pt2glu11}
\end{figure}
\section{The cancellation map for the example in \secref{sec:TwoGluonEG2} }\Label{app:CancellationMap}
\figref{Fig:4pt2glu11} presents the cancellation map between X- and BCJ-patterns for the example in \secref{sec:TwoGluonEG2}. The diagram (c3) in 
\figref{Fig:4pt2glu11} is the surviving term, corresponding to the second term of the second line in (\ref{Eq:0Ai2}).

%
%

\begin{figure}
\centering
\includegraphics[width=0.75\textwidth]{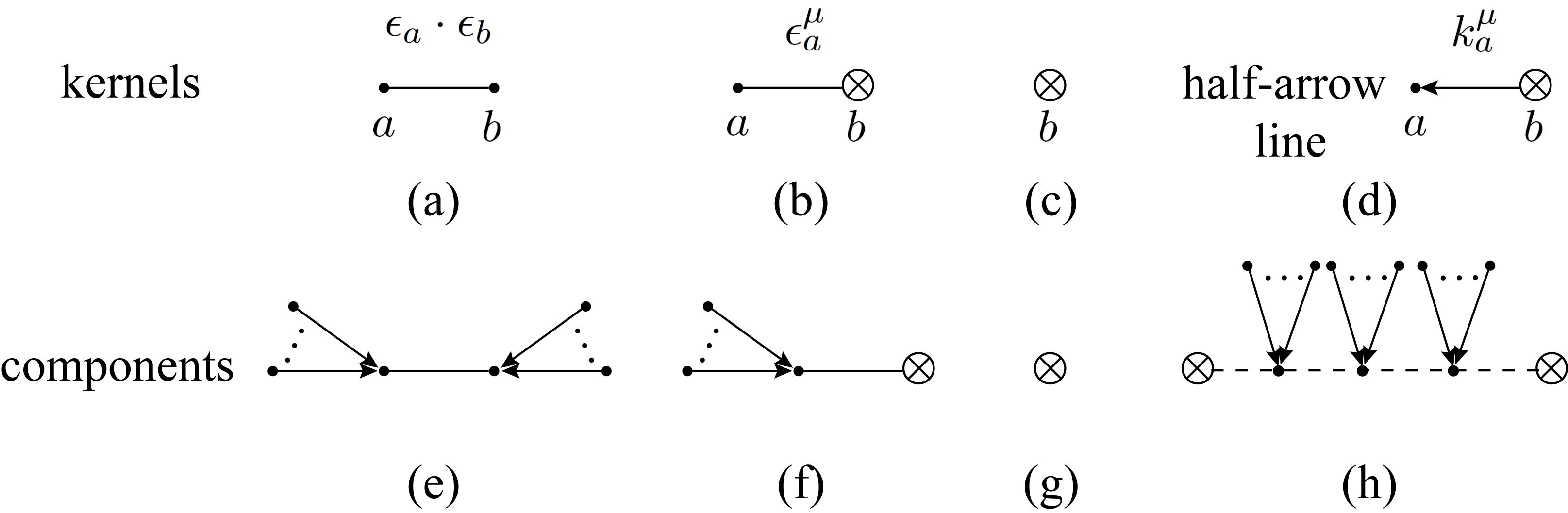}
\caption{Graphs (a), (b) and (c) are possible kernels in a subgraph. Graphs (e), (f), (g) and (h) are possible components in a subgraph. A line between an on-shell node and a cross node can either be a half line (with no arrow) or a half-arrow line, as shown by (b) and (d), respectively.}
\label{Fig:KernelSubgraph}
\end{figure}

\section{Two versions of graphic rules for a subgraph}\label{sec:Subgraph1}

When a partition $\{A_1, A_2, ..., A_j, ...\}$ and a topology for this partiton are fixed, we can construct subgraphs for a subset $A_j$ via graphic rule, which mostly inherits the rule for a full graph, see \secref{sec:RefinedGraph}. Some adjustments have to be introduced to adapt the fact that the subgraph contains off-shell nodes:
\newline
\newline
$~~$~~~~~~~~~~~~~~~~~~~~~~~~~~~~~~~~~~~~~~~~~~~~{\bf Version-1 rule for a subgraph}

\begin{itemize}
\item For a given subset $A_j$, we not only have the on-shell nodes in  $\mathcal{F}_j$, but also have off-shell nodes, i.e., the subtree structures attached to $\mathcal{F}_j$, as shown by \figref{Fig:GeneralPatternPatition} (c). Specifically, the off-shell nodes locating above $\mathcal{F}_j$ (e.g., $b_1$ and $b_2$ in \figref{Fig:GeneralPatternPatition} (c)) stand for the corresponding {\it full subtree structures attached to $\mathcal{F}_j$} whose distance from root are larger than that of  $\mathcal{F}_j$. The off-shell node below $\mathcal{F}_j$ only represents {\it the subgraph below $\mathcal{F}_j$}. The weights of off-shell nodes above $\mathcal{F}_j$ are defined by the highest-weight nodes in the corresponding subtree structure. The reference order between these nodes inherits the reference order for the full integrand.

\item   {\it If the subset  $A_j$ contains only gluons}, we construct $k$ kernels which can be obtained by (i). connecting a type-1 line between two on-shell gluons, as shown by \figref{Fig:KernelSubgraph} (a), (ii). connecting a half line (with no arrow) between an on-shell gluon and an off-shell node, as shown by \figref{Fig:KernelSubgraph} (b), or (iii). a single cross node, see \figref{Fig:KernelSubgraph} (c). All off-shell nodes are contained by kernels.  Connect other on-shell gluons towards the on-shell nodes in kernels (i) and (ii) via type-2 ($\epsilon\cdot k$) lines, together with possible single cross nodes, we get $k$ components, as shown by \figref{Fig:KernelSubgraph} (e), (f) and (g). In the component with both off-shell node above the subset and on-shell nodes, we require that the weight of the off-shell node is higher than those of on-shell nodes in this component.  {\it If the subset  $A_j$ contains scalars}, apart from the possible components which have been introduced previously, we also have a component involving scalars, as shown by  \figref{Fig:KernelSubgraph} (h). These scalars must live on the path from a cross node to another (both cross nodes also involve scalars) and are interconnected via type-4 lines, in accordance with the relative order of the scalar set. The gluons are connected to the scalars via type-2 lines.


\item Consider the component containing the off-shell node below $\mathcal{F}_j$ (e.g., the off-shell node representing $\mathcal{F}_1$ in \figref{Fig:GeneralPatternPatition} (c)) as root. Construct COC's following step-3 in \secref{sec:RefinedGraph} and require:
\begin{itemize}
\item (a). The lines between on-shell nodes of two components are type-3 ($k\cdot k$) lines. The line between an off-shell node and an on-shell node of two components should be a half-arrow line (shown by \figref{Fig:KernelSubgraph} (d)) pointing towards the on-shell node.


\item (b). An off-shell node above the subgraph can only live in the starting component (the  $C_{a}$ in (\ref{Eq:ChangeOfComponents})) of a COC.
\end{itemize}

\end{itemize}
An example for subgraphs constructed by version-1 rule is given by \figref{Fig:Version1EG} in \appref{app:ExampleSubgraphs}.

In order to cancel the nonlocalities following approach-1 and -2, we have to further introduce spurious graphs so that X- and BCJ-patterns are explicitly constructed. To achieve this, we provide the following equivalent construction rule.

\begin{figure}
\centering
\includegraphics[width=0.45\textwidth]{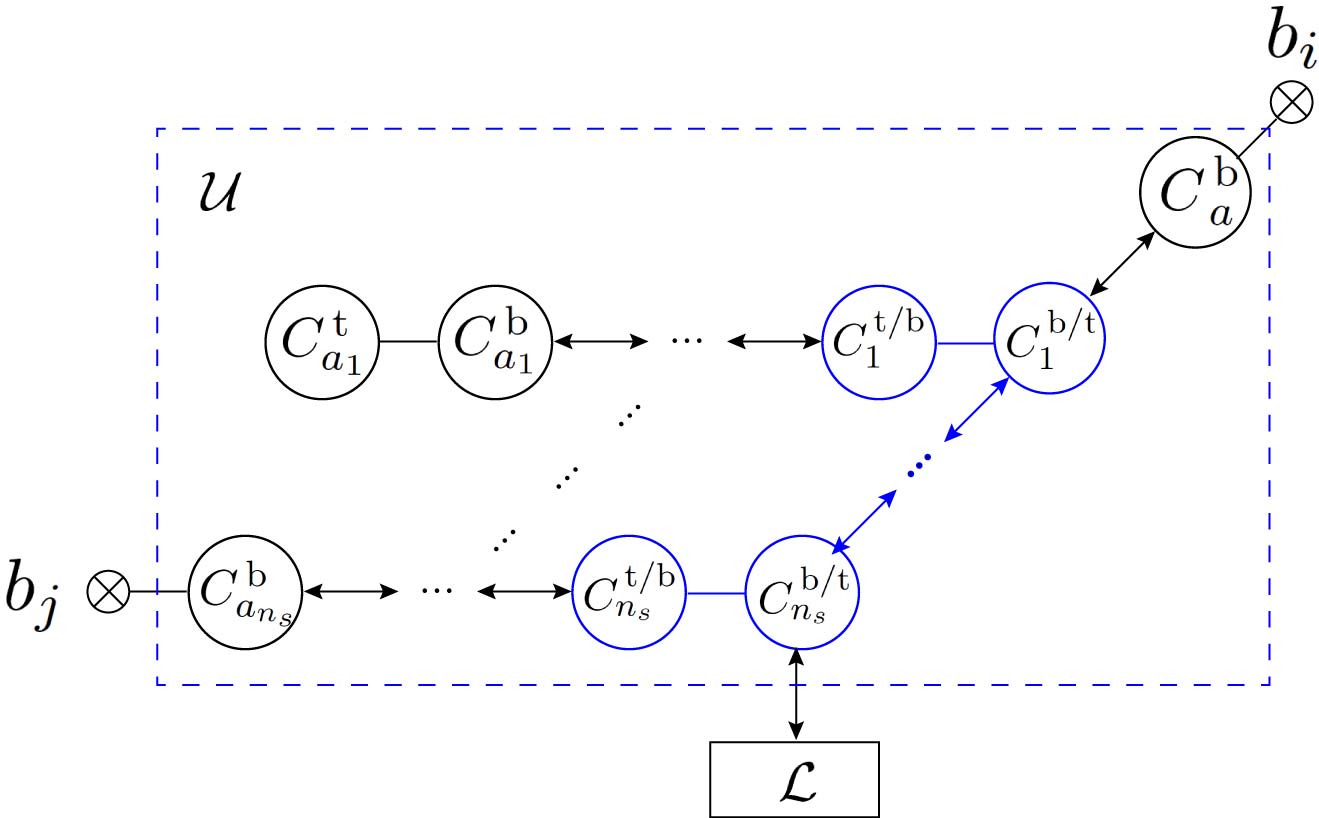}
\caption{In this spurious graph, the cross node $b_i$ is supposed to be the highest-weight node. In the upper block $\mathcal{U}$ attached to $b_i$, there are $n_s$ components (mentioned as {\it spurious components}) whose single sides are passed through by the path from $b_i$ to the lower component $
\mathcal{L}$. }
\label{Fig:Spurious}
\end{figure}

~~~~~~~~~~~~~~~~~~~~~~~~~~~~~~~~~~~~~~~{\bf Version-2 rule for a subgraph}

\begin{itemize}
\item {\bf Step-1}~Pick out a component led by an off-shell node $b_i$. Construct COC's, which are led by components with higher weights than that of $b_i$, towards root following version-1 rule.
\item {\bf Step-2}~Define the {\it lower block} $\mathcal{L}$ by the subgraph constructed in the above step, the {\it upper block} $\mathcal{U}$  by the component containing $b_i$ (the highest-weight cross node among the remaining on-shell and off-shell nodes). Construct a COC towards either $\mathcal{L}$ or  $\mathcal{U}$.

\item {\bf Step-3}~Redefine the upper and lower blocks $\mathcal{U}$ and  $\mathcal{L}$ by absorbing the COC which has been constructed in step-2. Repeat step-2 using the new defined  $\mathcal{U}$ and  $\mathcal{L}$ until all components have been used, we get the final configurations of $\mathcal{U}$ and  $\mathcal{L}$,  which are two subgraphs separated from each other.

\item {\bf Step-4}~  If both $\mathcal{U}$ and  $\mathcal{L}$ involve on-shell nodes, connect a type-3 line between them. If one of $\mathcal{U}$ and  $\mathcal{L}$ is a single cross node, connect a half-arrow line pointing to the other block.
\end{itemize}
Following these steps, we finally get fully connected tree graphs. {\it A part of them are those constructed according to version-1 rule, while the other part are pairs of spurious graphs with opposite signs \cite{Hou:2018bwm}.} The sign of a spurious graph is given by $(-1)^{n_s+n_a}$ where (i).  $n_s$ denotes the number of spurious components in the upper block, as shown by 
\figref{Fig:Spurious}, (ii). $n_a$  denotes the number of arrows pointing away from root.

\begin{figure}
\centering
\includegraphics[width=0.6\textwidth]{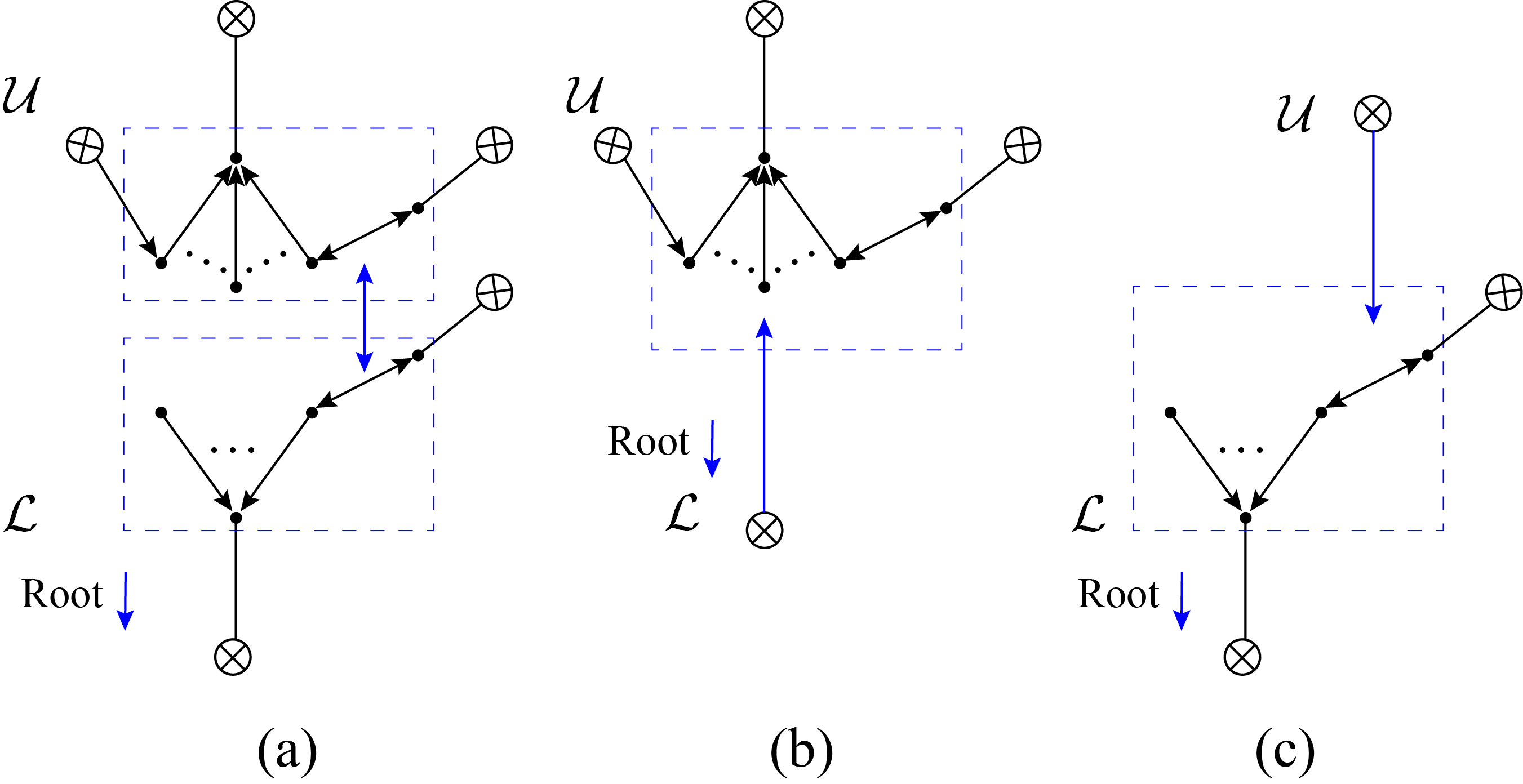}
\caption{Possible structures of subgraphs}
\label{Fig:SubgraphSituations0}
\end{figure}

A graph constructed by version-2 rule consists of the upper and lower blocks  $\mathcal{U}$ and $\mathcal{L}$, which are interconnected to each other. The structures of subgraphs can further be  classified according to whether the  $\mathcal{U}$ or $\mathcal{L}$ contains on-shell nodes:
\begin{itemize}
\item {\bf Both $\mathcal{U}$ and $\mathcal{L}$ contain on-shell nodes, as shown by \figref{Fig:SubgraphSituations0} (a).} The line between them is a type-3 line. When the kinematic coefficients of  $\mathcal{U}$ and $\mathcal{L}$ are extracted out, we get a BCJ-pattern with correct signs in each term \cite{Hou:2018bwm}. 
\item {\bf The lower block $\mathcal{L}$ contains no on-shell node, as shown by \figref{Fig:SubgraphSituations0} (b).} Denote such subgraph and the subgraph below by $\mathcal{F}_j$ and $\mathcal{F}_l$, respectively. If the upper block of $\mathcal{F}_l$ has no on-shell node (as shown by \figref{Fig:SubgraphSituations0} (c)),  $\mathcal{F}_j$ makes a contribution ($-k_{A_j}\cdot k_{A_l}$) to an X-pattern ($-k_{A_j}\cdot X_{A_j}$). Else, if the upper block of $\mathcal{F}_l$ has on-shell nodes (as shown by \figref{Fig:SubgraphSituations0} (a) and (b)), the $\mathcal{F}_j$ is neither a BCJ-pattern nor a part of the X-pattern with $-k_{A_j}\cdot X_{A_j}$.  Subgraph in this case should be treated differently according to different configurations of $\mathcal{F}_l$.

\item  {\bf The upper block $\mathcal{U}$ contains no on-shell node, as shown by \figref{Fig:SubgraphSituations0} (c).} In this case, the upper block has only one cross node which can further be extracted out as a $C\cdot k_{A_j}$ factor. When the cross node is extracted out, the subgraph becomes a physical graph constructed by version-1 rule.

\end{itemize}

\section{Constructing a subgraph in a given partition and topology}\label{sec:Subgraph2}

In this section, we construct subgraphs $\mathcal{F}_j$ for a given partition and a fixed topology. To achieve this, we can {\it combine the two versions of graphic rules in \appref{sec:Subgraph1} for subgraphs in various ways}. Different approaches are equivalent to each other since they must be reduced into the same set of physical graphs when pairs of spurious graphs are canceled out. Here, we provide two possible approaches to the construction of subgraphs for a given partition $\{A_1,A_2,...,A_j,...,A_i\}$ and a given topology.

\subsection{Approach-1}

We assume the cross nodes for the subgraph $A_j$ are $b_1$, ..., $b_s$ (the reference order is $b_1\prec ...\prec b_s$) whose highest-weight nodes are $c_1$, ..., $c_s$, respectively. The nodes $c_1$, ..., $c_s$ must belong to different subsets to the right side of $A_j$ in the partition. Supposing that the nodes  $c_1$, ..., $c_s$ in the partition from left to right are $c_{\sigma_1}$, ..., $c_{\sigma_s}$, we further generate all subgraphs properly in the following way:
\begin{itemize}
\item {\bf Step-1}~~ Generate all possible subgraphs according to version-1 rule.

\item {\bf Step-2}~~ Turn back to the step that generates the COC led by $b_{\sigma_s}$. Considering $b_{\sigma_s}$ as the upper block, the tree structure which has already been constructed as the lower block, we apply version-2 rule to generate possible graphs. As demonstrated before, such graphs may (a). contain X- or BCJ-pattern,  (b). provide a  $C\cdot k_{A_j}$ factor when $b_{\sigma_s}$ is extracted out, or (c). be neither the case (a) nor (b) and can be treated when the subgraph below is taken into account. Further treatment of cases (a) and (c) are given in \secref{sec:CancellationXBCJ}. For (b), we get a physical subgraph (a subgraph generated by version-1 rule) with $b_{\sigma_s}$ extracted out. 
\begin{figure}
\centering
\includegraphics[width=0.4\textwidth]{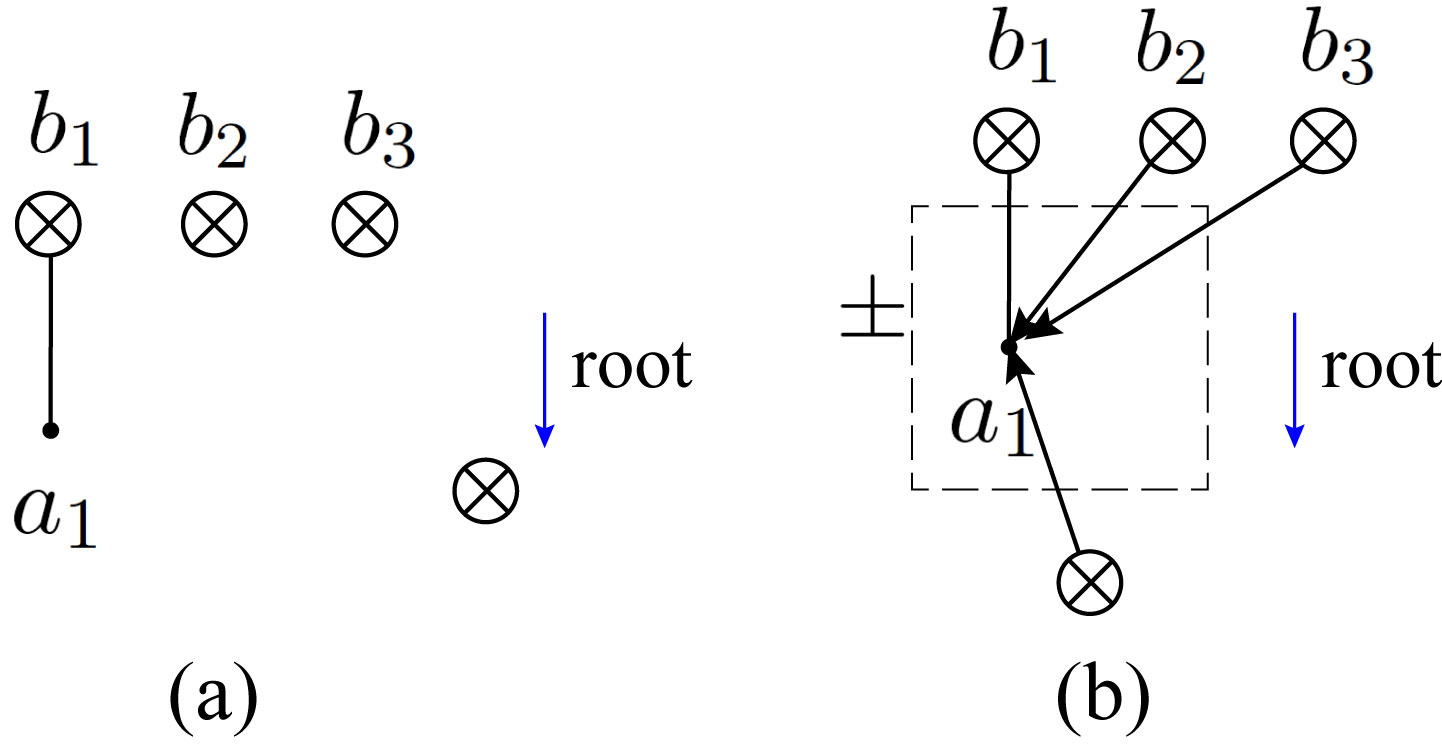}
\caption{Suppose the reference order is $a_1\prec b_1\prec b_2\prec b_3$. If the components of a subgraph are shown by (a), there does not exist physical graph constructed by step-1 in approach-1. However, we may introduce a pair of spurious graphs (b) with opposite signs so that the cancellations totally follow from \secref{sec:CancellationXBCJ}.  }
\label{Fig:Boundary}
\end{figure}
\begin{figure}
\centering
\includegraphics[width=0.4\textwidth]{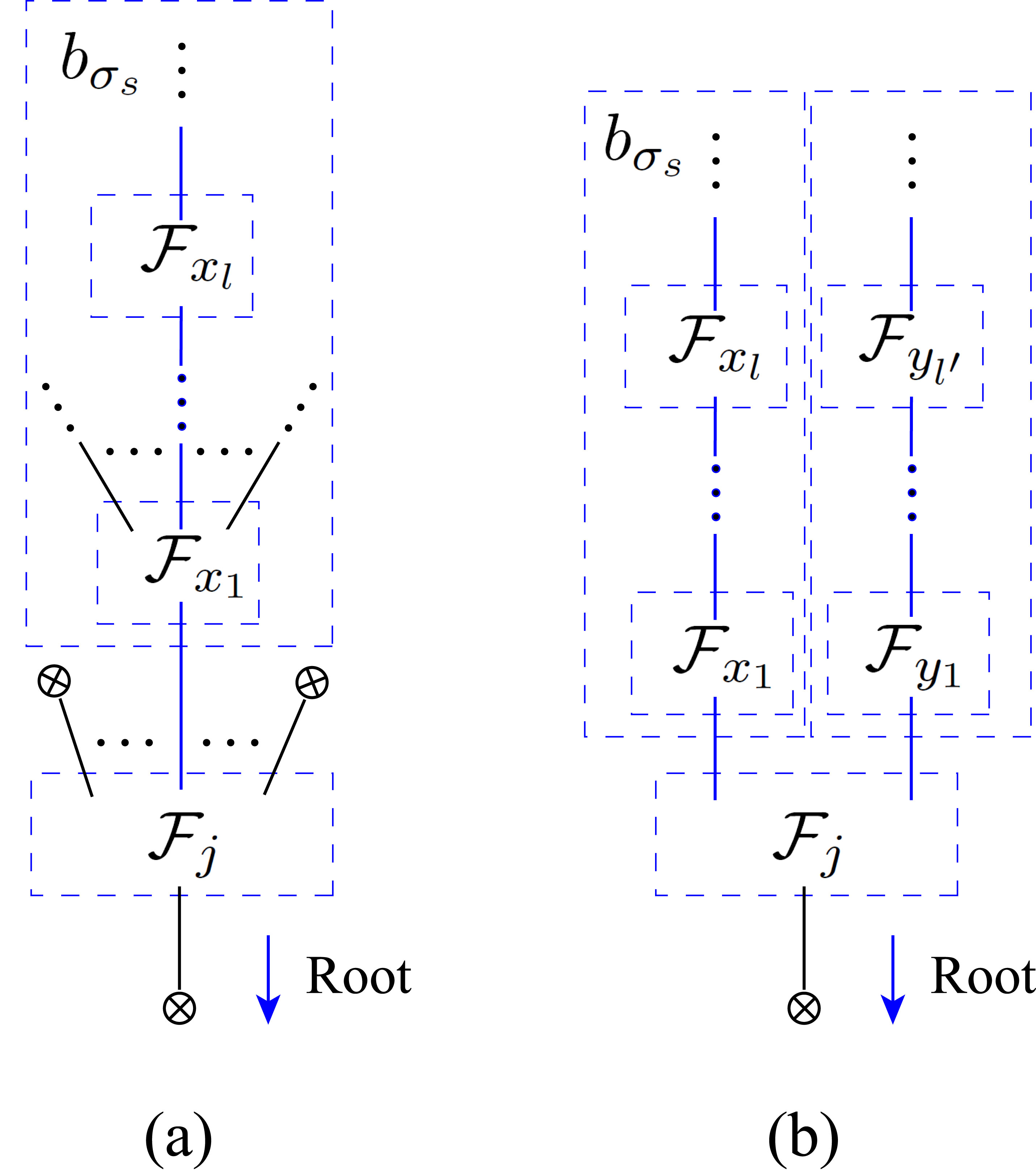}
\caption{Assuming that the $c_{\sigma_s}\in\mathcal{F}_{x_l}$, the subgraphs in (a) on the path from $c_{\sigma_s}$ to $\mathcal{F}_j$ are not separated by other subsets in the partition $\{A_1,...,A_j,...,A_{x_1},...,A_{x_l},...,A_i\}$ where $A_{x_1},...,A_{x_l}$ are the subsets corresponding to  $\mathcal{F}_{x_1},...,\mathcal{F}_{x_l}$.  In graph (b), the subgraphs on the path from  $c_{\sigma_s}$ to $\mathcal{F}_j$ are separated in the partition $\{...,A_j,...,A_{x_1},...,A_{x_{l-k}}, A_{y_1},...,A_{y_{l'}},A_{x_{l-k+1}},...,A_{x_l},...\}$.}
\label{Fig:SubgraphSituations}
\end{figure}

\item {\bf Step-3}~~For the physical graph in (b) with $b_{\sigma_s}$ extracted out, we repeat step-2 for cross nodes $b_{\sigma_{s-1}}$, ..., $b_{\sigma_2}$ in turn. 

\item {\bf Step-4}~~When all cross nodes $b_{\sigma_{s}}$, ..., $b_{\sigma_2}$  have been extracted out, we have a physical graph which has only one cross node above $\mathcal{F}_j$, i.e., $b_{\sigma_1}$. We finally consider $b_{\sigma_1}$ as the upper block, the component containing the cross node below $\mathcal{F}_j$ as the lower one, and apply version-2 rule. We again obtain subgraphs, each of which  (a). contains X- or BCJ-pattern,  (b). provides  a $C\cdot k_{A_j}$ factor with $b_{\sigma_1}$ extracted out, or (c). is neither the case (a) nor (b).
\end{itemize}

{\it Comment on the boundary}: If there does not exist physical subgraph constructed in the first step, as shown by \figref{Fig:Boundary} (a), the total contribution of this subset must be zero. Nevertheless, we may construct pairs of spurious graphs from step-2 (as shown by \figref{Fig:Boundary} (b)) so that all cancellations follow from the way pointed in \secref{sec:CancellationXBCJ}. 
\begin{itemize}

\item (i). If $b_{\sigma_s}$ is not the highest-weight node $b_s$ and the component with $b_{\sigma_s}$ contains on-shell nodes, we do not further construct subgraphs and this case has no contribution to the final result.

\item (ii). If $b_{\sigma_s}$ is not the highest-weight node and the component with $b_{\sigma_s}$ is a single cross node, the construction can  further be performed as:
\begin{itemize}
\item (a). If the subgraphs on the path from $c_{\sigma_s}$ to $\mathcal{F}_j$ (including the subgraph containing $c_{\sigma_s}$ but excluding $\mathcal{F}_j$) are not separated by other subsets in the partition (as demonstrated by \figref{Fig:SubgraphSituations} (a)), the cross node $b_{\sigma_s}$ can be extracted out and provides a $C\cdot k_{A_j}$ factor. We may have pairs of spurious graphs generated at the next step when $b_{\sigma_{s-1}}$ is considered as the upper block. 

\item (b).  If the subgraphs on the path from $c_{\sigma_s}$ to $\mathcal{F}_j$ are separated by other subsets in the partition (as shown by \figref{Fig:SubgraphSituations} (b)), we do not have any reasonable subgraph $\mathcal{F}_j$.
\end{itemize}

\item (iii). If $b_{\sigma_s}$ is the highest-weight node $b_s$, we consider it as the upper block and construct subgraphs (pairs of spurious graphs) from step-2. In the special case that no such subgraph can be produced, we follow the (a) and (b) in (ii).

\end{itemize}
In the cases with $b_{\sigma_s}$ extracted out, we return to step-3 and perform a further construction.

As an example of the boundary case, we assume that $A_{m_i}$ ($i=1,2,3$) is the subset containing the highest-weight node in \figref{Fig:Boundary} for each cross node $b_i$. For partition $\{...,A_j,...,A_{m_2},...,A_{m_3},...,A_{m_1},...\}$, we do not construct any subgraph according to (i). For partition $\{...,A_j,...,A_{m_1},...,A_{m_3},...,A_{m_2},...\}$, if the subgraphs on the path from $\mathcal{F}_{m_2}$ to $\mathcal{F}_j$ are not separated by other subsets, the cross node $b_2$ can be extracted out and produces a $C\cdot k_{A_j}$ factor, according to (a) in (ii). After this step, we must turn to $b_3$ and construct a pair of subgraphs, as shown by \figref{Fig:Boundary} (b).

\subsection{Approach-2}

Now we provide another possible approach to constructing subgraphs. Assuming the relative order of cross nodes above $\mathcal{F}_j$ is $b_1\prec...\prec b_k \prec ...\prec b_s$, there are two possible situations depending on different relative positions of the subgraphs contained by cross nodes of $\mathcal{F}_j$ in the partition:
\begin{itemize}
\item  {\bf Situation (i).} Assuming that the highest-weight node $c_k$ in the cross node $b_k$ is contained by subgraph $\mathcal{F}_{x_l}$, graphs on the path between $\mathcal{F}_j$ and $\mathcal{F}_{x_l}$ (including $\mathcal{F}_j$ and $\mathcal{F}_{x_l}$) are not separated by other subset in the partition
\bea
\{A_1,...,A_j,A_{x_1},...,A_{x_l},...,A_i\},\Label{Eq:Partition1}
\eea
where $A_{x_1},...,A_{x_l}$ are the subsets corresponding to $\mathcal{F}_{x_1}, ...,\mathcal{F}_{x_l}$,  as shown by \figref{Fig:SubgraphSituations} (a) (when $b_{\sigma_s}$ is replaced by $b_{k}$).
 
\item {\bf Situation (ii).} For any cross node $b_k$ above $\mathcal{F}_j$, the path from  $\mathcal{F}_j$  to $c_k$ in $b_k$ (including $\mathcal{F}_j$ and the subgraph containing $c_k$) is always separated by other subsets in the partition. For example in \figref{Fig:SubgraphSituations} (b) (when $b_{\sigma_s}$ is replaced by $b_{k}$), there are two cross nodes whose highest-weight nodes are involved in $\mathcal{F}_{x_l}$ and $\mathcal{F}_{y_{l'}}$, respectively. The subgraph $\mathcal{F}_j$ in \figref{Fig:SubgraphSituations} (b) with the partition
\bea
 \{...,A_j,A_{x_1},...,A_{x_{l-k}}, A_{y_1},...,A_{y_{l'-k'}},A_{x_{l-k+1}},...,A_{x_l},A_{y_{l'-k'+1}},...,A_{y_{l'}},...\}\Label{Eq:Partition2}
 \eea
 is a graph in this situation.

\end{itemize}

For  {\bf situation (i)}, we generate all possible subgraphs according to version-1 rule and then extract cross nodes according to the order $b_2$, ..., $b_{k-1}$, $b_{s}$, $b_{s-1}$, ..., $b_{k+1}$, $b_k$ in turn by version-2 rule. For {\bf situation (ii)}, we extract cross nodes $b_1$, $b_2$, ..., $b_s$ in turn. Similar with the treatment in approach-1, this approach also produces X- and BCJ-patterns. Thus the cancellations in \secref{sec:CancellationXBCJ} can also be applied while proper boundaries (pairs of spurious graphs) are further introduced.

\newpage

\begin{figure}[h]
\centering
\includegraphics[width=0.85\textwidth]{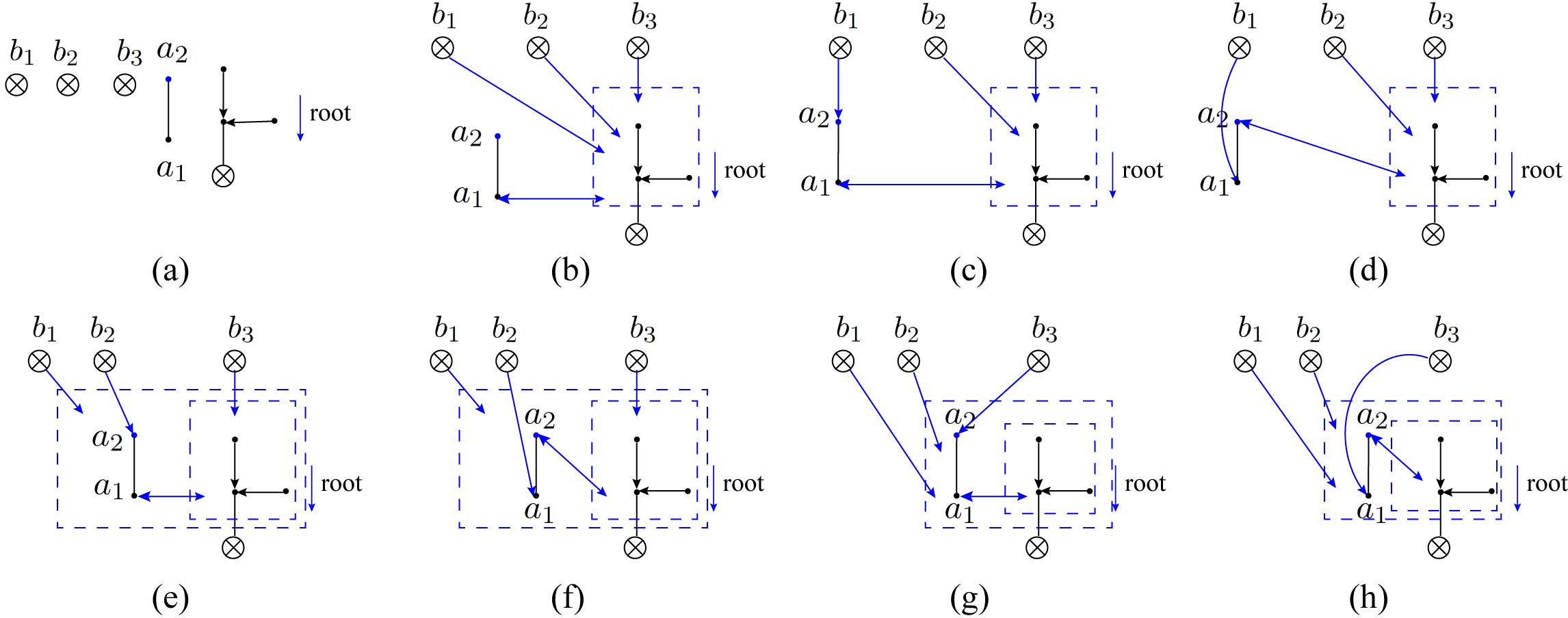}
\caption{Graph (a) present five components including three off-shell nodes $b_1$, $b_2$, $b_3$, the component consisting of two on-shell nodes $a_1$, $a_2$, and the component below the subgraph. The reference order of nodes is supposed to be $a_1\prec a_2\prec b_1\prec b_2\prec b_3$. Graphs (b)-(h) are all possible graphs constructed by these components, via version-1 rule.}
\label{Fig:Version1EG}
\end{figure}
\begin{figure}[h]
\centering
\includegraphics[width=1\textwidth]{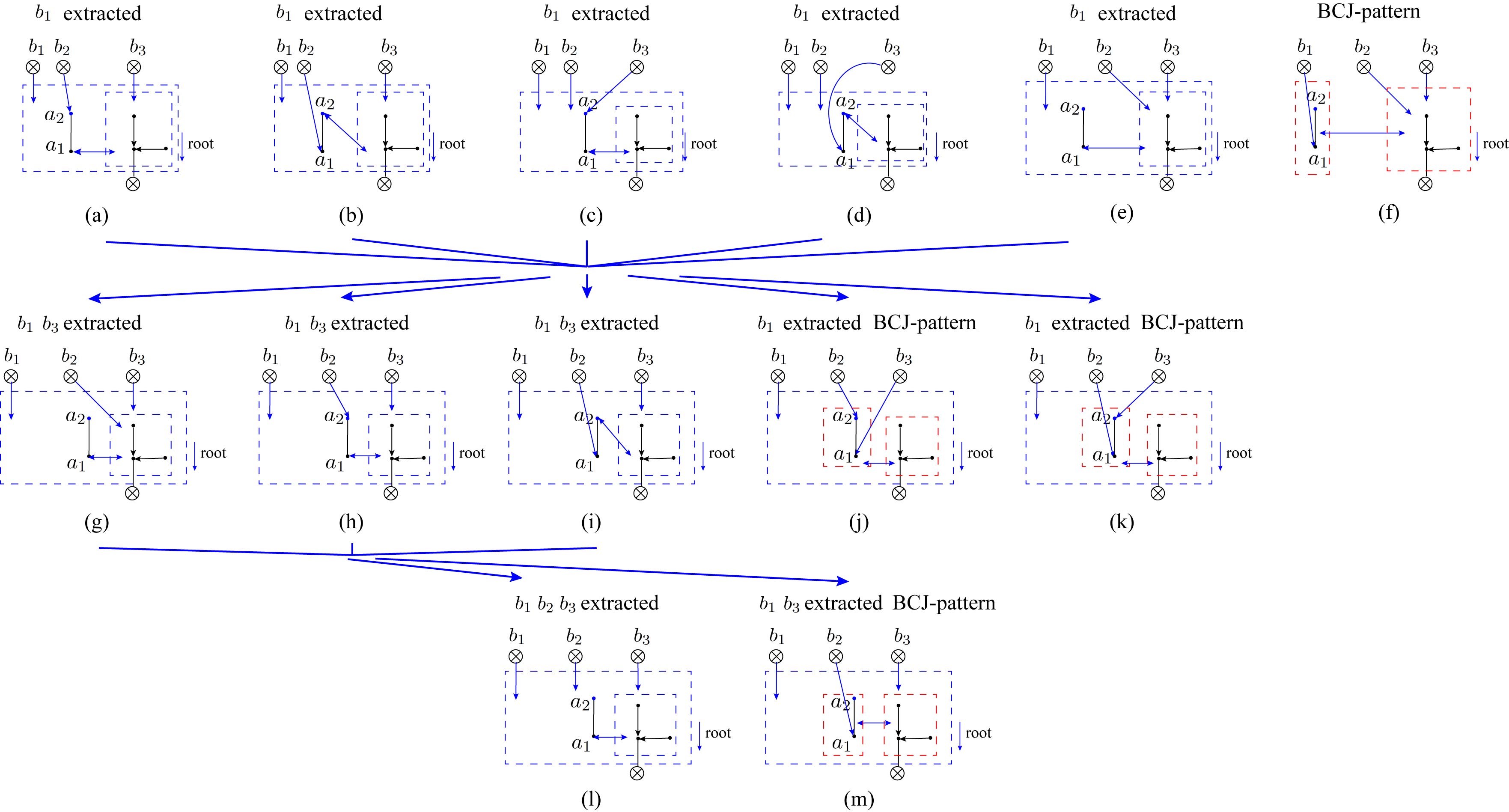}
\caption{Subgraphs constructed by combining version-1 and version-2 rules, using the components in \figref{Fig:Version1EG} (a) with the same reference order $a_1\prec a_2\prec b_1\prec b_2\prec b_3$. These graphs not only reproduce the physical graphs \figref{Fig:Version1EG}, but also introduce pairs of spurious graphs. Each of the final constructed subgraphs either contains BCJ-pattern or has all cross nodes extracted out.}
\label{Fig:Version2EG}
\end{figure}

\newpage

\begin{figure}[h]
\centering
\includegraphics[width=0.5\textwidth]{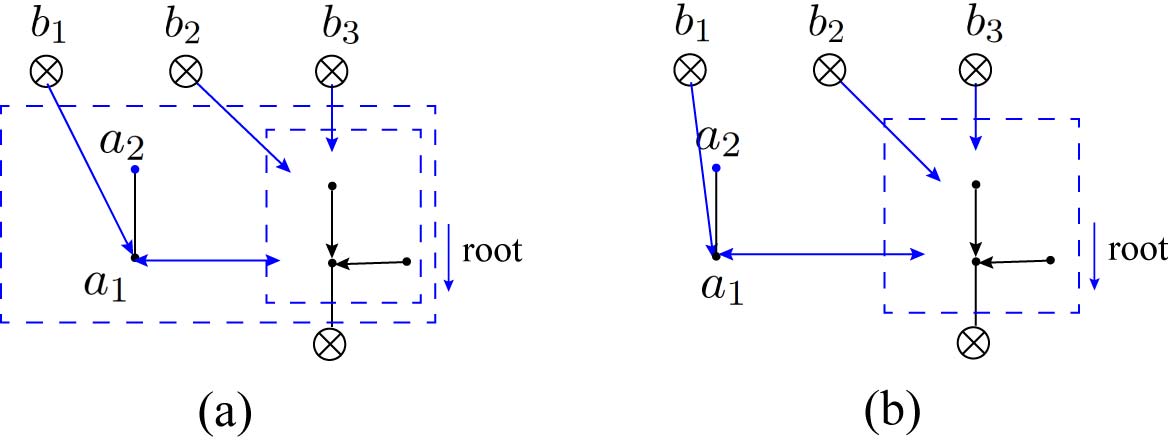}
\caption{Graphs (a) and (b) are the same spurious graph coming from \figref{Fig:Version2EG} (e) and (f), respectively. The sign for (b) is minus, while that for (a) is plus. Thus they cancel each other.}
\label{Fig:Version2EGa}
\end{figure}

\section{Explicit examples for the construction of subgraphs}\label{app:ExampleSubgraphs}

\figref{Fig:Version1EG} (b)-(h) present an example for graphs constructed by five components as shown by \figref{Fig:Version1EG} (a), via version-1 rule in \appref{sec:Subgraph1}. Graphs in \figref{Fig:Version2EG} are generated  by combining version-1 and -2 rules, with the same components in \figref{Fig:Version1EG} (a). These graphs can be considered as those constructed according to approach-1 in \appref{sec:Subgraph2} for the partition $\{...,A_j,...,A_{m_2},...,A_{m_3},...,A_{m_1},...\}$, where the highest-weight nodes in $b_1$, $b_2$, $b_3$ are respectively contained by $A_{m_1}$, $A_{m_2}$, $A_{m_3}$. On another hand, graphs in \figref{Fig:Version2EG} can be obtained by approach-2 when partition is given as $\{...,A_j,A_{j+1},...,A_{m_2},...,A_{m_1},...,A_{m_3},...\}$ or $\{...,A_j,A_{j+1},...,A_{m_2},...,A_{m_3},...,A_{m_1},...\}$. Here, all the subsets $A_{j+1}$, ..., $A_{m_2}$ correspond to the subgraphs which belong to $b_2$ and live on the path from $\mathcal{F}_{m_2}$ to $\mathcal{F}_j$. In \figref{Fig:Version2EG}, there are pairs of spurious graphs with opposite signs, e.g., \figref{Fig:Version2EGa} (a) and (b). When all pairs of spurious graphs are canceled out, graphs in \figref{Fig:Version2EG} become \figref{Fig:Version1EG} (b)-(h).

\newpage

\begin{figure}
\centering
\includegraphics[width=0.7\textwidth]{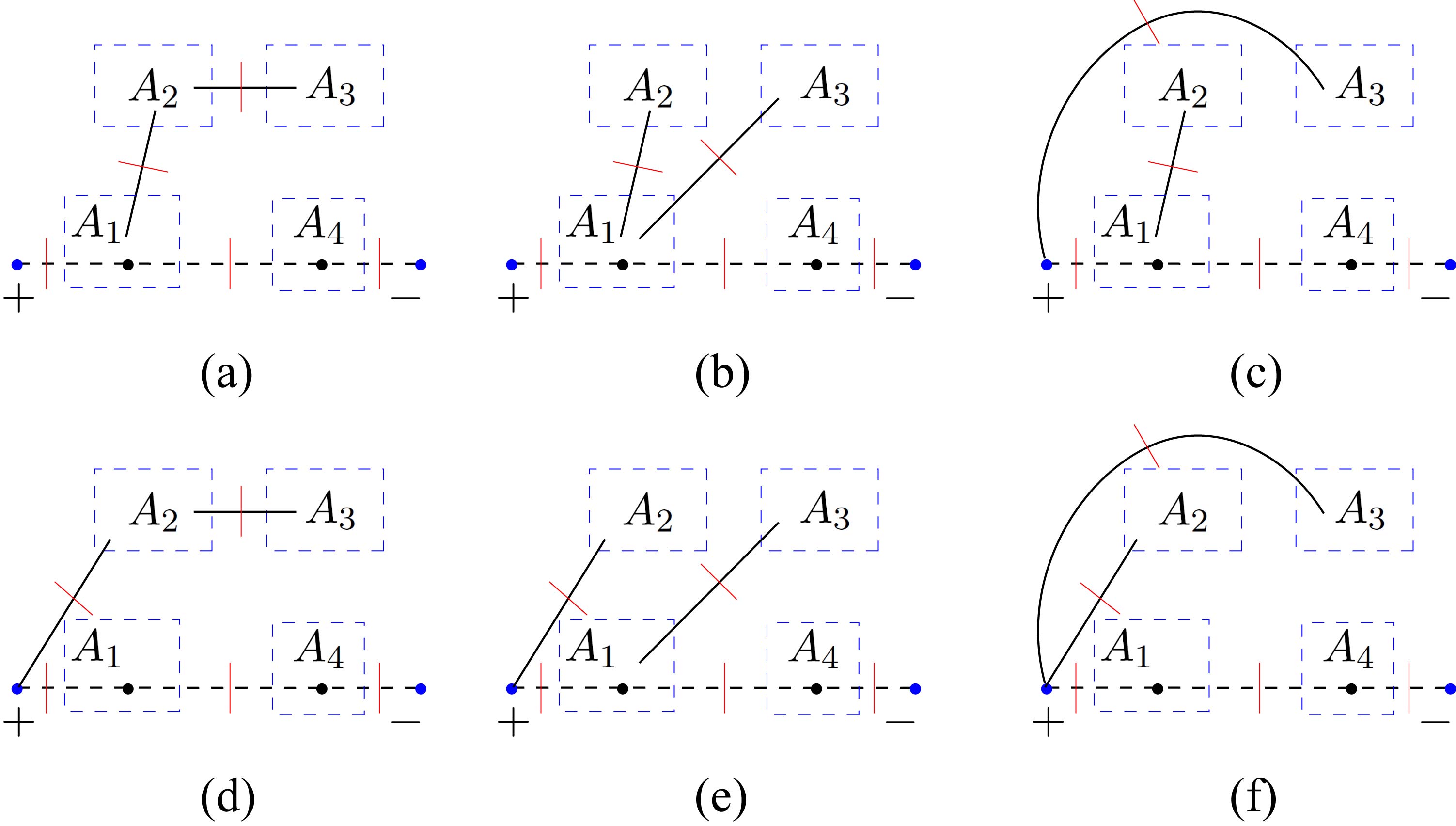}
\caption{All topologies for the partition $\{A_1,A_2,A_3,A_4\}=\{x_1,\{p,q\},\{r,s\},x_2\}$.}
\label{Fig:4GluonEgNewA}
\end{figure}

\begin{figure}
\centering
\includegraphics[width=0.6\textwidth]{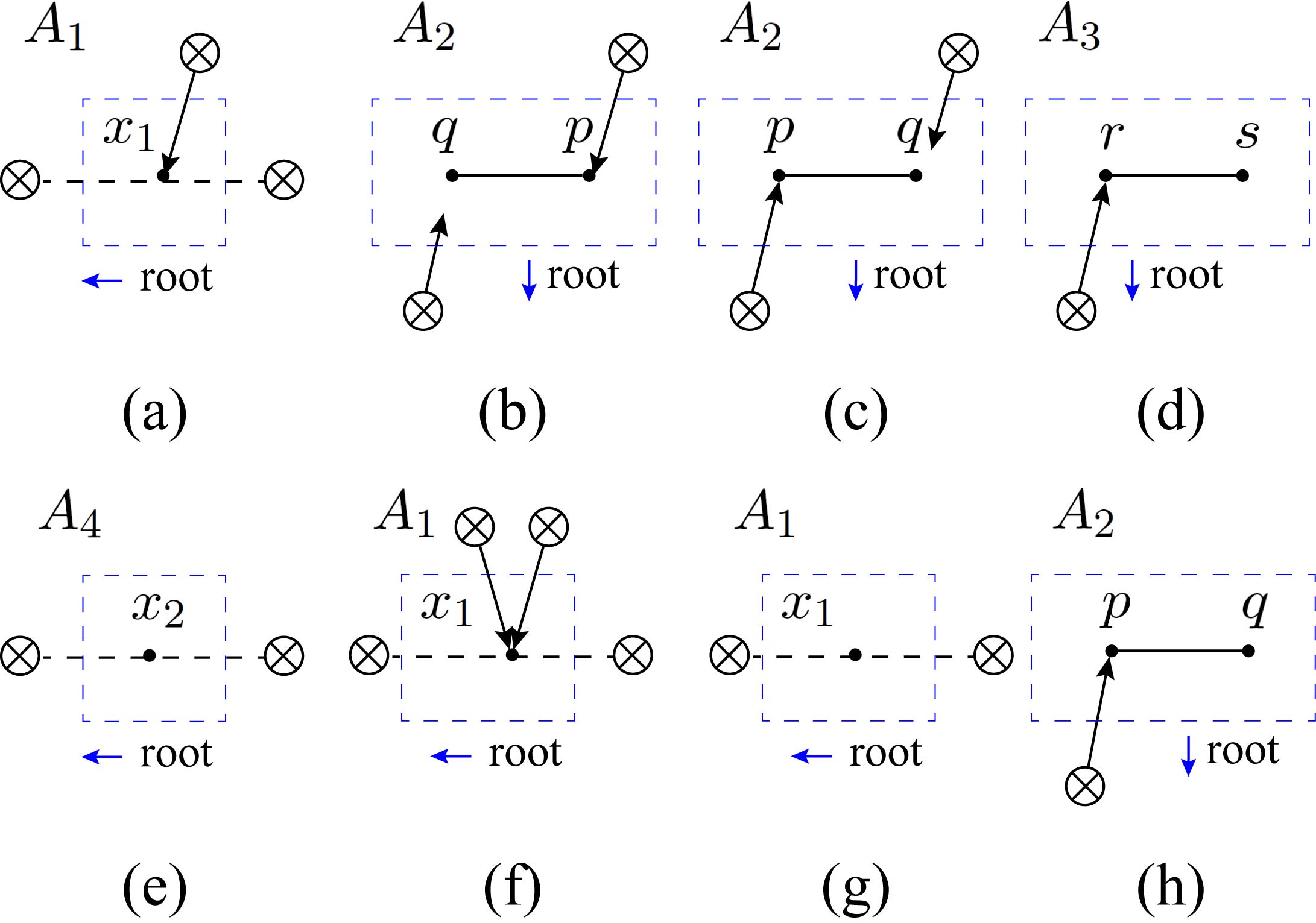}
\caption{All possible structures of subgraphs for topologies in \figref{Fig:4GluonEgNewA}, which contribute a factor $(\epsilon_p\cdot \epsilon_q)(\epsilon_r\cdot\epsilon_s)$. The reference order is assumed to be $p\prec q\prec r\prec s$. }
\label{Fig:4GluonEgNewB}
\end{figure}


\begin{figure}
\centering
\includegraphics[width=0.8\textwidth]{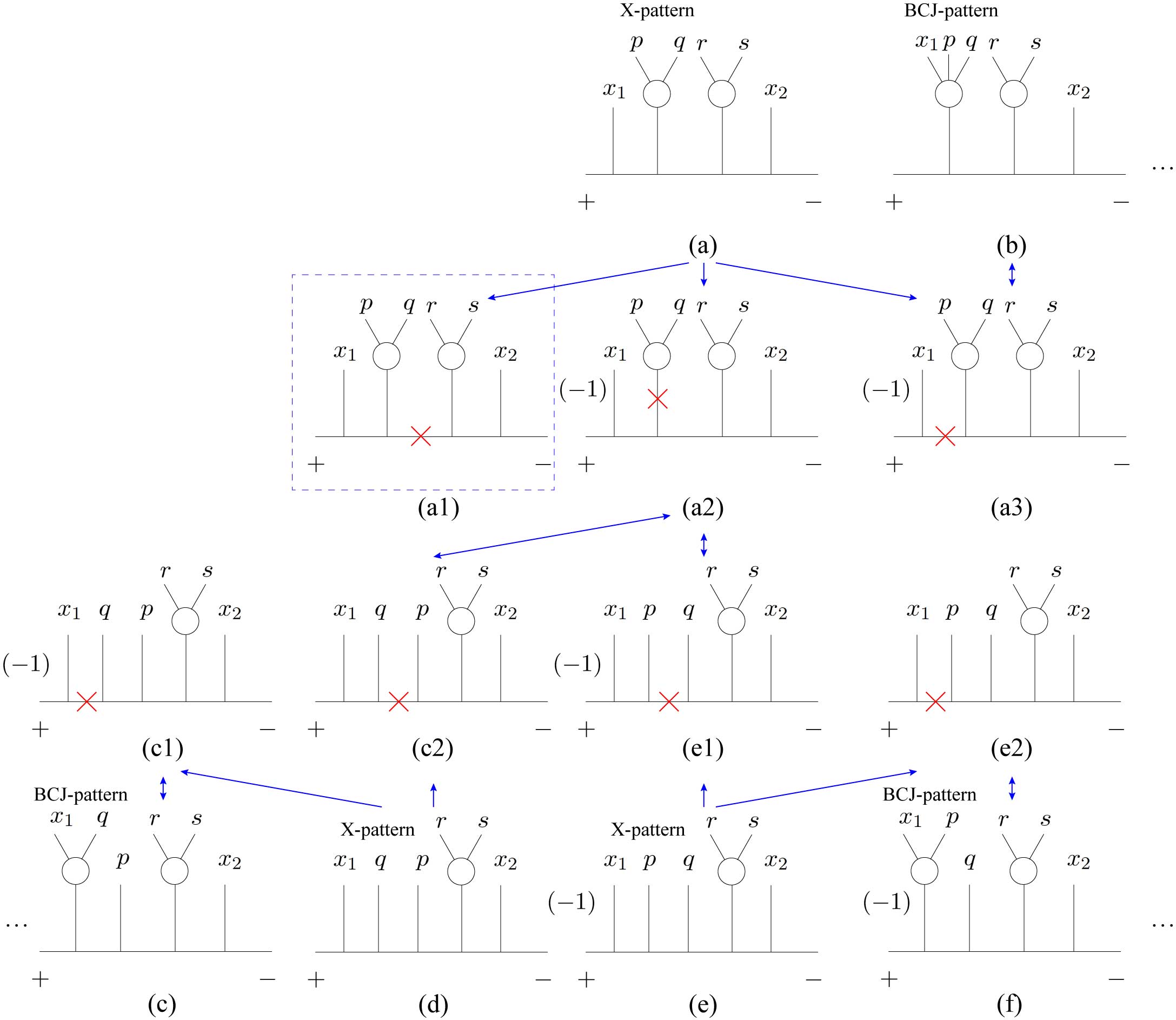}
\caption{Cancellation map related to the topology \figref{Fig:4GluonEgNewA} (a) when the subgraph of $A_2$ is chosen as \figref{Fig:4GluonEgNewB} (b). The `X-pattern' or `BCJ-pattern' for a diagram  means the graph accompanying to this diagram contains the corresponding pattern. The minus signs in (a2), (a3) and (c1) come from the decomposition of an  X-pattern, see \figref{Fig:4pt2glu9}. The minus in (e) and (f) are induced by spurious graph. The signs of (e1) and (e2) are resulted by both X-pattern and the spurious graph associating with (e).  }
\label{Fig:4GluonEgNewD}
\end{figure}

\section{Quadratic propagators from refined graphic rule: an example with four gluons}\label{sec:MoreComplicatedExample}
By an example with four gluons and two scalars, we show the cancellation between diagrams and the final quadratic propagator form,  using the general method in \secref{Sec:GenYMS}. We focus on the partition $\{x_1,\{p,q\},\{r,s\},x_2\}$ and only consider terms with a factor $(\epsilon_p\cdot\epsilon_q)(\epsilon_r\cdot\epsilon_s)$. All possible topologies for the partition $\{x_1,\{p,q\},\{r,s\},x_2\}$ are given by  \figref{Fig:4GluonEgNewA}, where $A_1=\{x_1\}$, $A_2=\{p,q\}$, $A_3=\{r,s\}$ and $A_4=\{x_2\}$. All structure of subgraphs are displayed in \figref{Fig:4GluonEgNewB}. The choice of  a subgraph for a subset relies on topology. For example, corresponding to the topology \figref{Fig:4GluonEgNewA} (a), subgraphs of $A_1$, $A_3$ and $A_4$ are shown by \figref{Fig:4GluonEgNewB} (a), (d) and (e), while the subgraph of $A_2$ has two distinct structures \figref{Fig:4GluonEgNewB} (b) and (c). Another example is given by the topology \figref{Fig:4GluonEgNewA} (b), possible subgraphs are \figref{Fig:4GluonEgNewB} (d), (e), (f), (h).

For the topology \figref{Fig:4GluonEgNewA} (a), if the subgraph for $A_2$ is chosen as \figref{Fig:4GluonEgNewB} (b), the cancellation map is shown by  \figref{Fig:4GluonEgNewD}. After cancellation,  the remaining diagram resulted by \figref{Fig:4GluonEgNewD} (a) is \figref{Fig:4GluonEgNewD} (a1), which is expressed as 
\bea
\phi_{x_1}\Big[{-{1\over 2}}(\epsilon_q\cdot\epsilon_p)(-k_p\cdot k_r)(\epsilon_r\cdot\epsilon_s)\phi_{qp}\phi_{rs}\Big]\phi_{x_2}{1\over l^2}{1\over s_{x_1,l}}{1\over s_{x_1pqrs,l}}.
\eea
When all cyclic permutations of $(x_1x_2)$ and cyclic permutations of all elements in the right half integrand are considered, graphic rule allows all cyclic permutations of $A_1,A_2,A_3,A_4$. Altogether provides a term with quadratic propagators
\bea
\phi_{x_1}\Big[{-{1\over 2}}(\epsilon_q\cdot\epsilon_p)(-k_p\cdot k_r)(\epsilon_r\cdot\epsilon_s)\phi_{qp}\phi_{rs}\Big]\phi_{x_2}{1\over l^2}{1\over l^2_{x_1}}{1\over l^2_{x_1pqrs}},
\eea
where the expression inside the squarebrackets contributes one term to $\left(-{1\over 2}\right)\W J(p,q)\cdot \W J(r,s)$, agrees with (\ref{Eq:GenResultNewJ2}). The $\phi_{x_1}$ and  $\phi_{x_2}$ are $\W J(x_1)$ and $\W J(x_2)$ which are the currents containing scalars (defined by  (\ref{Eq:GenResultNewJ1})).

When the subgraph of $A_2$ is chosen as \figref{Fig:4GluonEgNewB} (c), the topology  \figref{Fig:4GluonEgNewA} (a) and other  topologies \figref{Fig:4GluonEgNewA} (b) and (c) together provide
\bea
\phi_{x_1}\Big[(\epsilon_p\cdot\epsilon_q)(-k_p\cdot x_1)\phi_{pq}\Big]\Big[(\epsilon_r\cdot\epsilon_s)(-k_r\cdot X_r)\phi_{rs}\Big]\phi_{x_2}{1\over l^2}{1\over s_{x_1,l}}{1\over s_{x_1pq,l}}{1\over s_{x_1pqrs,l}},\Label{Eq:4GluonNew1}
\eea
where, $X_r^{\mu}=l^{\mu}+k_{x_1}^{\mu}+k_{p}^{\mu}+k_{q}^{\mu}$. We have used the fact that the subgraphs for the topology \figref{Fig:4GluonEgNewA} (b) are  \figref{Fig:4GluonEgNewB} (d), (e), (f), (h) and the subgraphs for \figref{Fig:4GluonEgNewA} (c) are \figref{Fig:4GluonEgNewB} (a), (d), (e), (h). Similarly, the topologies \figref{Fig:4GluonEgNewA} (d), (e) and (f) produce 
\bea
\phi_{x_1}\Big[(\epsilon_p\cdot\epsilon_q)(-k_p\cdot l)\phi_{pq}\Big]\Big[(\epsilon_r\cdot\epsilon_s)(-k_r\cdot X_r)\phi_{rs}\Big]\phi_{x_2}{1\over l^2}{1\over s_{x_1,l}}{1\over s_{x_1pq,l}}{1\over s_{x_1pqrs,l}}.\Label{Eq:4GluonNew2}
\eea
The sum of (\ref{Eq:4GluonNew1}) and (\ref{Eq:4GluonNew2}) gives rise 
\bea
\phi_{x_1}\Big[(\epsilon_p\cdot\epsilon_q)(-k_p\cdot X_p)\phi_{pq}\Big]\Big[(\epsilon_r\cdot\epsilon_s)(-k_r\cdot X_r)\phi_{rs}\Big]\phi_{x_2}{1\over l^2}{1\over s_{x_1,l}}{1\over s_{x_1pq,l}}{1\over s_{x_1pqrs,l}},\Label{Eq:4GluonNew3}
\eea
where $X_p^{\mu}=l^{\mu}+k_{x_1}^{\mu}$. After summing over cyclic permutations $A_1,A_2,A_3,A_4$, we get the quadratic propagator form
\bea
\W J(x_1) \Big[(\epsilon_p\cdot\epsilon_q)(-k_p\cdot X_p)\phi_{pq}\Big]\Big[(\epsilon_r\cdot\epsilon_s)(-k_r\cdot X_r)\phi_{rs}\Big]\W J(x_2) {1\over l^2}{1\over l^2_{x_1}}{1\over l^2_{x_1pq}}{1\over l^2_{x_1pqrs}}.\Label{Eq:4GluonNew4}
\eea
The expressions in the squarebrackets contribute to $\W J(p,q)\cdot X_{\{p,q\}}$, $\W J(r,s)\cdot X_{\{r,s\}}$, agree with (\ref{Eq:GenResultNewJ2}). The definition of subcurrent with scalars (\ref{Eq:GenResultNewJ1}) was applied again.

In the above, we only presented the remaining terms with $(\epsilon_p\cdot\epsilon_q)(\epsilon_r\cdot\epsilon_s)$ after cancellation, for the partition $\{x_1,\{p,q\},\{r,s\},x_2\}$ with topology \figref{Fig:4GluonEgNewA} (a). In fact, all surviving terms in the cancellations for all partitions and all topologies can be obtained in a similar way. Terms expressed by effective currents are further demonstrated in the next section.

\section{Expressions of terms in (\ref{Eq:GenResultNew}): demonstrated by explicit examples }\label{app:ExplicitFourGluons}

We display terms in (\ref{Eq:GenResultNew}) for the integrand $I^{\text{1-loop}}(x_1,x_2||\{p,q,r,s\}|x_1,x_2,p,q,r,s)$ with four gluons $p$, $q$, $r$, $s$ and two scalars $x_1$, $x_2$, by partitions with four to six subsets. All partitions with four to six subsets are given by
\bea
&&\{x_1,x_2,p,q,r,s\},~~~~~~\,\,\{x_1,\{x_2,p\},q,r,s\},~~~~~\{x_1,x_2,\{p,q\},r,s\},~~~~~\{x_1,x_2,p,\{q,r\},s\},\nn
&&\{x_1,x_2,p,q,\{r,s\}\},~~~\,\,\{\{s,x_1\},x_2,p,q,r\},~~~~~\{x_1,\{x_2,p,q\},r,s\},~~~~~\{x_1,x_2,\{p,q,r\},s\},\nn
&&\{x_1,x_2,p,\{q,r,s\}\},~~~\,\,\{\{r,s,x_1\},x_2,p,q\},~~~~~\{x_1,\{x_2,p\},\{q,r\},s\},~~\{x_1,\{x_2,p\},q,\{r,s\}\},\nn
&&\{\{s,x_1\},\{x_2,p\},q,r\},~\{x_1,x_2,\{p,q\},\{r,s\}\},~~\{\{s,x_1\},x_2,\{p,q\},r\},~~\{\{s,x_1\},x_2,p,\{q,r\}\},
\eea
where those partitions with $x_1,x_2$ in the same subset vanishes due to $U(1)$-decoupling identity, partitions related by cyclic permutations of subsets are equivalent to each other.

The expression accompanying to the partition $\{x_1,x_2,\{p,q,r\},s\}$ is explicitly given by
\bea
&&{1\over l^2}\W J(x_1){1\over l_{x_1}^2}\W J(x_2){1\over l_{x_1x_2}^2}\bigg\{\W J(p,q,r)\cdot l_{x_1x_2}\nn
&&~~~~~~~~~~~~~~~~~+\Big(-{1\over 2}\Big)\Big[\W J(p)\cdot \W J(q,r)+\W J(q)\cdot \W J(p,r)+\W J(p,q)\cdot \W J(r)\Big]\bigg\}{1\over l_{x_1x_2pqr}^2}\W J(s)\cdot l_{x_1x_2pqr},
\eea
in which, $\W J(x_1)=\phi_{x_1|x_1}$,  $\W J(x_2)=\phi_{x_2|x_2}$, $\W J^{\mu}(p)$, $\W J^{\mu}(q)$ and $\W J^{\mu}(r)$ are respectively given by $\epsilon_p^{\mu}\phi_{p|p}$, $\epsilon_q^{\mu}\phi_{q|q}$ and $\epsilon_r^{\mu}\phi_{r|r}$. The effective current $\W J^{\mu}(p,q,r)$ is given by (\ref{Eq:EffCurrentpqr}), while the currents $\W J^{\mu}(q,r)$, $\W J^{\mu}(p,r)$, $\W J^{\mu}(p,q)$ are expressed as
\bea
\W J^{\mu}(q,r)&=&(\epsilon_r\cdot F_q)^{\mu}\phi_{qr|qr}+(\epsilon_q\cdot k_r)\epsilon_r^{\mu}\phi_{rq|qr},\nn
\W J^{\mu}(p,r)&=&(\epsilon_r\cdot F_p)^{\mu}\phi_{pr|qr}+(\epsilon_p\cdot k_r)\epsilon_r^{\mu}\phi_{rp|qr},\nn
\W J^{\mu}(p,q)&=&(\epsilon_q\cdot F_p)^{\mu}\phi_{pq|pq}+(\epsilon_p\cdot k_q)\epsilon_q^{\mu}\phi_{qp|pq},\nonumber
\eea
where the current $\W J^{\mu}(p,r)$ finally vanishes since $\phi_{pr|qr}=\phi_{rp|qr}=0$.
The expression for the partition $\{x_1,\{x_2,p\},\{q,r\},s\}$ is written as 
\bea
&&{1\over l^2}\W J(x_1){1\over l_{x_1}^2}\W J(x_2,p){1\over l_{x_1x_2p}^2}\bigg[\W J(q,r)\cdot l_{x_1x_2p}+\Big(-{1\over 2}\Big)\W J(q)\cdot \W J(r)\bigg]{1\over l_{x_1x_2pqr}^2}\W J(s)\cdot l_{x_1x_2pqr},
\eea
where the current $\W J(x_2,p)$ is explicitly written as 
\bea
\W J(x_2,p)=(\epsilon_p\cdot k_{x_2})\phi_{x_2p|x_2p}.
\eea

\newpage

\begin{figure}
\centering
\includegraphics[width=0.75\textwidth]{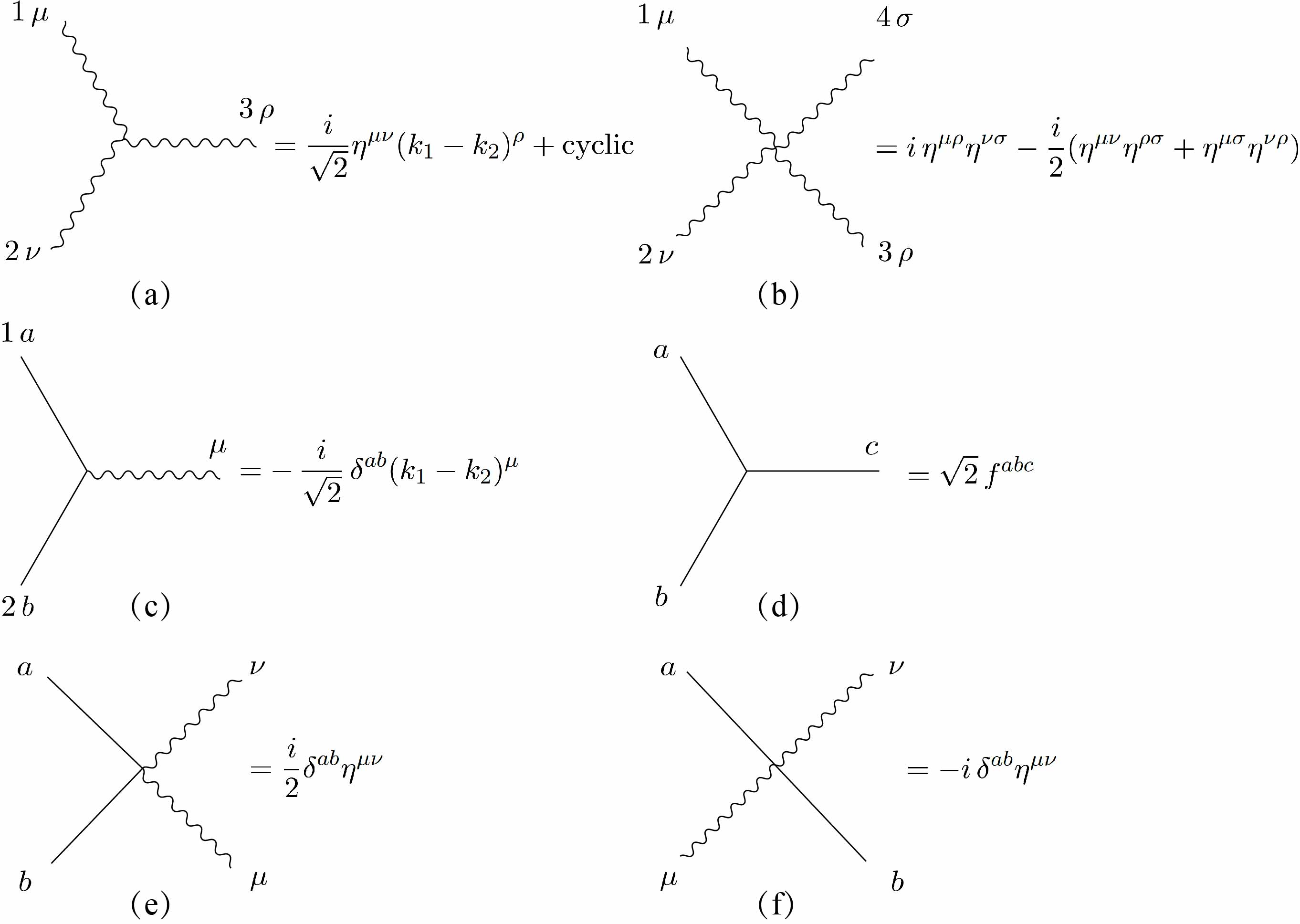}
\caption{Vertices in the Feynman rule for single-trace YMS amplitudes (see \cite{Bern:1999bx})}
\label{Fig:Feynman0}
\end{figure}

\section{Feynman rule in YMS}

The vertices for single-trace YMS amplitudes are given by \figref{Fig:Feynman0}. This Feynman rule evaluates the color-ordered YMS amplitude of gluons and color-scalars. When the color factors for color-scalars are further extracted out, the kinematic part becomes the YMS amplitudes discussed in this paper.

\section{KK relation for one-loop half integrands}\label{App:KKOneLoop}
As pointed in \cite{Porkert:2022efy}, there are three types of one-loop CHY half integrands $I_{\,\text{L}}^{\,\text{1-loop}}\big(\pmb{\sigma}_1;...;\pmb{\sigma}_m||\pmb{G}\big)$, which depend on different orders in the couplings $g$ and $\lambda$, in the YMS theory.  Specifically
\bea
I_{\,\text{L}}^{\,\text{1-loop}}\big(\pmb{\sigma}_1;...;\pmb{\sigma}_m||\pmb{G}\big)|_{g^n\lambda^{r-2m+4}}&=& \Sl_{1\leq i<j\leq m}\Sl_{\substack{Q\in \text{cyc}\,\pmb{\sigma}_i\\ R\in \text{cyc}\,\pmb{\sigma}_j}} I_{\,\text{L}}^{\text{tree}}(+,Q,-,R;\pmb{\sigma}_1;...;\cancel{\pmb{\sigma}_i};...;\cancel{\pmb{\sigma}_j};...;\pmb{\sigma}_m||\pmb{G})+(+\leftrightarrow-),\nn
I_{\,\text{L}}^{\,\text{1-loop}}\big(\pmb{\sigma}_1;...;\pmb{\sigma}_m||\pmb{G}\big)|_{g^n\lambda^{r-2m+2}}&=& N\Sl_{j=1}^m\,\Sl_{Q\in \text{cyc}\,\pmb{\sigma}_i}I_{\,\text{L}}^{\text{tree}}(+,Q,-;\pmb{\sigma}_1;...;\cancel{\pmb{\sigma}_i};...;\pmb{\sigma}_m||\pmb{G})+(+\leftrightarrow-),  \nn
I_{\,\text{L}}^{\,\text{1-loop}}\big(\pmb{\sigma}_1;...;\pmb{\sigma}_m||\pmb{G}\big)|_{g^n\lambda^{r-2m}}&= &N\,c_2\,I_{\,\text{L}}^{\text{tree}}(\pmb{\sigma}_1;...;\pmb{\sigma}_m;+,-||\pmb{G})+I_{\,\text{L}}^{\text{tree}}(\pmb{\sigma}_1;...;\pmb{\sigma}_m||\pmb{G},+,-),\nn
&&~~~+\Sl_{j=1}^m\Bigl\{\Sl_{\substack{\pmb{\sigma}_j=QR\\ |Q|,|R|\neq 0}}I_{\,\text{L}}^{\text{tree}}(Q,+;R,-;\pmb{\sigma}_1;...;\cancel{\pmb{\sigma}_j};...;\pmb{\sigma}_m||\pmb{G})+\text{cyc}\,\,\pmb{\sigma}_j\Bigr\},\Label{Eq:mTraceSector}
\eea
where $N$ refered to the $U(N)$ or $SU(N)$ group for the color scalar, while the coefficient $c_2$ is $N$ for $U(N)$ group and ${{N^2-1\over N}}$ for $SU(N)$ group. The sets $Q$ and $R$ are two ordered sets of scalars, the set $\pmb{G}$ denotes the gluon set, and $\cancel{\pmb{\sigma}_i}$ represents that  the $i$-th trace $\pmb{\sigma}_i$ is removed. The number of total particles and scalars are supposed to be $n$ and $r$, respectively. In the above expression, the RHS of the first and the second equations as well as the first term on the RHS of the third equation correspond to the case that there are only scalar propagators on the loop. On the contrary, the second and the third terms on the RHS of the third equation contribute to loops with at least one gluon propagator. 

The one-loop YMS integrands (in both single- and multi-trace cases) with quadratic propagators, which come from the forward limit of the tree-level amplitudes whose particles $+$ and $-$ are in the same trace and adjacent to each other, are directly obtained from (\ref{Eq:GenResultNew}) and the discussions in \secref{sec:FeynDiagrams}. Now we show that the tree-level CHY half integrands $I_{\,\text{L}}^{\,\text{tree}}(+,Q,-,R;...||\pmb{G})$ with particles $+$ and $-$ separated by the scalars belonging to the same trace, can be rewritten as 
\bea
&&\Sl_{\substack{Q\in \text{cyc}\,\pmb{\sigma}_i\\ R\in \text{cyc}\,\pmb{\sigma}_j}} I_{\,\text{L}}^{\text{tree}}(+,Q,-,R;\pmb{\sigma}_1;...;\cancel{\pmb{\sigma}_i};...;\cancel{\pmb{\sigma}_j};...;\pmb{\sigma}_m||\pmb{G})\nn
&=&\Sl_{\text{cyc}\,\pmb{\rho}}\,\Sl_{R^{\text{T}}\in \text{cyc}\,\pmb{\sigma}^{\text{T}}_j}(-1)^{|\pmb{\sigma}_j|} I_{\,\text{L}}^{\text{tree}}(+,Q_1,\{Q_2,...\}\shuffle R^{\text{T}},-;\pmb{\sigma}_1;...;\cancel{\pmb{\sigma}_i};...;\cancel{\pmb{\sigma}_j};...;\pmb{\sigma}_m||\pmb{G})\,,\Label{Eq:OneLoopKK}
\eea
where the ordered sets $\pmb{\sigma}_i$ and $\pmb{\rho}$ in the above summations are defined by $\pmb{\sigma}_i=Q_1,Q_2,...$, and $\pmb{\rho}=Q_1,\{Q_2,...\}\shuffle R^{\text{T}}$. The $|\pmb{\sigma}_j|$, $\pmb{\sigma}^{\text{T}}_j$ denote the number of elements and the inverse of the permutation $\pmb{\sigma}_j$, respectively. To prove \eqref{Eq:OneLoopKK}, we arrange the tree-level PT factors as
\bea
\Sl_{\substack{ \text{cyc}\,\pmb{\alpha}\\  \text{cyc}\,\pmb{\beta}}}{\text{PT}}(+,\pmb{\alpha},-,\pmb{\beta})&&=\Sl_{\substack{ \text{cyc}\,\pmb{\alpha}\\  \text{cyc}\,\pmb{\beta}^{\text{T}}}}(-1)^{|\pmb{\beta}|}\,{\text{PT}}(+,\pmb{\alpha}\shuffle\pmb{\beta}^{\text{T}},-)\nn
&&=\Sl_{\text{cyc}\,\pmb{\rho}}\Sl_{\text{cyc}\,\pmb{\beta}^{\text{T}}}(-1)^{|\pmb{\beta}|}\,{\text{PT}}(+,\alpha_1,\{\alpha_2,...,\alpha_s\}\shuffle\pmb{\beta}^{\text{T}},-)\,,
\eea
where $\pmb{\alpha}=\alpha_1,\alpha_2,...,\alpha_s$, $\pmb{\rho}=\alpha_1,\{\alpha_2,...,\alpha_s\}\shuffle\pmb{\beta}^{\text{T}}$. The first equality is the KK relation, while the second  can be checked by comparing the permutations on both sides of this equation. For any permutation on the LHS, one can always find it on the RHS of the second equality. Both sides of this equation have the  number of permutations ${(s+t)!\over (s-1)!(t-1)!}$, where $s$ and $t$ are the number of elements of $\pmb{\alpha}$ and $\pmb{\beta}$. For example, we choose $\pmb{\alpha}=1,2$ and $\pmb{\beta}=3,4$, then both the LHS of the first line and the RHS of the second line have the following $24$ permutations: 1234, 2341, 3412, 4123, 1324, 3241, 2413, 4132, 1342, 3421, 4213, 2134, 1243, 2431, 4312, 3124, 1423, 4231, 2314, 3142, 1432, 4321, 3214, 2143. This observation can be straightforwardly extended to general permutations, therefore \eqref{Eq:OneLoopKK} is proven.

This relation of PT factors can be considered as the CHY version of the one-loop KK relation (the one for YM amplitudes was proposed in \cite{DelDuca:1999rs})
\bea
A^{\text{1-loop}}(\pmb{\alpha};\pmb{\beta})=\Sl_{\pmb{\rho} \in\text{COP}(\pmb{\alpha}\,\cup\,\pmb{\beta}^{\text{T}})}(-1)^{|\pmb{\beta}|}\,A^{\text{1-loop}}(\pmb{\rho})\,,
\eea
where $A^{\text{1-loop}}(\pmb{\alpha};\pmb{\beta})$ refers to a one-loop multi-trace amplitude  and $A^{\text{1-loop}}(\pmb{\rho})$ denote a one-loop amplitude with  one less trace. The $\text{COP}(\pmb{\alpha}\,\cup\,\pmb{\beta}^{\text{T}})$  denotes the summation of all possible permutations that keep $\pmb{\alpha}$, $\pmb{\beta}^{\text{T}}$ in their own cyclic order.

\appendix

\bibliographystyle{JHEP}
\bibliography{reference.bib}

\end{document}